\documentclass{aa} 
\usepackage{times,amssymb,amsmath} 
\usepackage{epsfig} 

\title{
The Building the Bridge survey for z=3 Ly$\alpha$ emitting galaxies~I: \\
method and first results
\thanks{Based on observations made with ESO Telescopes at the
Paranal Observatory under programmes ID 67.A-0033 and 69.A-0380}
}

\author{
	J.P.U. Fynbo \inst{1,2,3}
        \and C. Ledoux \inst{4}
	\and P. M\o ller \inst{3} 
	\and B. Thomsen \inst{1}
        \and I. Burud \inst{5}
			}

\institute{
	   Department of Physics and Astronomy,
	   University of {\AA}rhus,
	   Ny Munkegade, DK--8000 {\AA}rhus C, Denmark
           \and
	   Astronomical Observatory, University of Copenhagen,
	   Juliane Maries Vej 30, DK-2100 Copenhagen {\O},  Denmark
           \and
           European Southern Observatory,  
           Karl-Schwarzschild-Stra\ss e 2,
	   D-85748, Garching by M\"unchen, Germany
           \and
           European Southern Observatory, Casilla 19001, Santiago 19,
	   Chile
	   \and
	   Space Telescope Science Institute, 3700 San Martin Drive,
           Baltimore, MD21218, U.S.A.
           }
\offprints{jfynbo@phys.au.dk}
\mail{jfynbo@phys.au.dk}

\date{Received  / Accepted }

\begin{document} 

\titlerunning{Building the bridge between DLAs and LBGs} 

\abstract{We present the first results of an observational programme at
the ESO Very Large Telescope aimed at detecting a large sample
of high-redshift galaxies fainter than the current spectroscopic limit of
R=25.5 for Lyman-Break galaxies. In this paper, we describe the results of
deep narrow and broad-band imaging and subsequent follow-up multi-object
spectroscopy of faint high-redshift galaxies in the fields of the QSOs
BRI\,1346$-$0322 and Q\,2138$-$4427. These QSOs have intervening high neutral
hydrogen column density absorbers, at redshifts z=2.85 and z=3.15
respectively, for which redshifted Ly$\alpha$ emission falls within less than
a few \AA\ from the central wavelengths of existing VLT ($\sim$60 \AA -wide)
narrow-band filters. We selected 37 and 27 candidate emission-line galaxies in
the two fields respectively. Most ($\sim$85\%) of the candidates have R-band
magnitudes fainter than R=25.5. The first spectroscopic follow-up of a
sub-sample of the candidates resulted in 41 confirmed candidates and 4
foreground emission line galaxies (three [\ion{O}{ii}] emitters and one
\ion{C}{iv} emitter). The confirmation rate for Ly$\alpha$ emitters is
82\% and 68\% in the field of BRI\,1346$-$0322 and Q\,2138$-$4427 respectively.
In addition, we serendipitously detect a number of other
emission-line sources on some of the slitlets not used for candidates. Of
these, 9 are also most likely Ly$\alpha$ emitters with redshifts ranging from
1.98 to 3.47. The redshift distribution of confirmed candidates in the field
of BRI\,1346$-$0322 is consistent with being drawn from a uniform distribution
weighted by the filter response curve, whereas the galaxies in the field of
Q\,2138$-$4427 have redshifts clustering very close to the redshift of the
damped Ly$\alpha$ absorber. This latter fact indicates the existence of a
large `pancake'-like structure confirming the earlier suggestions of
Francis \& Hewitt (1993).
}

\maketitle

\section{Introduction}

Our knowledge of the properties of high-redshift galaxies (here z$\approx$3)
currently mainly comes from two very different kinds of studies,
namely, {\it i)} the study of Lyman-Break Galaxies (LBGs) selected by the
Lyman-limit break in their spectrum (e.g. Steidel et al. 1996, 2000;
Fontana et al. 2000; Papovich et al. 2001; Vanzella et al. 2002),
and {\it ii)} the study of the chemical and kinematical properties of
(proto)galaxies, so called Damped Ly$\alpha$ Absorbers (DLAs), intervening
the lines of sight to QSOs (e.g. Wolfe et al. 1986; Pettini et al. 1997;
Prochaska \& Wolfe 2002; Ledoux et al. 2002). However, there is strong
evidence that there is only a small  overlap between the galaxies in 
the current LBG samples and the DLAs (Fynbo, M\o ller \& Warren 1999;
Haehnelt et al. 2000; Schaye 2001; M{\o}ller et al. 2002;
Adelberger et al. 2003). The reason for this is that Lyman-Break galaxy
samples are continuum-flux limited and that the current flux limit
corresponding to R$\approx$25.5 is not deep enough to reach the level of
typical damped Ly$\alpha$ absorption selected galaxies. However, due to
the very steep faint-end slope of the z=3 galaxy luminosity function
(Adelberger \& Steidel 2000; Poli et al. 2001a, 2001b) $\sim$70\% of the
observer rest-frame R-band flux from z=3 galaxies is emitted by galaxies
fainter than R=25.5. Therefore, besides accounting for the bulk of the damped
Ly$\alpha$ absorption in QSO spectra, the R$>$25.5 galaxies could dominate 
the integrated star-formation rate and the metal enrichment and heating of 
the intergalactic medium at z$\approx$3.

In 2000, we started the programme ``Building the Bridge
between Damped Ly$\alpha$ Absorbers and Lyman-break Galaxies: Ly$\alpha$
Selection of Galaxies'' at the European Southern Observatory's Very Large
Telescope (VLT). This project aims at bridging the gap between
absorption- and emission-line selected galaxy populations by creating a
database of z$\approx$3 galaxies that are fainter than R=25.5, and to study
the properties, such as morphology, star-formation rates, clustering and
continuum colours, of these faint, very numerous, and so far little studied
high-redshift galaxies.

Our method is to obtain deep narrow-band Ly$\alpha$ observations of the fields
of z$\approx$3 QSO absorbers whose redshifts match existing VLT narrow-band
filters. We chose to observe the fields of QSO absorbers to anchor our fields
to already known structures at the target redshift and, hence, minimise the
risk of observing a void. Ly$\alpha$ narrow-band imaging can, with
comparatively short integration times, lead to the detection of a significant
sample of z$\approx$3 galaxies having much fainter continuum fluxes than LBGs,
as demonstrated by, e.g., Cowie et al. (1998), Kudritzki et al. (2000), and
Fynbo et al. (2000, 2001, 2002). In the following, we shall use the acronym
LEGO for Ly$\alpha$-Emitting Galaxy-building Objects (M\o ller \& Fynbo 2001)
to refer to the Ly$\alpha$ emitters. In this first paper, we describe the
results of deep narrow and broad-band imaging and subsequent follow-up
multi-object spectroscopy of the fields of the QSOs BRI\,1346$-$0322 and
Q\,2138$-$4427. These QSOs have intervening high neutral hydrogen column
density Ly$\alpha$ absorbers at redshifts z=3.15 
($\log(N(\ion{H}{i}))\sim19.9$) 
and z=2.85 ($\log(N(\ion{H}{i}))=20.9$) respectively
(Francis \& Hewett 1993; Storrie-Lombardi et al. 1996). The paper is organized
in the following way. First, in Sect. 2 we describe the imaging and the 
selection of LEGO candidates. Then, in Sect. 3 we describe the observations and
results from the first spectroscopic run. Finally, in Sect. 4. we discuss our 
findings. Throughout this paper we assume a cosmology with
$H_0$=65 km s$^{-1}$ Mpc$^{-1}$, $\Omega_m$=0.3 and $\Omega_\Lambda$=0.7.
In this model, a redshift of 3.15(2.85) corresponds to a luminosity
distance $d_{lum}$ = 29.05(25.76) Gpc and a distance modulus of 47.31(47.05).
One arcsecond on the sky corresponds to 8.18(8.42) proper kpc and the
look-back time is 12.3(12.1) Gyr (roughly 85\% of the time since the event
commonly referred to as the Big Bang).

\section{Imaging}
\subsection{Observations and data reduction}
\label{obs}
\begin{figure}
\begin{center}
\epsfig{file=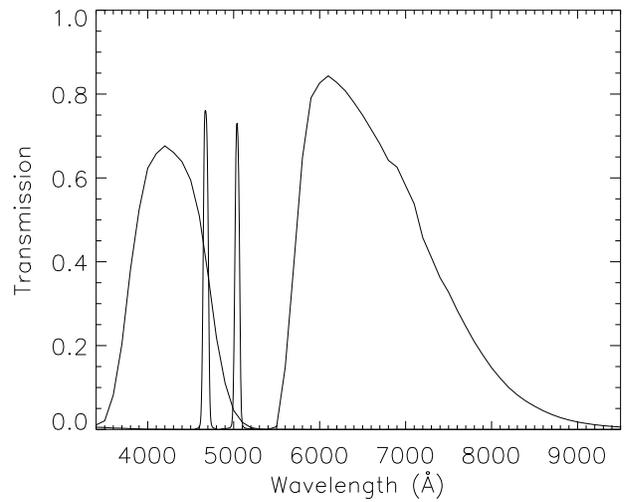,width=8.5cm}
\caption{Transmission curves of the two narrow-band and the broad-band Bessel
B and R filters used in this study.}
\label{filters}
\end{center}
\end{figure}

The fields of BRI\,1346$-$0322 and Q\,2138$-$4427 were observed in service
mode at the VLT 8.2 m telescopes, units Antu and Melipal, during the months of
May to September 2001 and May to June 2002 respectively. The wavelength of
Ly$\alpha$ at the redshift of the z=3.15 absorber in the spectrum of
BRI\,1346$-$0322 corresponds to the central wavelength of a 59 \AA -wide
[\ion{O}{iii}] off-band VLT filter. Similarly, the wavelength of Ly$\alpha$
at the redshift of the z=2.85 absorber in the spectrum of Q\,2138$-$4427 is
2.5 \AA\ off from the central wavelength of a 66 \AA -wide \ion{He}{ii} VLT
filter. Both fields were observed in one of the two narrow-band filters
and also in the Bessel B and R broad-band filters. The transmission curve of
each of the filters is shown in Fig.~\ref{filters}. The integration times in
the B and R filters were set by the criterion that the broad on-band B imaging
should reach about half a magnitude deeper than the narrow-band imaging
for objects having a pure continuum in the narrow filter so as to get
a reliable selection of objects with {\it excess} emission in the narrow
filter. For the broad off-band R imaging, we reached a magnitude deeper
(at the 5$\sigma$ significance level) than the spectroscopic limit of
R(AB)=25.5 for LBGs. The total integration times, the seeing (FWHM) of the
combined images and the 5$\sigma$ detection limits are given for each filter
and both fields in Table~\ref{journal}.

\begin{table}
\begin{center}
\caption{Log of imaging observations with FORS1. The 5$\sigma$ detection limit
was measured in a circular aperture of $3\arcsec$ diameter and is given in the
AB system.}
\begin{tabular}{@{}llllccc}
\hline
filter & total exp. & combined seeing & 5$\sigma$ detection\\
       & time (hr)  & ($\arcsec$)     & limit              \\
\hline
{\it BRI\,1346$-$0322} \\
narrow &  8.9    & 0.93  & 26.0 \\
B      &  2.5    & 1.02  & 26.5 \\
R      &  1.7    & 0.94  & 26.2 \\
{\it Q\,2138$-$4427} \\
narrow &  10.0   &  0.96  & 26.5 \\
B      &  2.5    &  1.04  & 26.9 \\
R      &  1.7    &  0.93  & 26.5 \\
\hline
\label{journal}
\end{tabular}
\end{center}
\end{table}

The images were reduced (de-biased and corrected for CCD pixel-to-pixel
variations) using the FORS1 pipeline (Grosb{\o}l et al. 1999). The
individual reduced images in each filter were combined using a code that
optimizes the Signal-to-Noise (S/N) ratio for faint, sky-dominated sources
(see M\o ller \& Warren 1993 for details on this code).

\begin{figure*}
\begin{center}
\epsfig{file=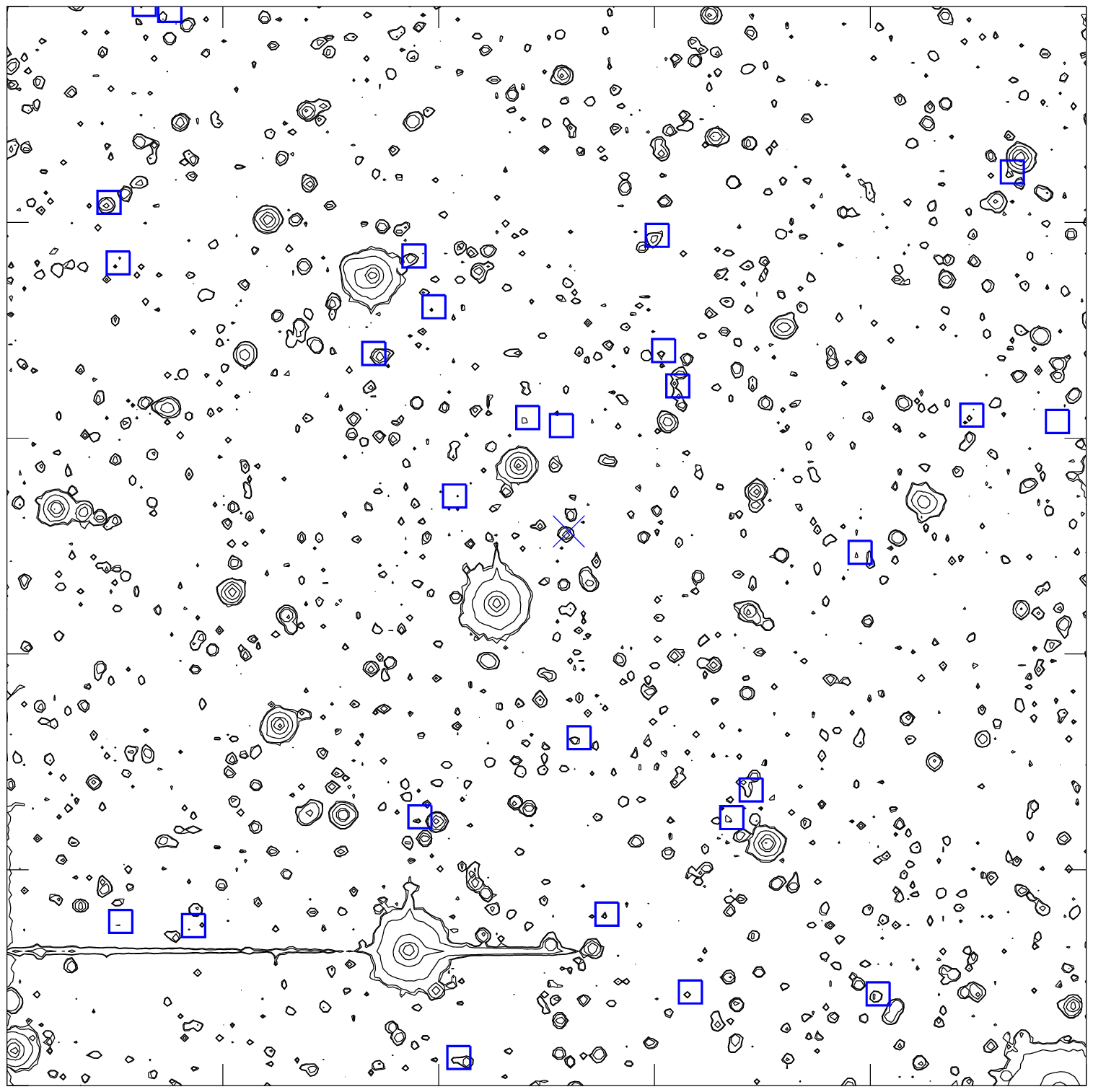,width=8.8cm}
\epsfig{file=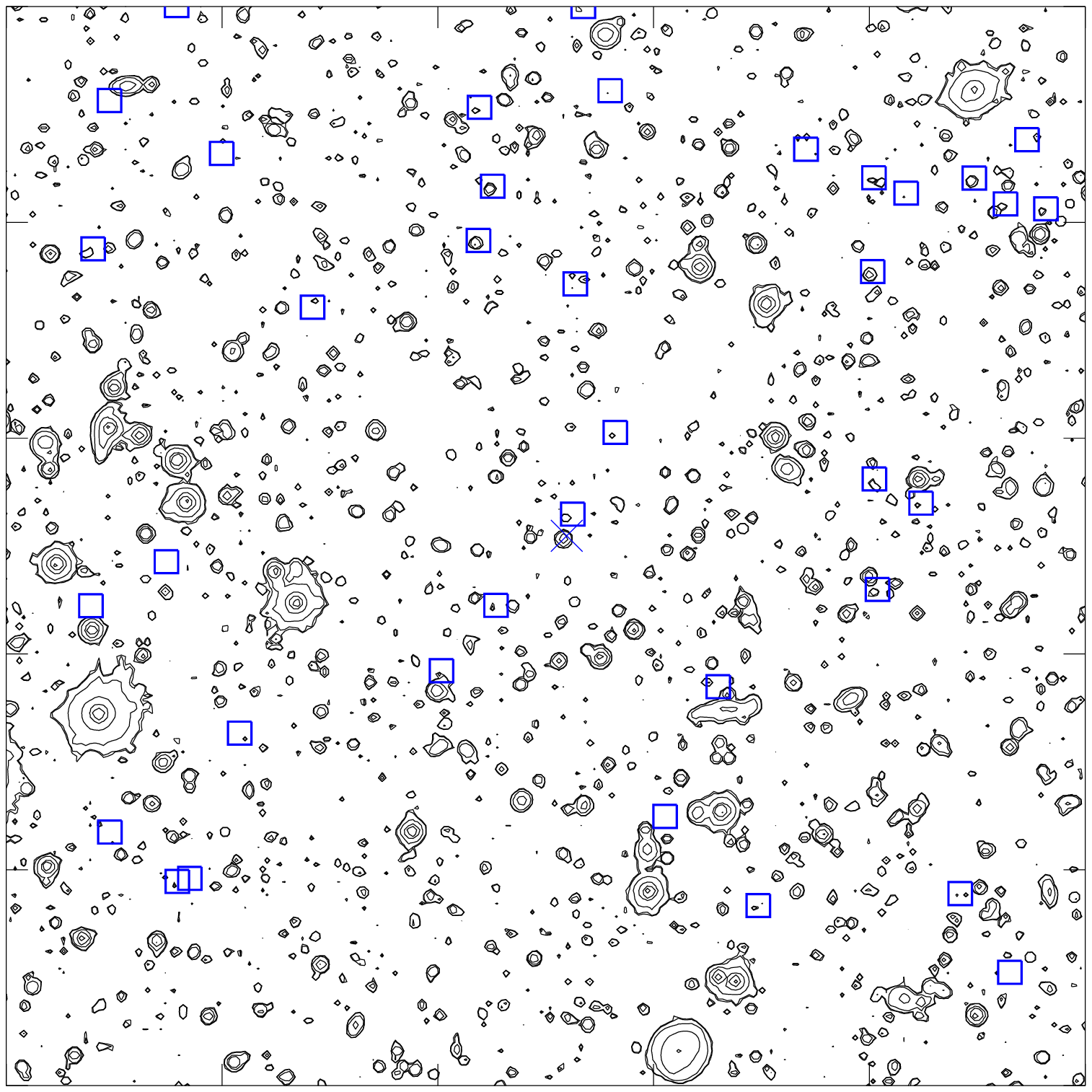,width=8.8cm}
\caption{The 400$\times$400 arcsec$^2$ fields surrounding the QSOs
BRI\,1346$-$0322 (left panel) and Q\,2138$-$4427 (right panel) as observed in
the narrow-band filters. Both QSOs are identified by an ``$\times$'' at the
field centre and the positions of selected LEGO candidates (see
Sect.~\ref{LEGOs}) are shown with boxes.}
\label{field}
\end{center}
\end{figure*}

The broad-band images were calibrated as part of the FORS1 calibration plan
via observations of Landolt standard stars (Landolt 1992). As the data were
collected over a time span of months, we have data from several photometric
nights which we used to determine independent zeropoints for the combined
images. These zero-points are consistent with each other within 0.02 mag. We
transformed the zero-points to the AB system using the relations given by
Fukugita et al. (1995): B(AB)=B$-0.14$ and R(AB)=R$+0.17$. The Q\,2138$-$4427
images are deeper than the BRI\,1346$-$0322 ones by nearly half a magnitude as
the BRI\,1346$-$0322 data were obtained at higher airmass and with slightly
larger lunar illumination than the Q\,2138$-$4427 data. For the calibration of
the narrow-band images, we used observations of the spectrophotometric
standard stars LTT6248 and LDS749-B for, respectively, the BRI\,1346$-$0322
and Q\,2138$-$4427 fields.

Contour images of the combined narrow-band images of the 400$\times$400
arcsec$^2$ fields surrounding the QSOs BRI\,1346$-$0322 and Q\,2138$-$4427 are
shown in Fig.~\ref{field}. Both QSOs are identified by an ``$\times$'' at the
field centre and the positions of selected LEGO candidates (see
Sect.~\ref{LEGOs}) are shown with boxes.

\subsubsection{Selection of LEGOs in the fields}
\label{LEGOs}
For object detection and photometry, we used the software package SExtractor
(Bertin \& Arnouts 1996). As a detection image, we used the weighted sum
of the combined narrow- and B-band images, with 80\% weight for the
narrow-band image and 20\% weight for the B-band one to secure an
optimal selection of objects with excess in the narrow filter. However, the
selection was not very sensitive to the exact weights given to the 
narrow-band and B-band images. Before object detection we convolved the 
detection image 
with a Gaussian filter function having a FWHM equal to that of point sources. 
We used a detection threshold
of 1.1 times the background sky-noise in the unfiltered detection image and a
minimum area of 5 connected pixels above the detection threshold in the
filtered image. Isophotal
apertures were defined on the detection image and those same isophotal
apertures are used in the different bands (narrow, B and R). In our final
catalogue, we include only objects with total S/N$>$5 in the isophotal
aperture in either the narrow- or the B-band images. In total, we detect 1993
and 2499 such objects within the 400$\times$400 arcsec$^2$ fields around
BRI\,1346$-$0322 and Q\,2138$-$4427 respectively. We measured the flux of all
objects in the isophotal aperture (for precise colour measurement) and in an
ellipsoidal aperture with minor axis $b=5.0 r_1 \epsilon$ and major axis
$a=5.0 {r_1 \over \epsilon}$, where $r_1$ is the first moment of the light
distribution and $\epsilon$ is the ellipticity (to get a measure of the total 
magnitudes). As a minimum aperture radius,
we used $1.5\arcsec$. The fluxes measured in the large aperture are used to
measure the total magnitudes of the objects. We also derived error bars on
colours as described in Fynbo et al. (2002).

\begin{figure*}
\begin{center}
\epsfig{file=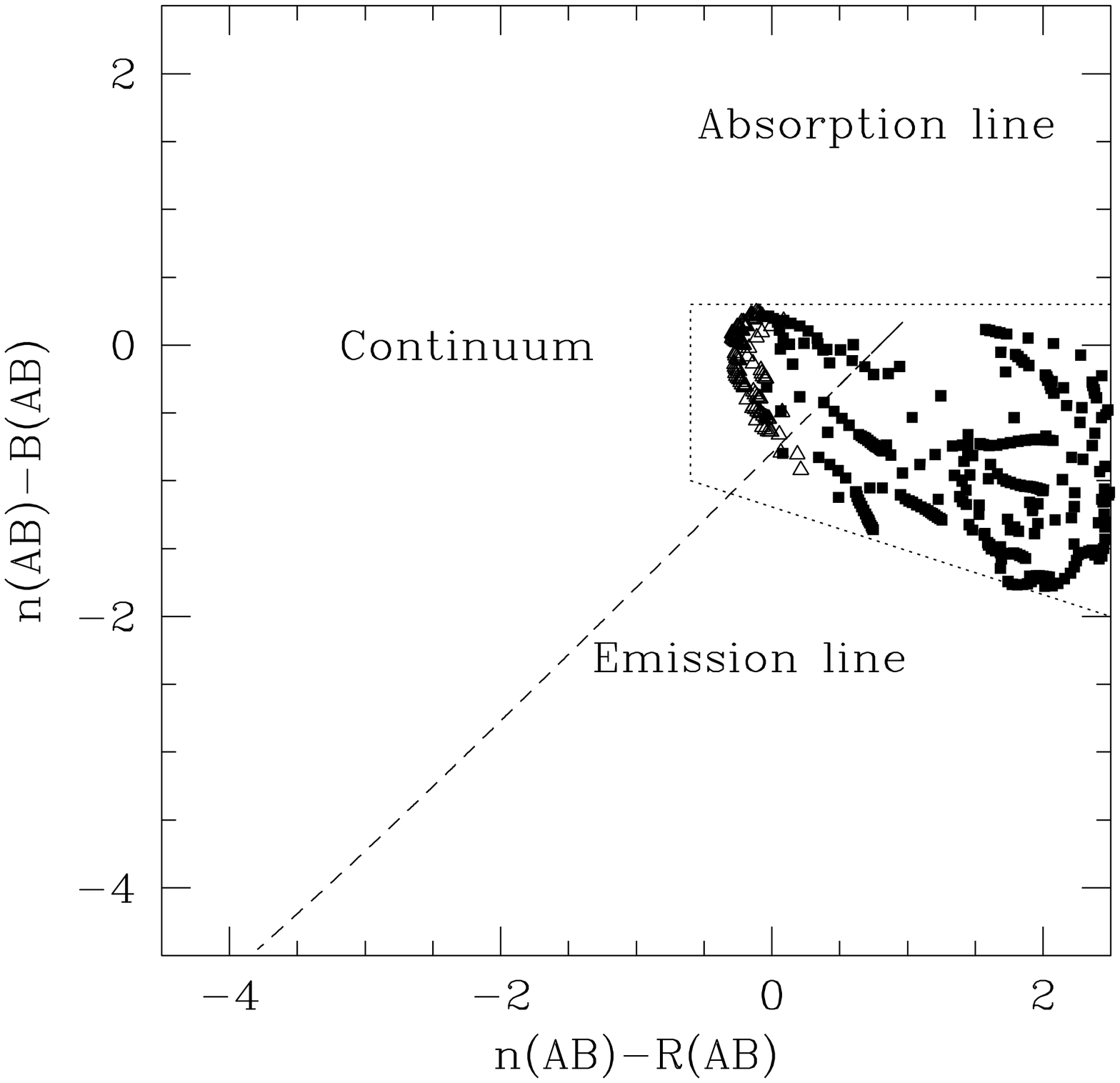,width=5.7cm}
\epsfig{file=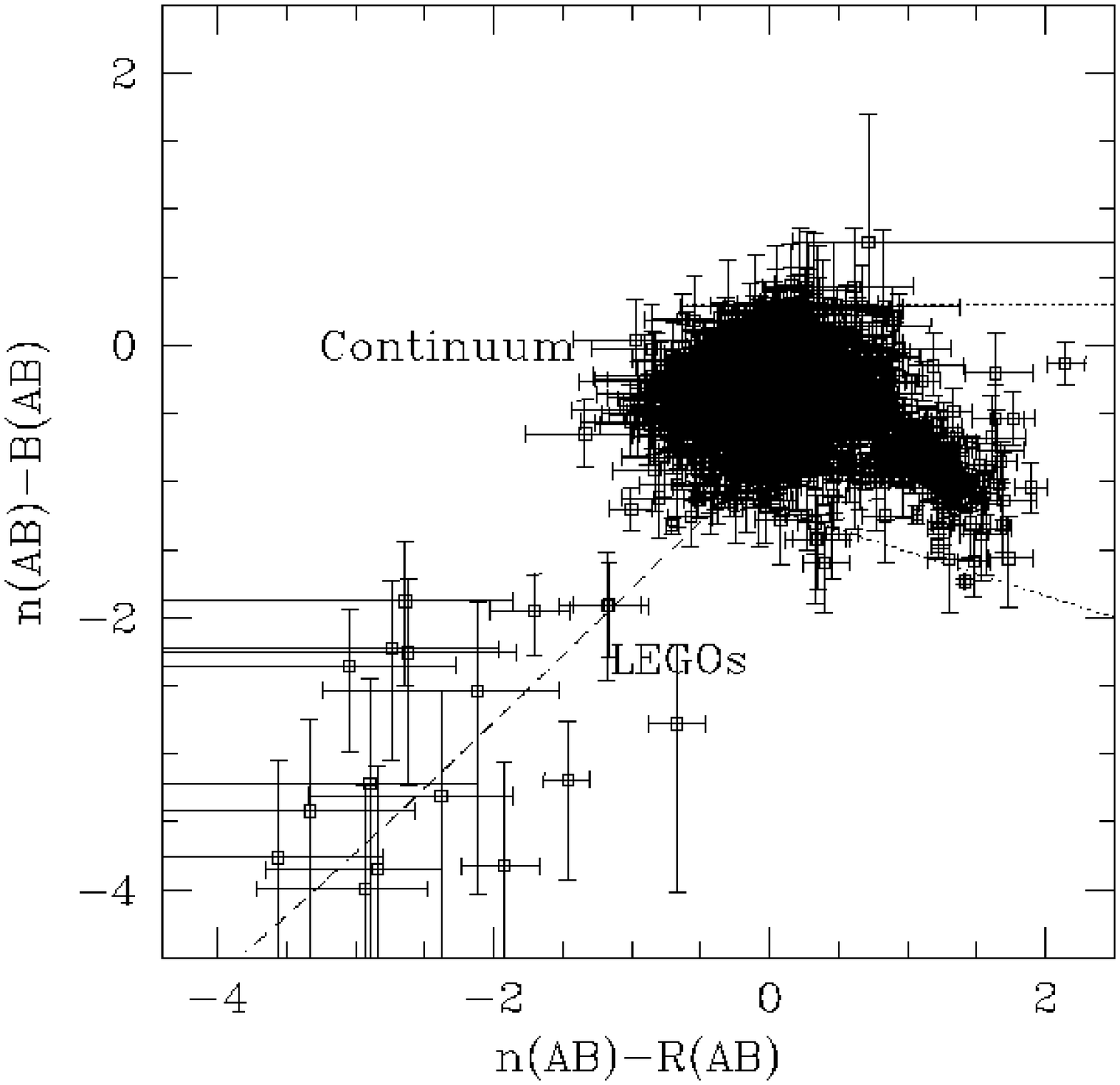,width=5.7cm}
\epsfig{file=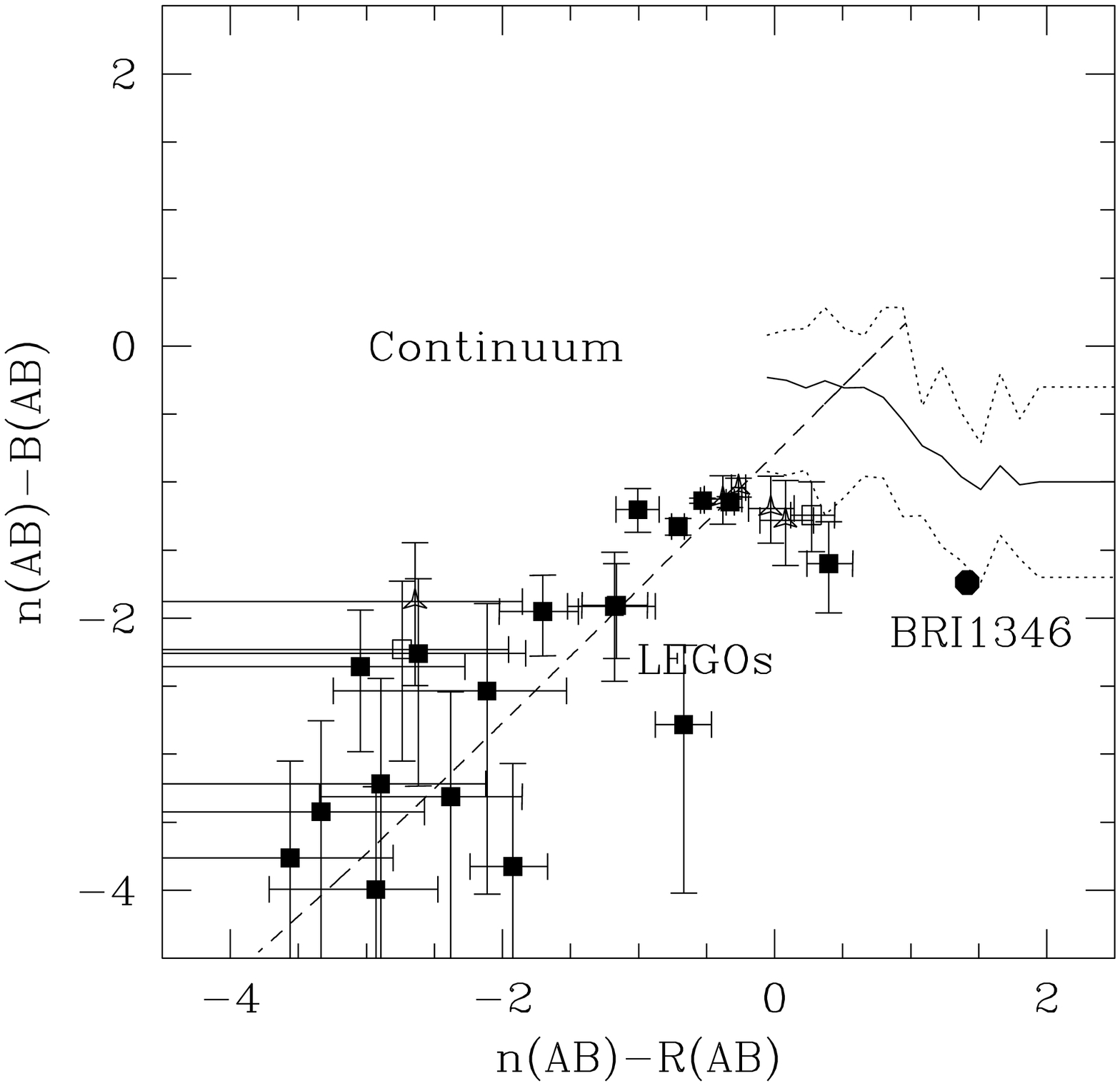,width=5.7cm}\\
\epsfig{file=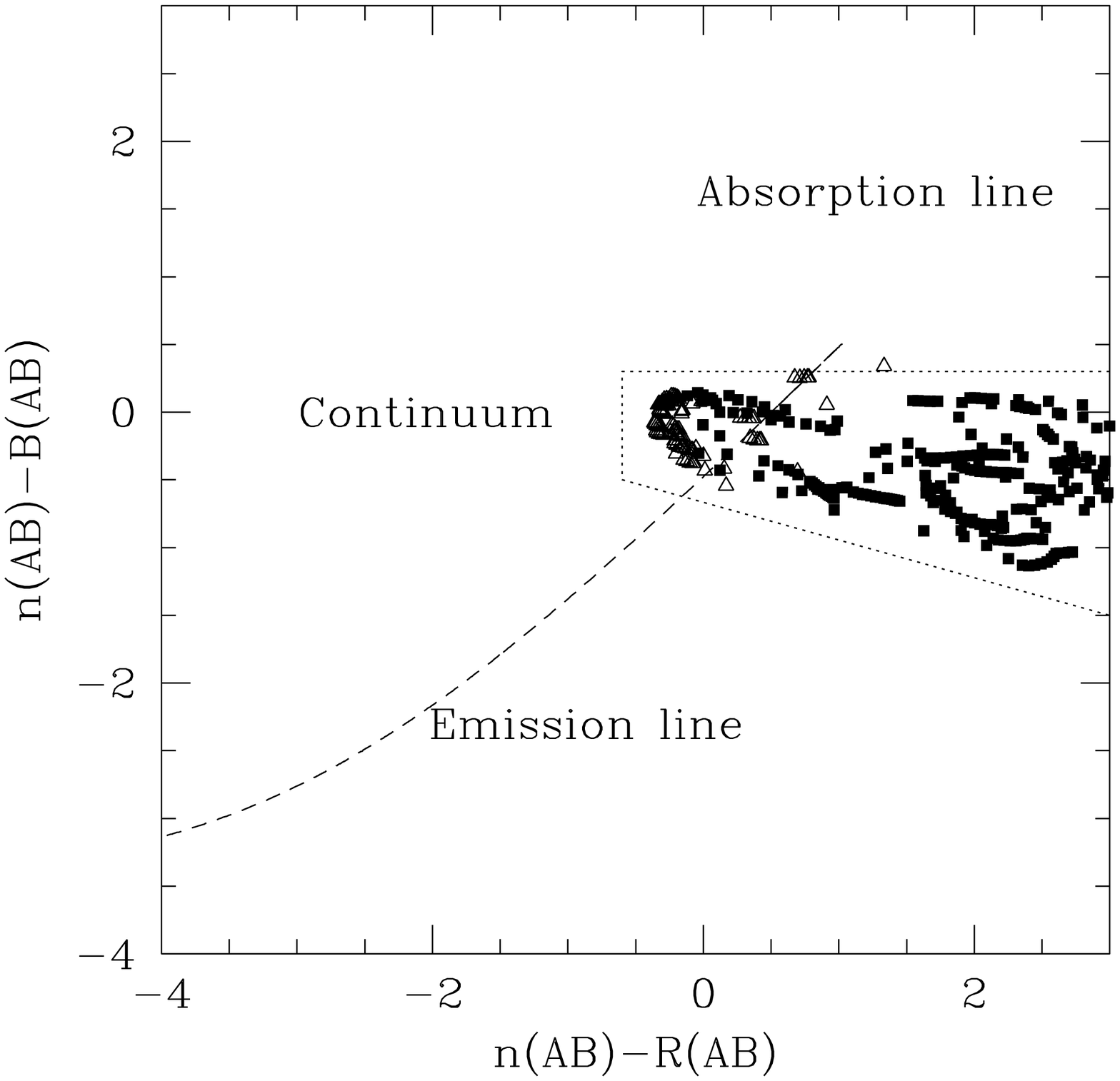,width=5.7cm}
\epsfig{file=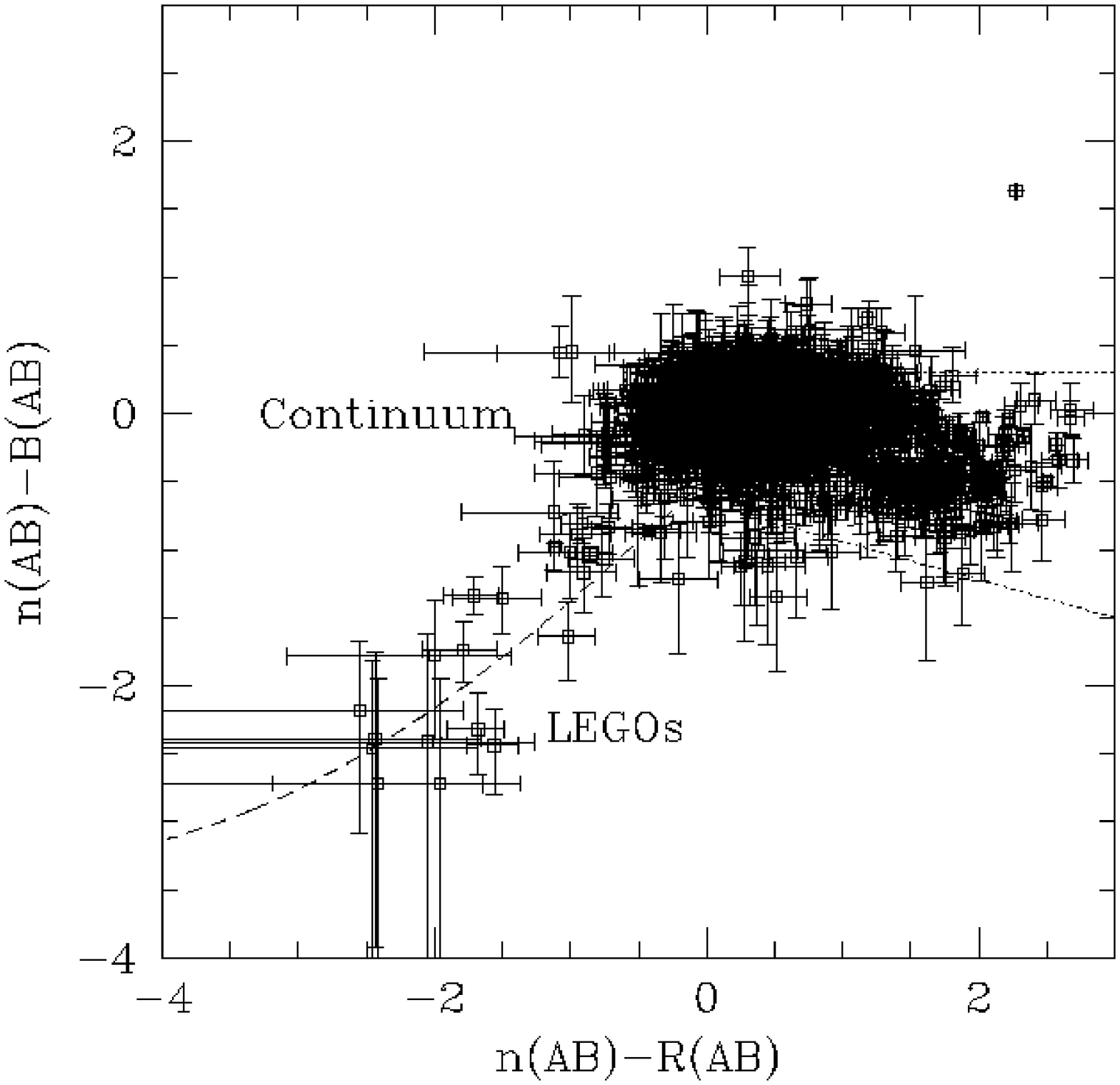,width=5.7cm}
\epsfig{file=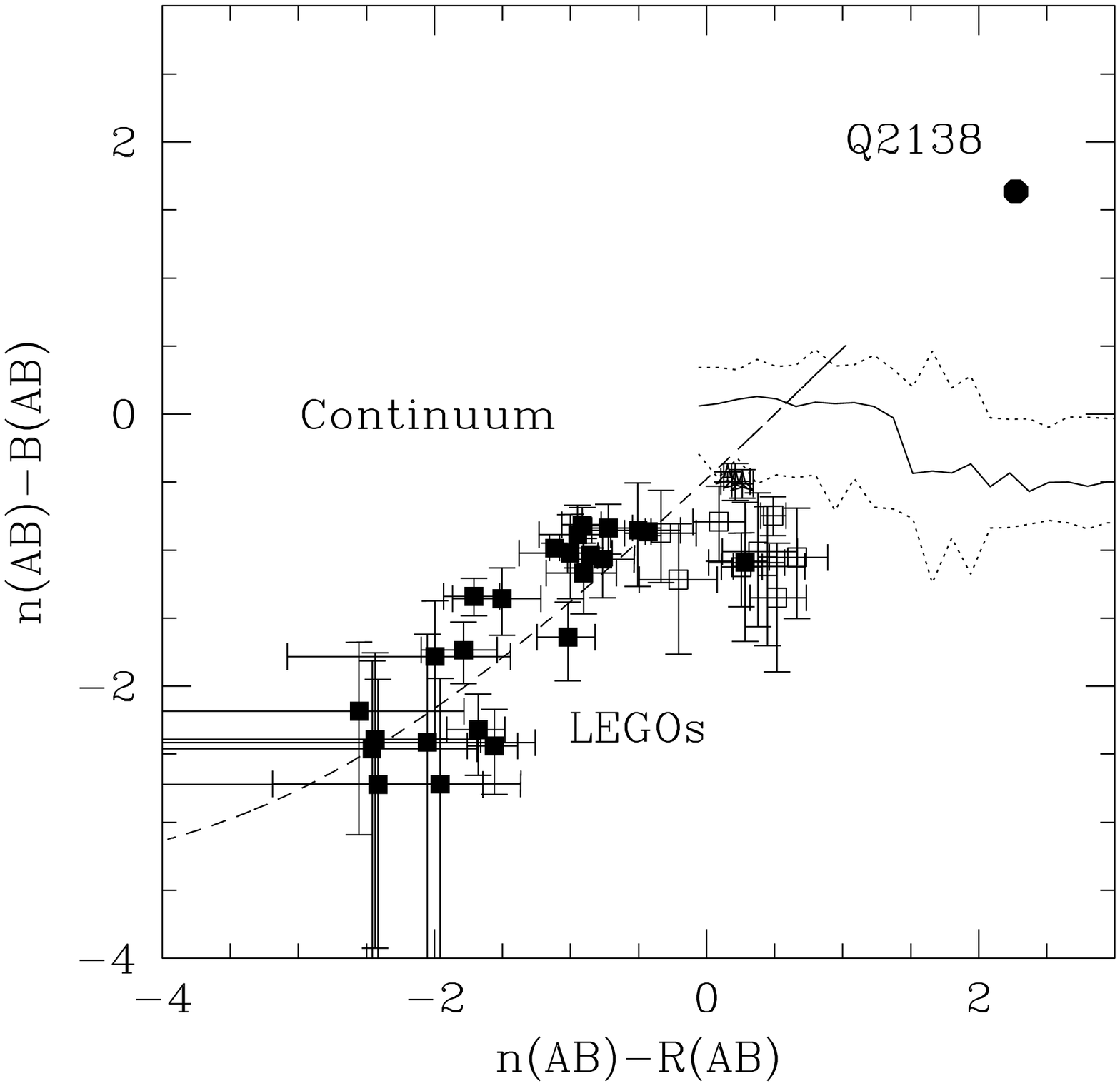,width=5.7cm}
\caption{The top three panels concern the BRI\,1346$-$0322 field and the
lower three panels the Q\,2138$-$4427 field. {\it Left panels:} Simulated
Colour-colour diagram based on Bruzual \& Charlot galaxy
SEDs. The filled squares are 0$<$z$<$1.5 galaxies with ages from a
few to 15 Gyr and the open triangles are 1.5$<$z$<$3.0 galaxies
with ages from a few Myr to 1Gyr. The dotted box encloses the
simulated galaxy colours. The dashed line indicates colours of
objects with a particular broad-band colour and with an SED in
the narrow-band filter ranging from an absorption line in the upper
right part to an strong emission line in the lower left part of the
diagram. {\it Middle panels:}
Colour-colour diagrams for all objects detected at S/N$>$5 in either
the narrow band or the B band. As expected, most objects have colours
consistent with being in the dotted box. Due to the damped Ly$\alpha$
line Q\,2138$-$4427 has a large positive n(AB)$-$B(AB) colour.
BRI\,1346$-$0322 has
a lower n(AB)$-$B(AB) colour due to Ly$\alpha$ blanketing affecting the
observed B-band flux.  In the lower
left part of the diagram there is a large number of objects with excess
emission in the narrow filter. Candidate LEGOs are those objects with a
1$\sigma$ upper limit on n(AB)$-$B(AB) below the 98\% percentile.
{\it Right panel:}
The colours of the QSOs and of the LEGO candidates. Candidates confirmed
to be emission line objects based on our spectroscopic observations
described in Sect.~\ref{MOS} are shown with filled symbols. Un-confirmed
candidates are shown with open circles and un-observed candidates are
shown with open triangles.
The solid line indicates the median n(AB)$-$B(AB) colour as a function
of n(AB)$-$R(AB) and the dotted lines indicate the 2\% and 98\%
percentiles in the n(AB)$-$B(AB) colour.
}
\label{colcol}
\end{center}
\end{figure*}

For the final selection of LEGO candidates, we used the ``narrow minus
on-band broad'' versus ``narrow minus off-band broad'' colour/colour
plot technique (M{\o}ller \& Warren 1993; Fynbo et al. 1999, 2000, 2002 and
Fig.~\ref{colcol}). In order to constrain where objects with no special
spectral features in the narrow filter are in the diagram, we calculated
colours based on synthetic galaxy SEDs taken from the
Bruzual \& Charlot (1995) models. We have used models with ages ranging from a
few Myr to 15 Gyr and with redshifts from 0 to 1.5 (open squares in
Fig.~\ref{colcol}) and models with ages ranging from a few Myr to 1 Gyr with
redshifts from 1.5 to 3.0 (open triangles). For the colours of high-redshift
galaxies, we included the effect of Ly$\alpha$ blanketing
(M{\o}ller \& Jakobsen 1990; Madau 1995). Fig.~\ref{colcol} shows the
n(AB)$-$B(AB) versus n(AB)$-$R(AB) colour diagram for the simulated galaxy
colours (left panels) and for the observed sources in the two target fields
(middle and right panels). The dashed line indicates where objects with a
particular broad-band colour and either absorption (upper right) or emission
(lower left) in the narrow filter will fall.

In the middle panel, we show the colour-colour diagram for all of the objects
detected in the two fields. Due to the damped Ly$\alpha$ line, Q\,2138$-$4427 
has a large positive n(AB)$-$B(AB) colour and is hence seen in the upper right
corner. Due to Lyman-forest blanketing, the B-band flux of BRI\,1346$-$0322 is
suppressed hence decreasing its n(AB)-B(AB) colour. In the lower left part of
the diagram, a large group of objects are found to lie significantly away from
the locus of continuum objects. In the right panel, we show with the solid
line the median n(AB)$-$B(AB) colour in the range 0$<$n(AB)$-$B(AB)$<$3. The
two dotted lines show the 98\% and 2\% percentiles in the n(AB)$-$B(AB)
colour. We selected as emission line sources objects with S/N$>$5 in the
narrow-band image and whose 1$\sigma$ upper limit on n(AB)$-$B(AB) is below
the 98\% line. We detect 27 and 37 such objects in the BRI\,1346$-$0322 and
Q\,2138$-$4427 fields respectively which we consider as LEGO candidates in the
following. The colours of the candidates are shown again in the two right
panels.

\section{Multi-Object Spectroscopy}
\label{MOS}
\subsection{Observations and data reduction}
Follow-up Multi-Object Spectroscopy (MOS) of the previously described sample
of LEGO candidates was carried out in visitor mode in July 10 -- 13, 2002,
with FORS1 installed at the VLT telescope, unit Melipal.
The mask preparation was done using the {\it FORS Instrumental Mask Simulator}.
The field of BRI\,1346$-$0322 was only visible at the beginning of the
nights so we only had time for using three masks for the
observations of this field. Nevertheless, all but five candidates could be put
on to slits. For the field of Q\,2138$-$4427, we did observe all but three
candidates using five masks. In the following, we will refer to these 8 masks
as mask1346A to C and mask2138A to E.
The observing conditions were generally mediocre due to very strong wind
coming from the North. As a result, the seeing was always above $1\arcsec$.
The seeing deteriorated during the nights so that the spectra in the field of
BRI\,1346$-$0322 were obtained under better conditions than the ones in the
other field. We used MOS slitlets having a width of $1.4\arcsec$ on the sky.
All spectra were obtained with the G600B grism covering the wavelength
range from 3600 \AA\ to 6000 \AA\ at a resolution of about 800. The detector
pixels were binned 2$\times$2 for all observations through all masks. The
journal of spectroscopic observations is given in Table~\ref{spec-journal}.

\begin{table}
\begin{center}
\caption{Log of spectroscopic observations with FORS1.
}
\begin{tabular}{@{}lllcccc}
\hline
mask & Exp.time & Date   & Effective seeing \\
     &    (hr)  & (2002) &    ($\arcsec$)   \\
\hline
mask1346A & 3.0 & July 10 & 1.1 \\
mask1346B & 3.0 & July 11 & 1.3 \\
mask1346C & 3.0 & July 12 & 1.4 \\
mask2138A & 3.0 & July 10 & 1.4 \\
mask2138B & 4.0 & July 10 & 1.4 \\
mask2138C & 3.0 & July 11 & 1.4 \\
mask2138D & 4.0 & July 11 & 1.1 \\
mask2138E & 5.5 & July 12 & 2.2 \\
\hline
\label{spec-journal}
\end{tabular}
\end{center}
\end{table}

\subsection{Results}

\begin{figure*}
\begin{center}
\epsfig{file=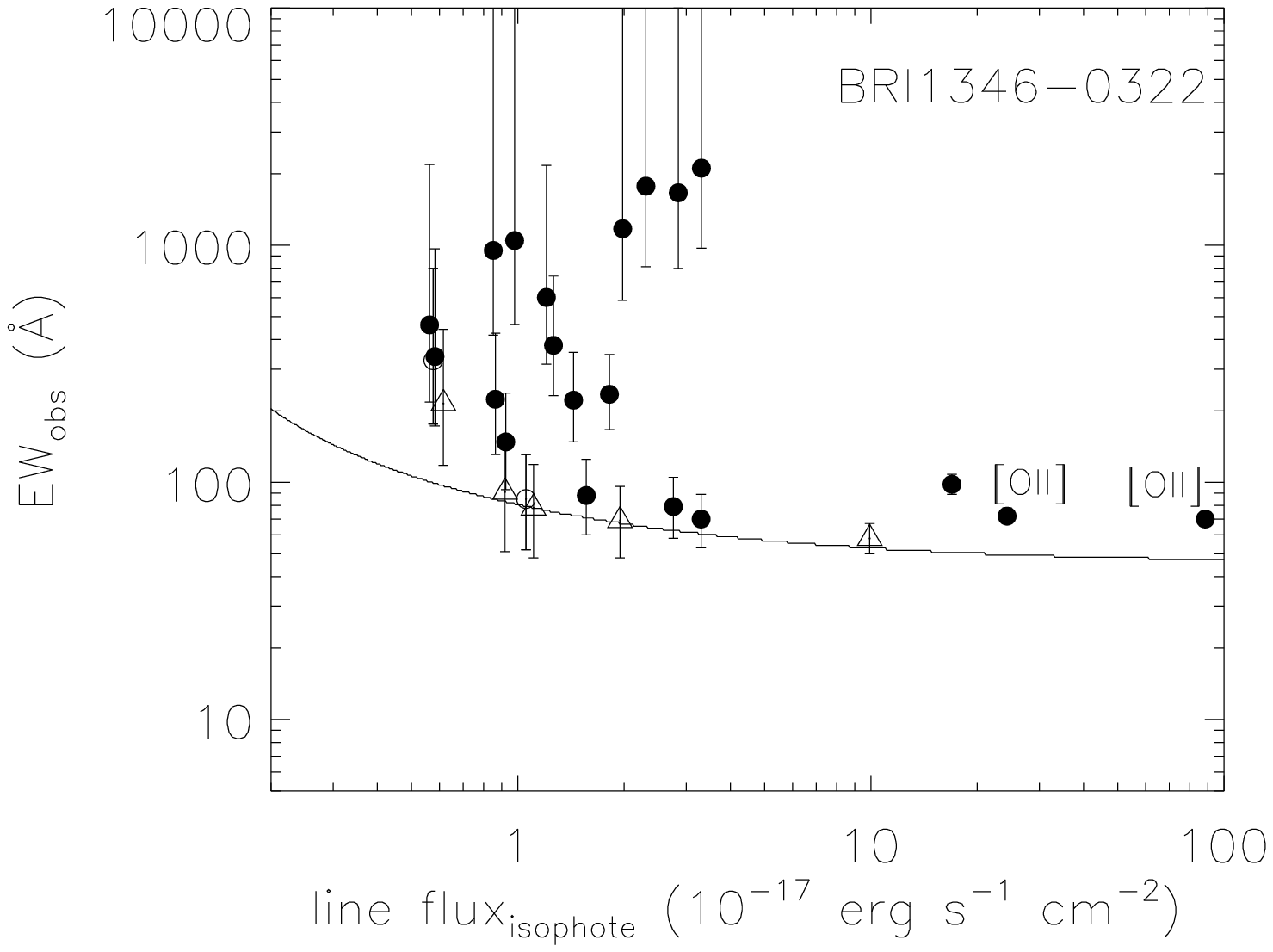, width=8.5cm}
\epsfig{file=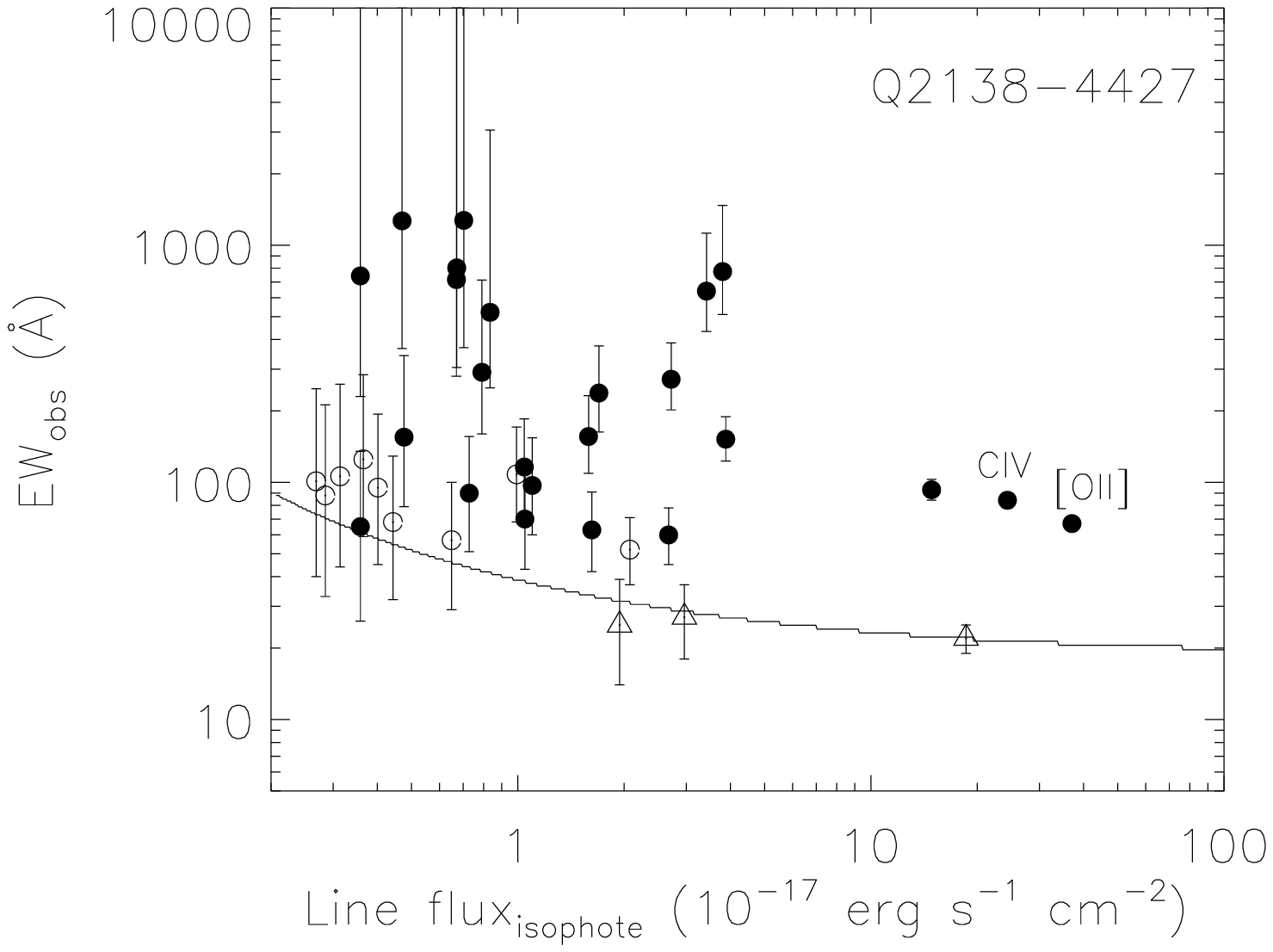, width=8.5cm}
\caption{Our selection criterion is illustrated by plotting the line
equivalent width against the Ly$\alpha$ flux in the isophotal aperture for the
LEGO candidates in the two fields. The continuous line shows our n(AB)$-$B(AB)
colour selection criterion converted into equivalent width. Filled symbols
indicate objects subsequently confirmed by spectroscopy to be emission-line
sources. Open circles indicate objects that were observed, but not confirmed,
and open triangles candidates not yet observed. The nature of the four
foreground emission-line sources is also indicated.
}
\label{select}
\end{center}
\end{figure*}

The MOS data were reduced and combined as described in Fynbo et al. (2001).
The accuracy in the wavelength calibration is about $\pm 0.1$ pixel for
a spectral resolution $R=900$, which translates to $\Delta$z=0.0002.
Average object extraction was performed within a variable window size
matching the spatial extension of the emission line(s). Therefore, the
flux should be conserved. When two or more individual exposures on the
same target have been obtained through different masks, the spectra
were appropriately combined with rescaling and weights, and using a mask for
rejecting cosmic ray impacts.

\begin{figure*}
\begin{center}
\epsfig{file=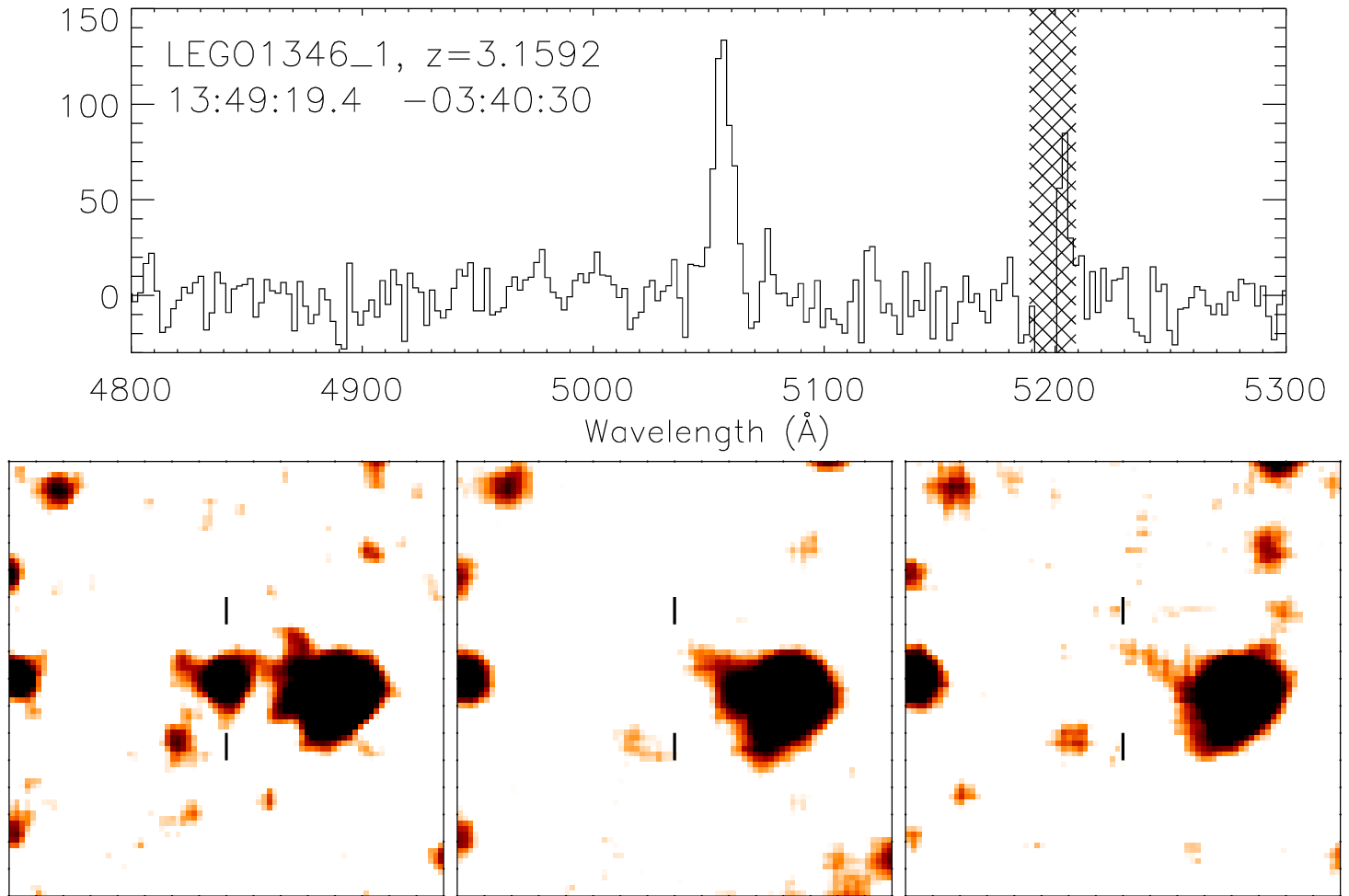, width=5.5cm}
\epsfig{file=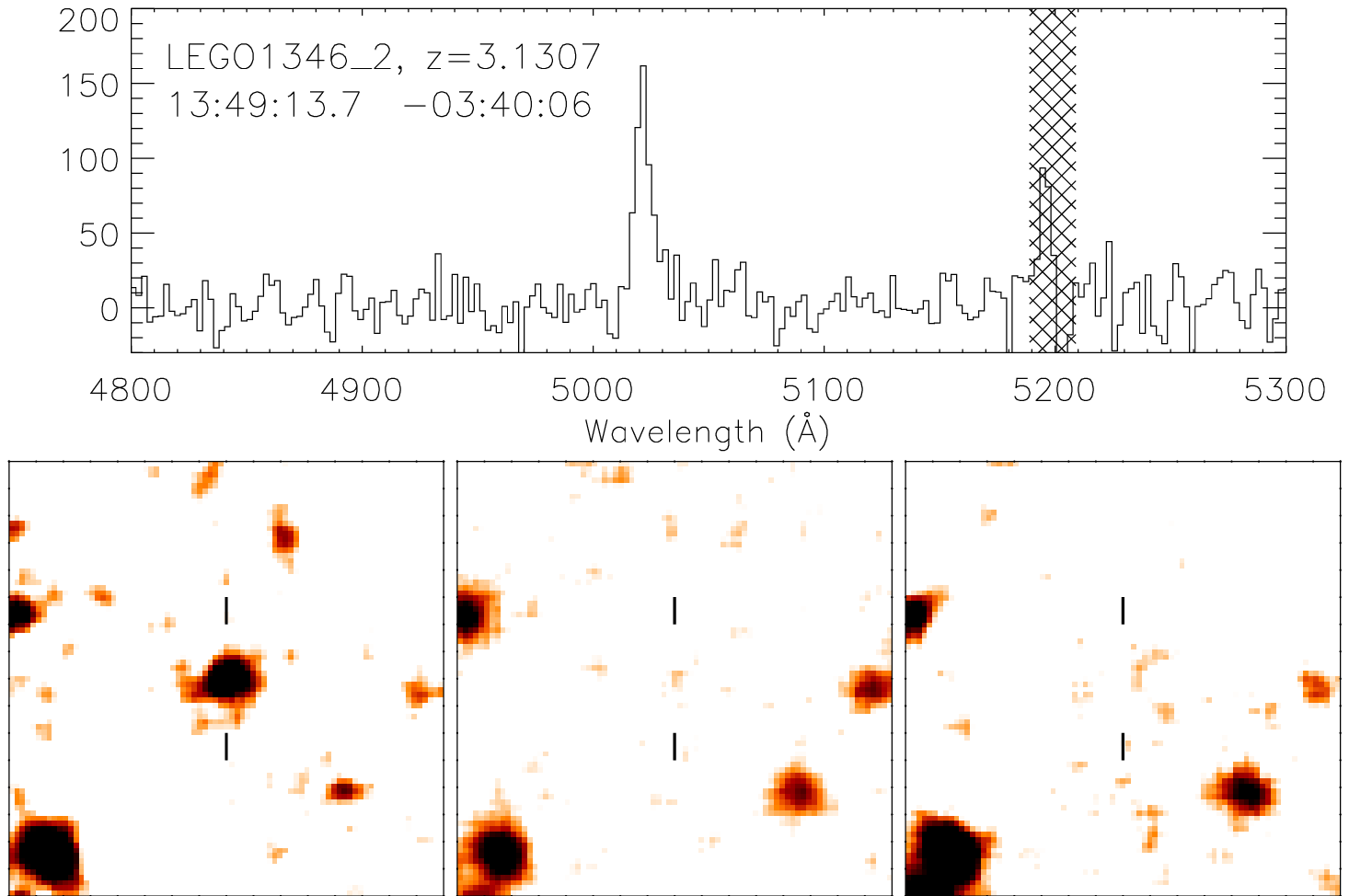, width=5.5cm}
\epsfig{file=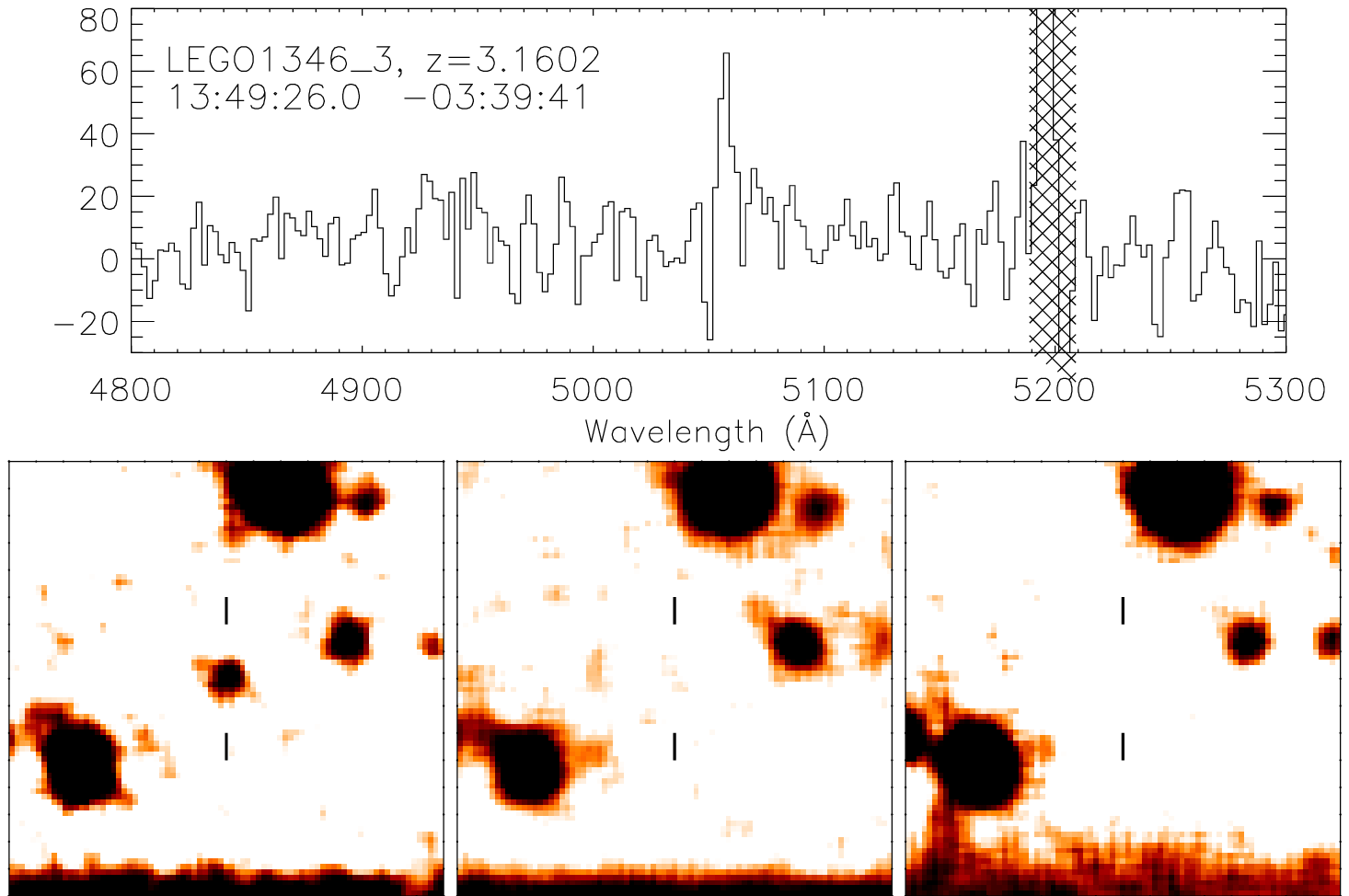, width=5.5cm}\\
\vskip 0.2cm
\epsfig{file=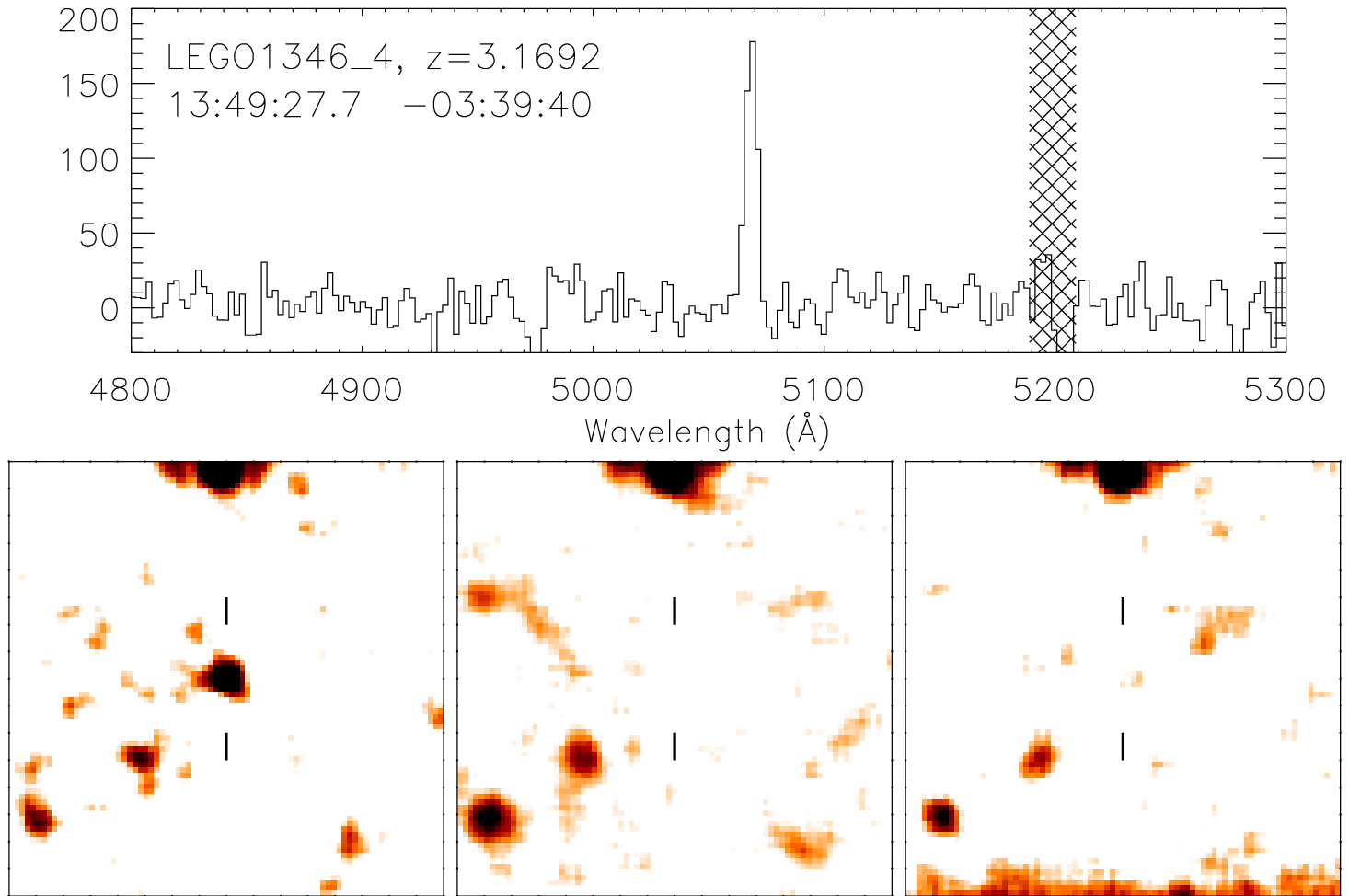, width=5.5cm}
\epsfig{file=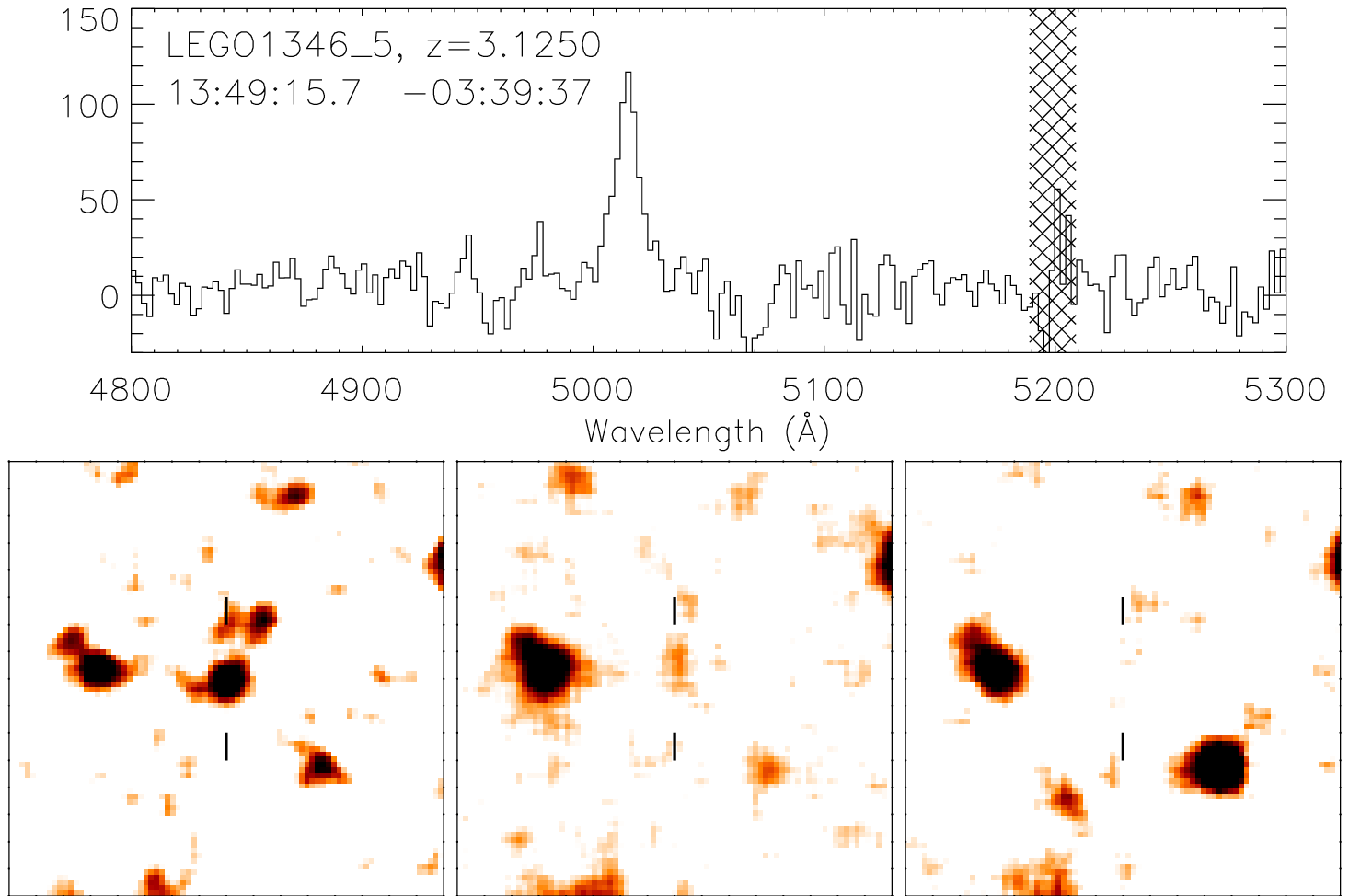, width=5.5cm}
\epsfig{file=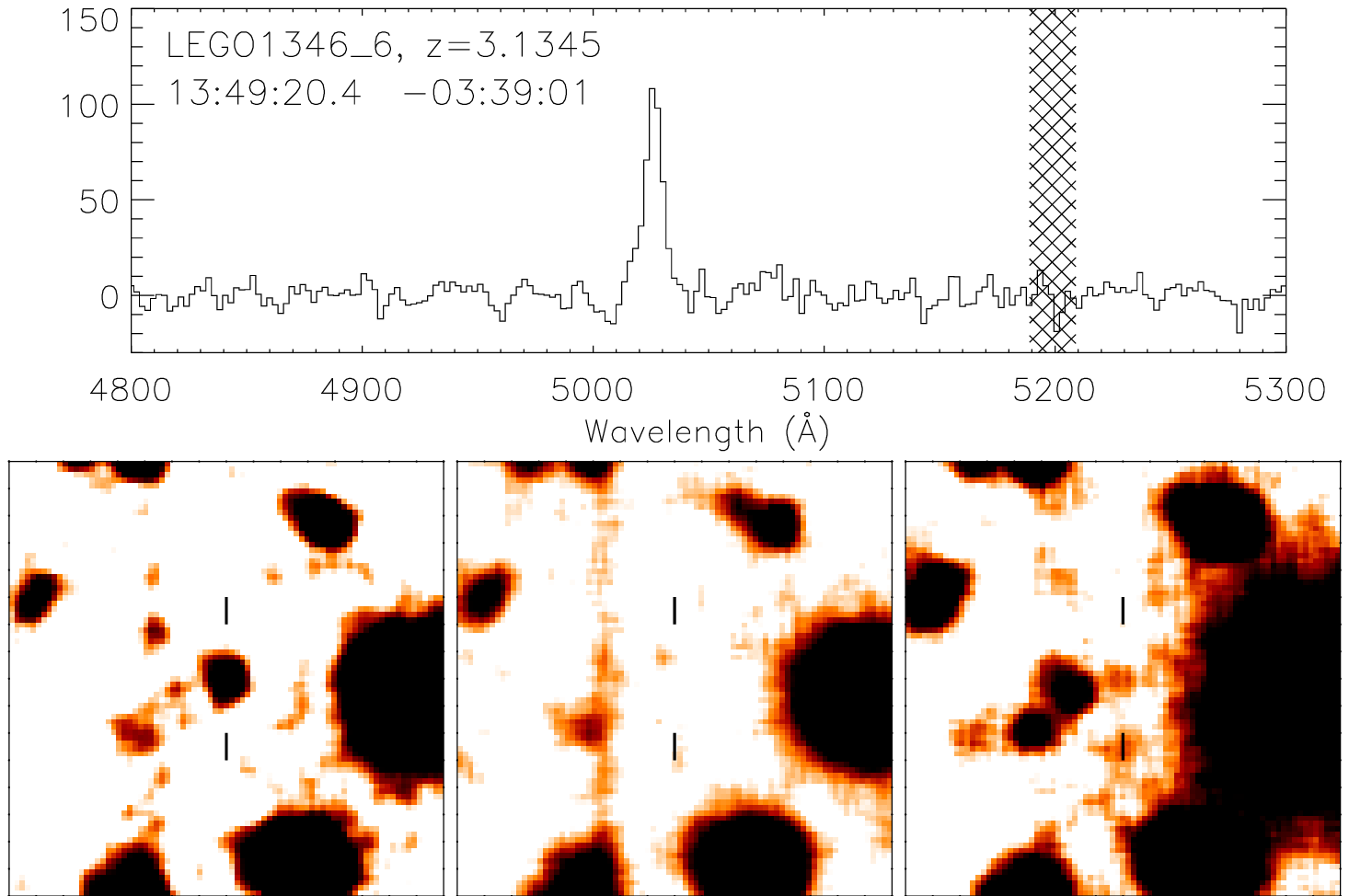, width=5.5cm}\\
\vskip 0.2cm
\epsfig{file=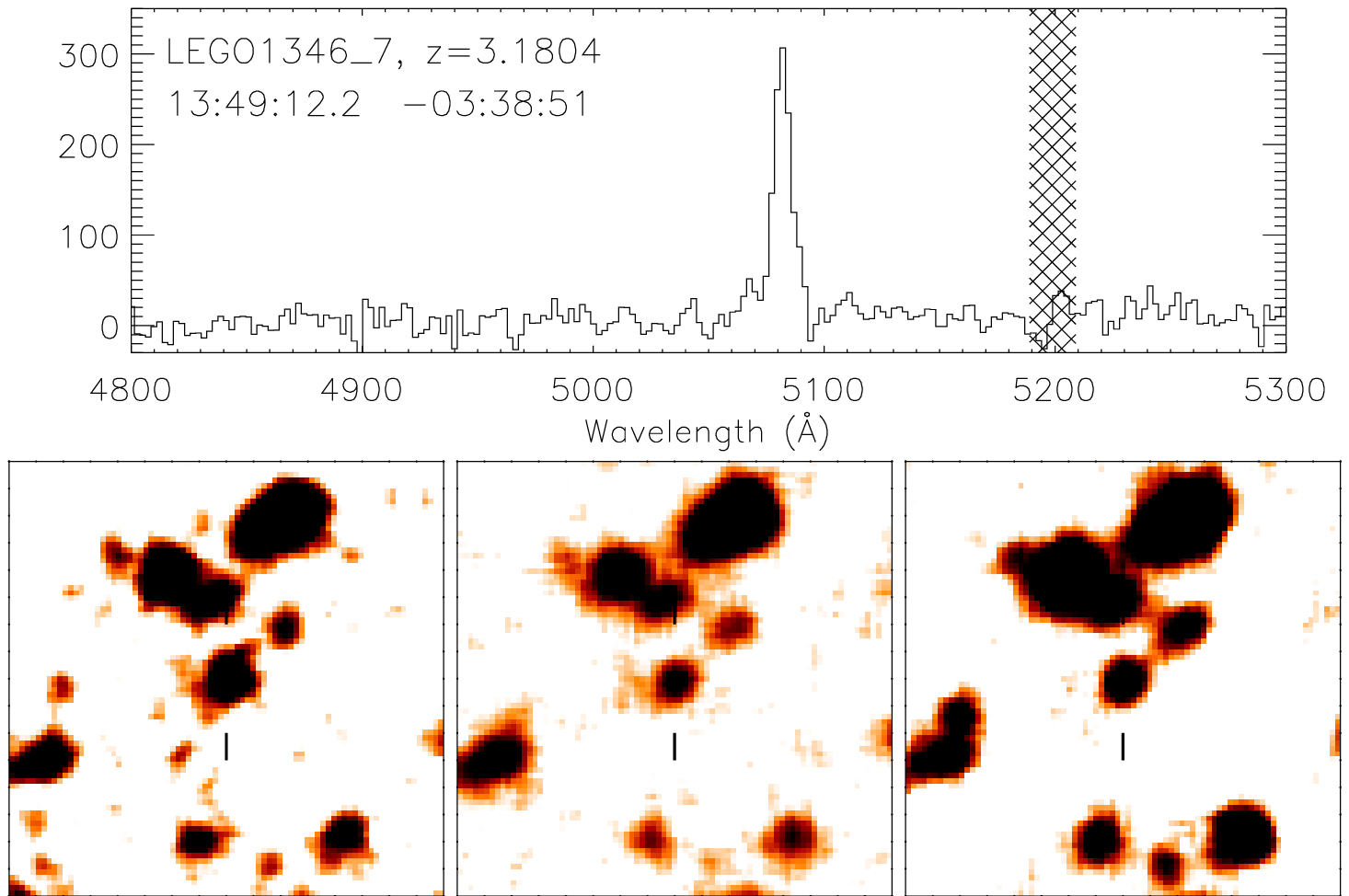, width=5.5cm}
\epsfig{file=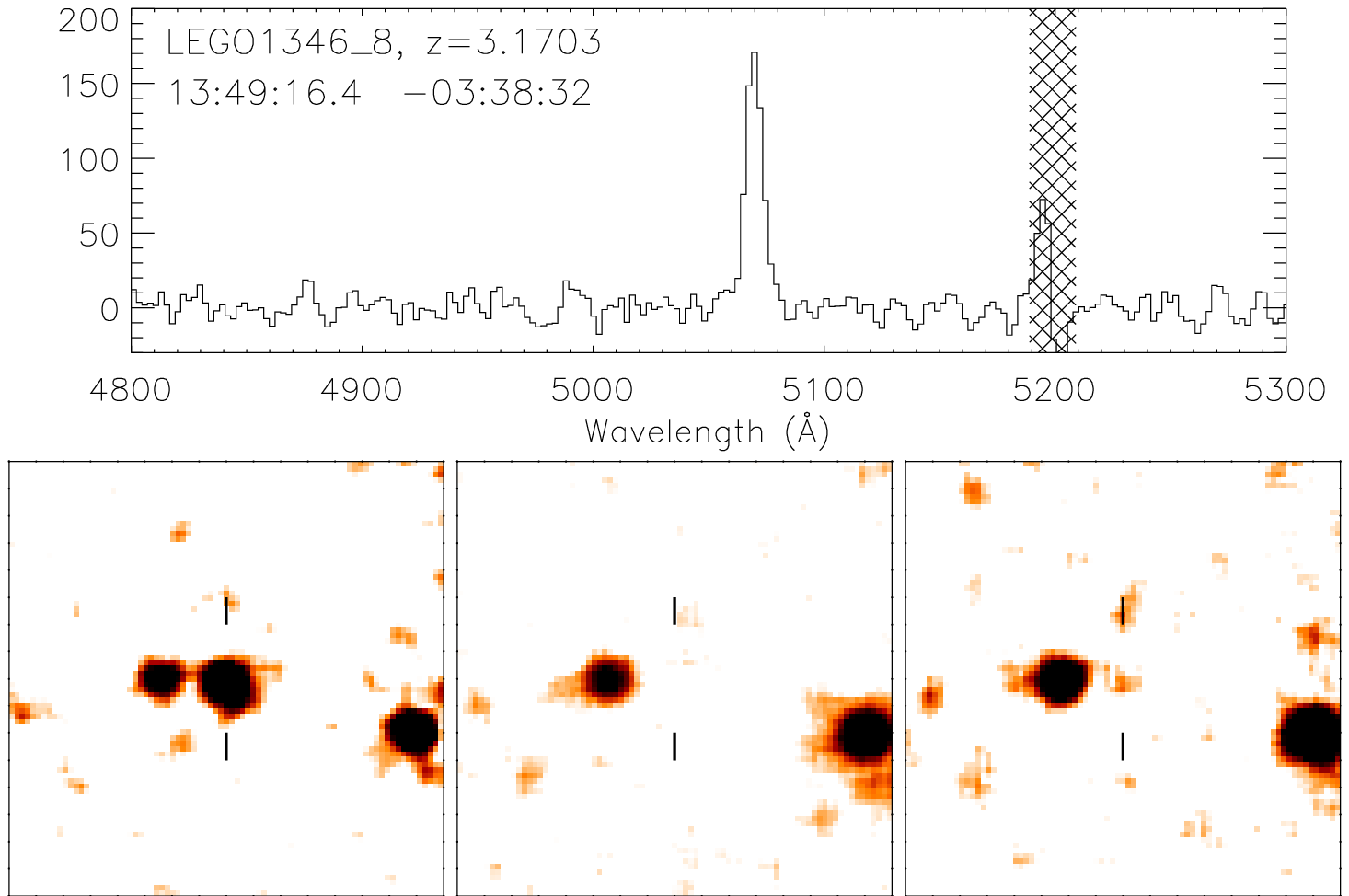, width=5.5cm}
\epsfig{file=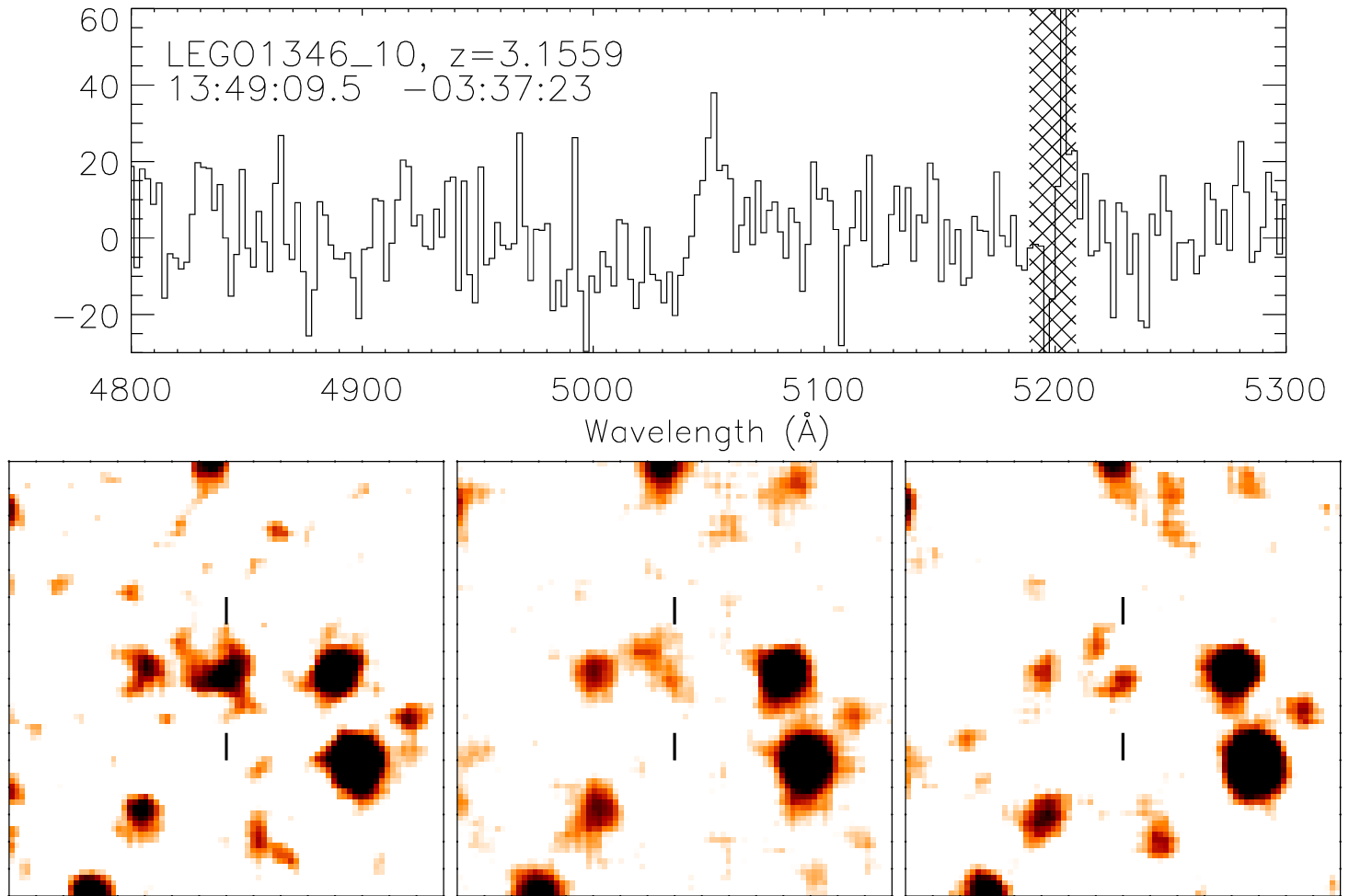, width=5.5cm}\\
\vskip 0.2cm
\epsfig{file=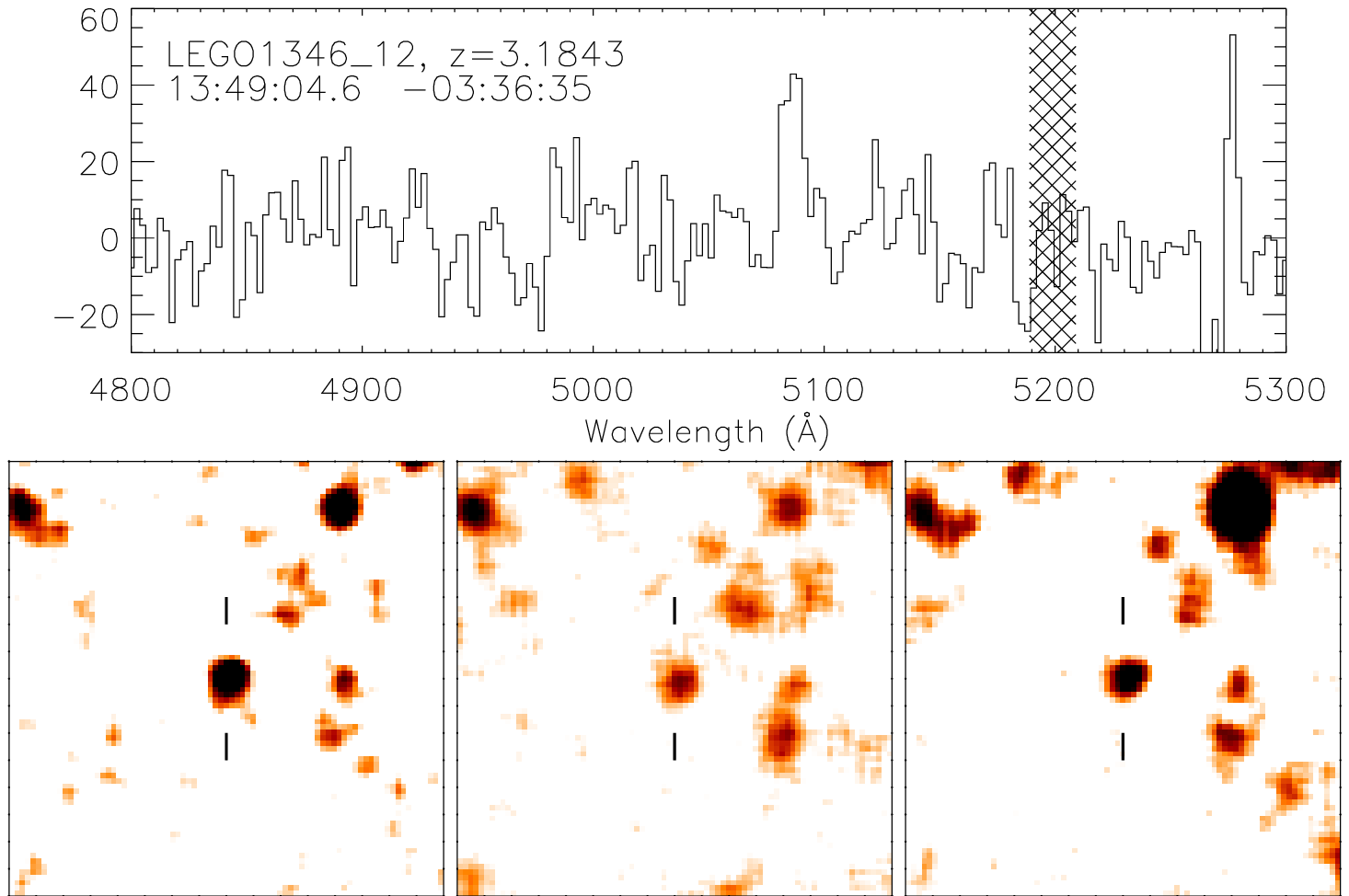, width=5.5cm}
\epsfig{file=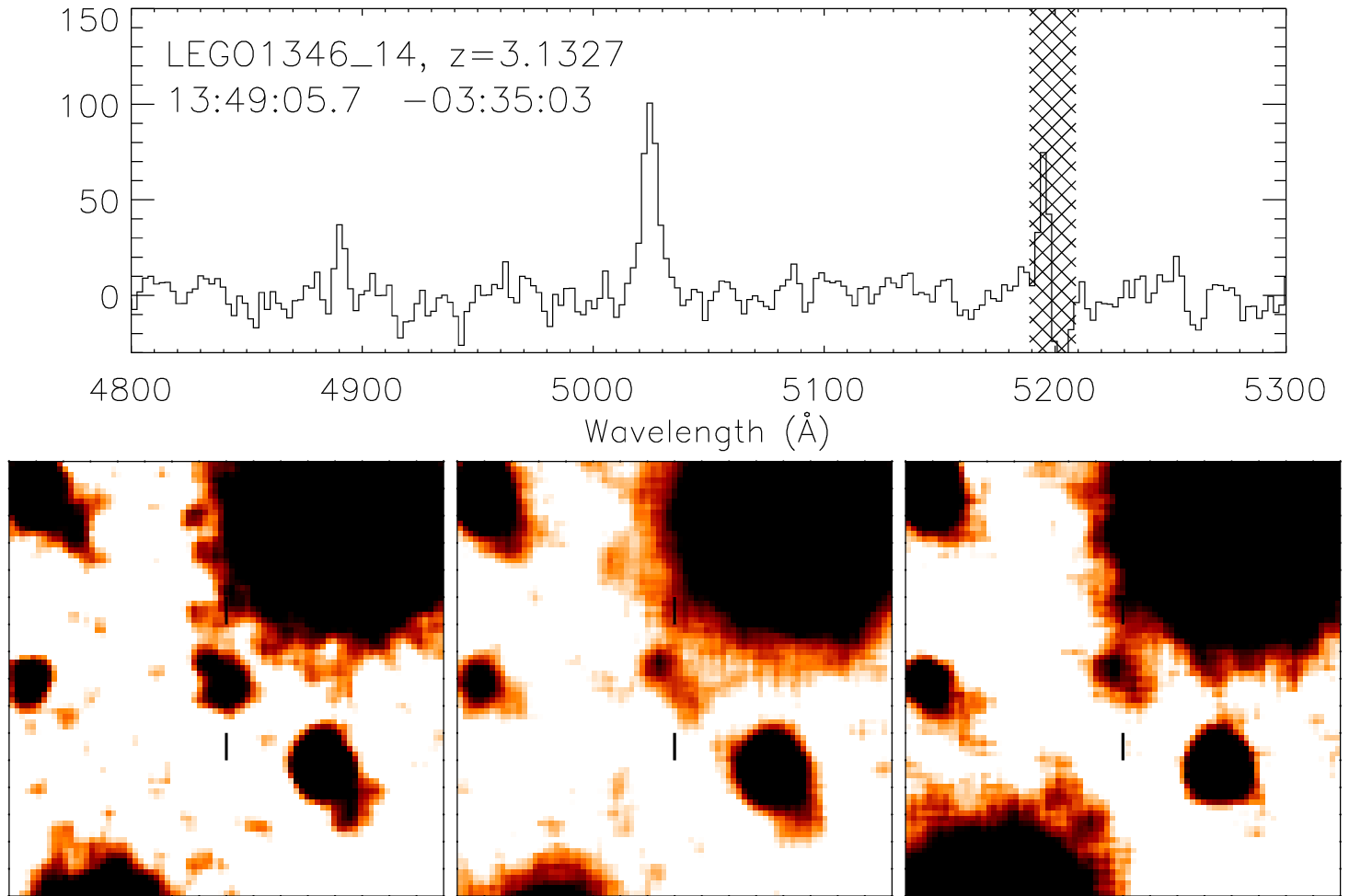, width=5.5cm}
\epsfig{file=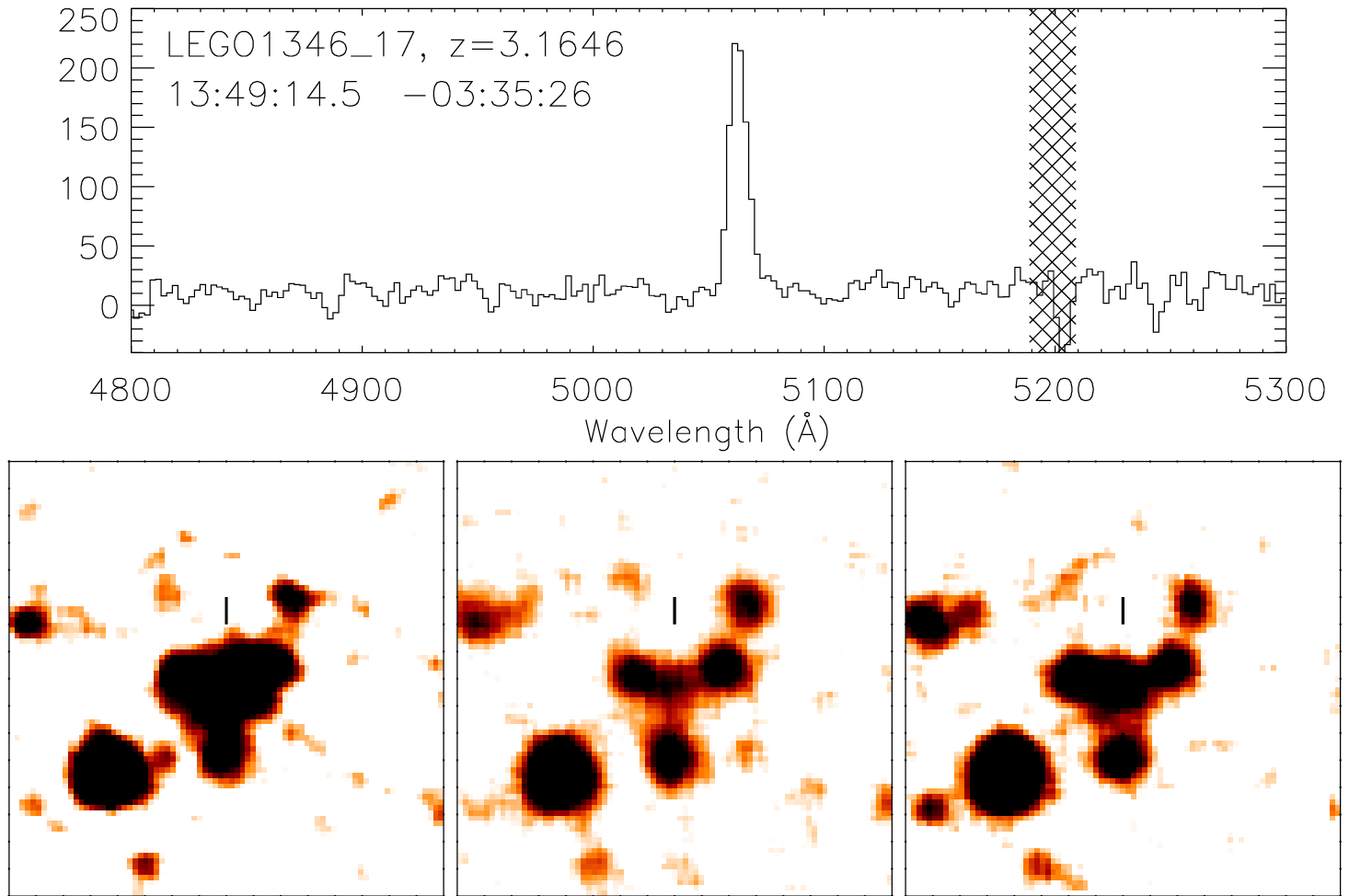, width=5.5cm}\\
\vskip 0.2cm
\epsfig{file=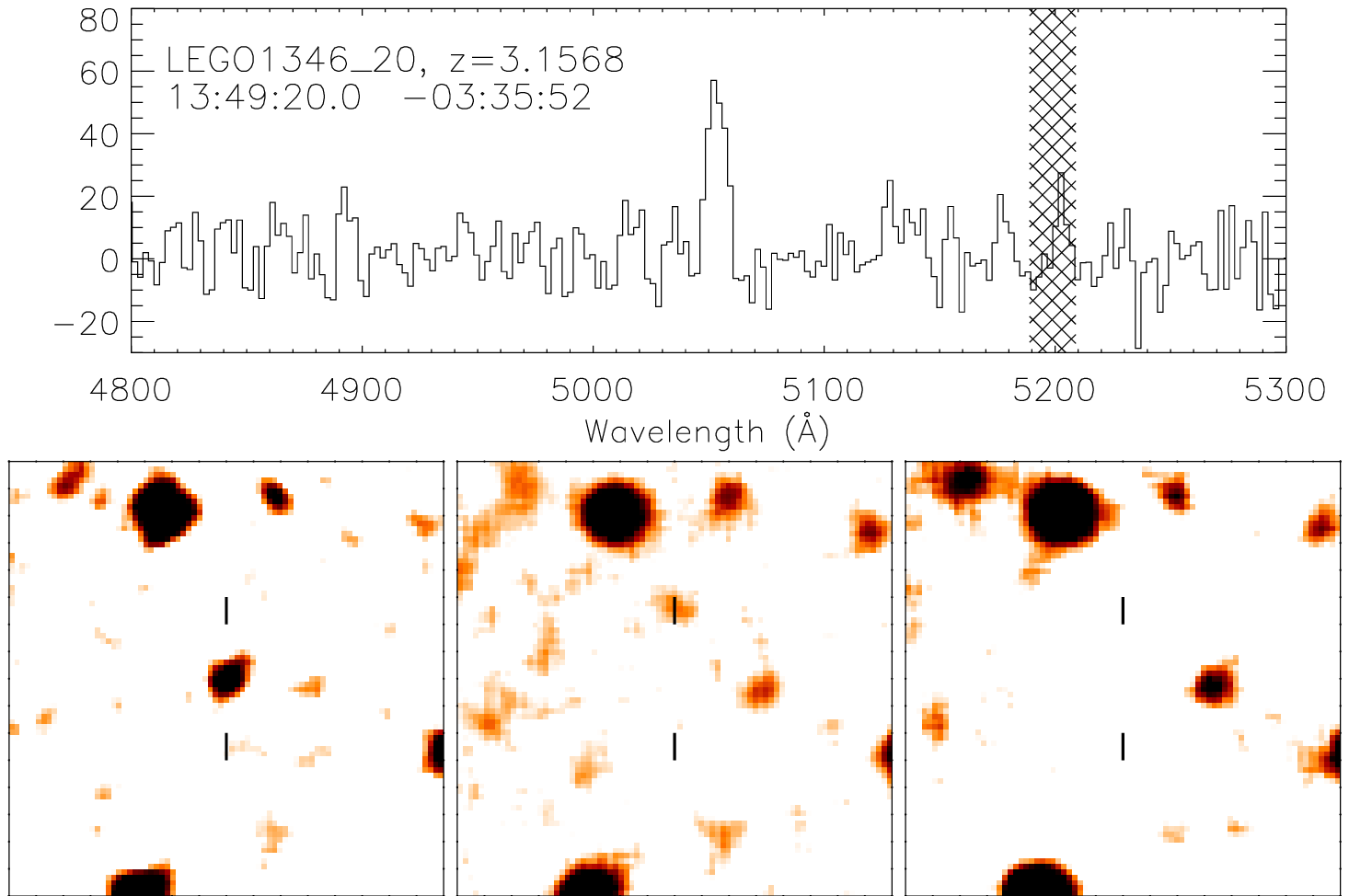, width=5.5cm}
\epsfig{file=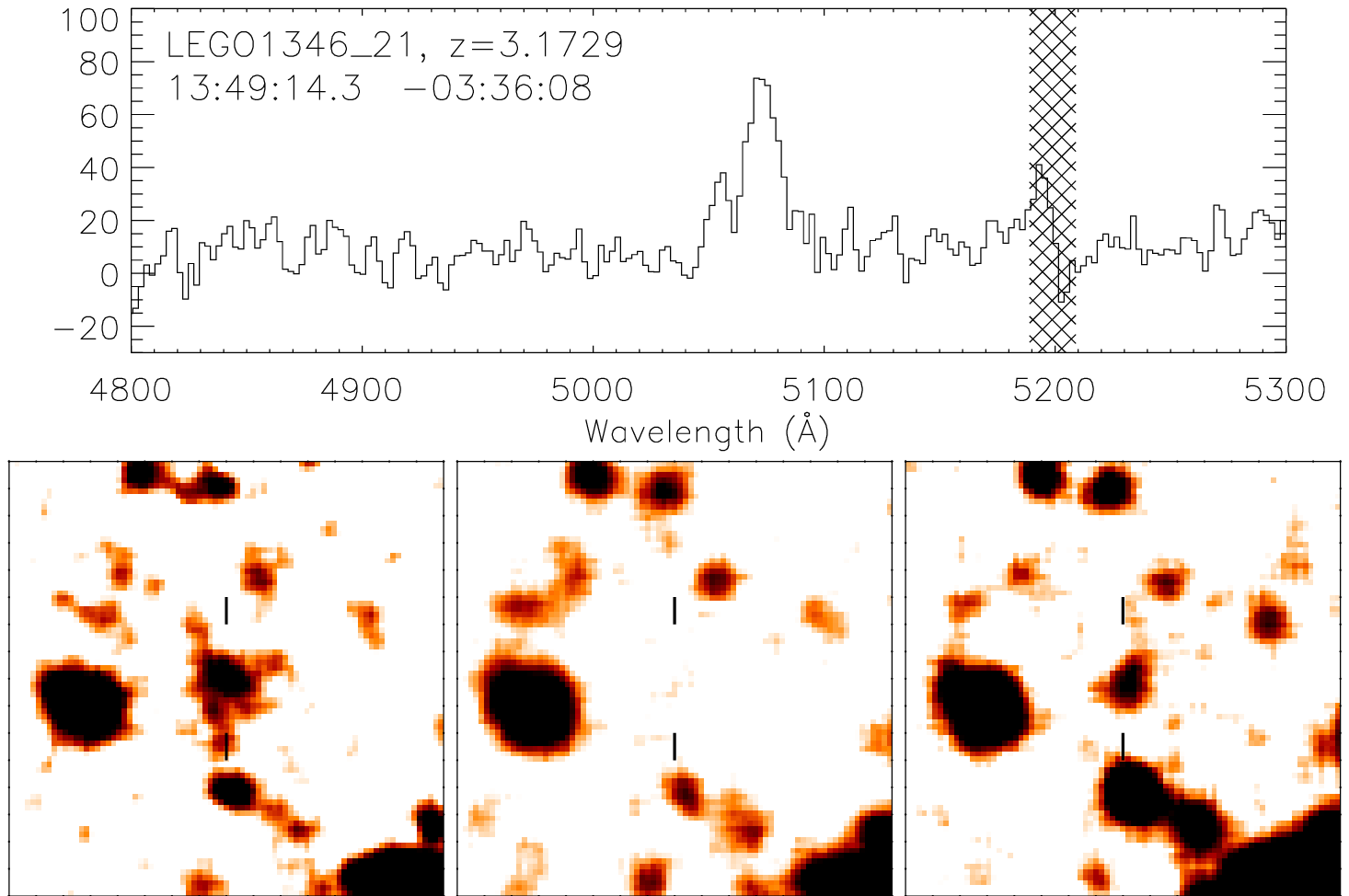, width=5.5cm}
\epsfig{file=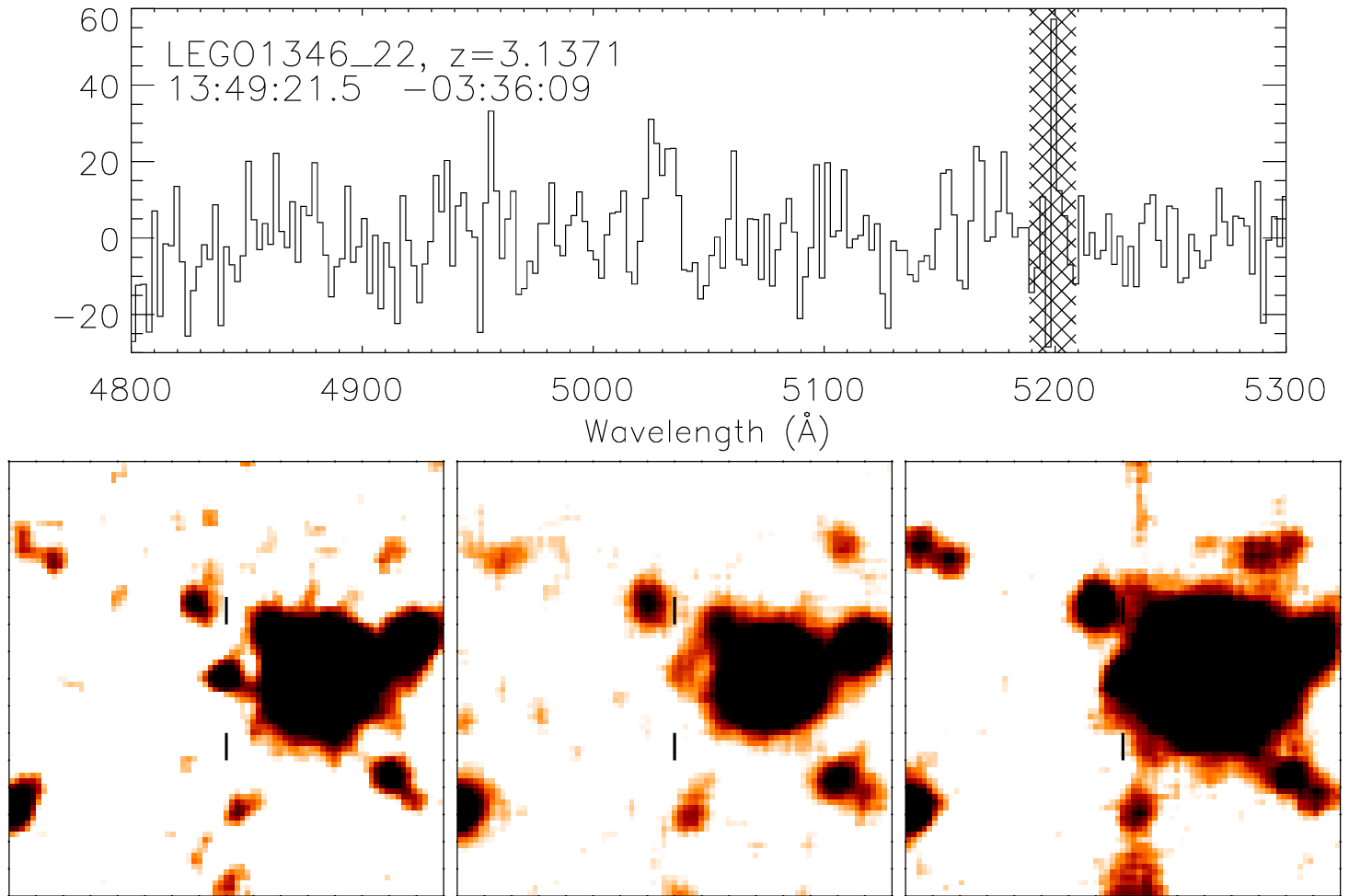, width=5.5cm}\\
\vskip 0.2cm
\epsfig{file=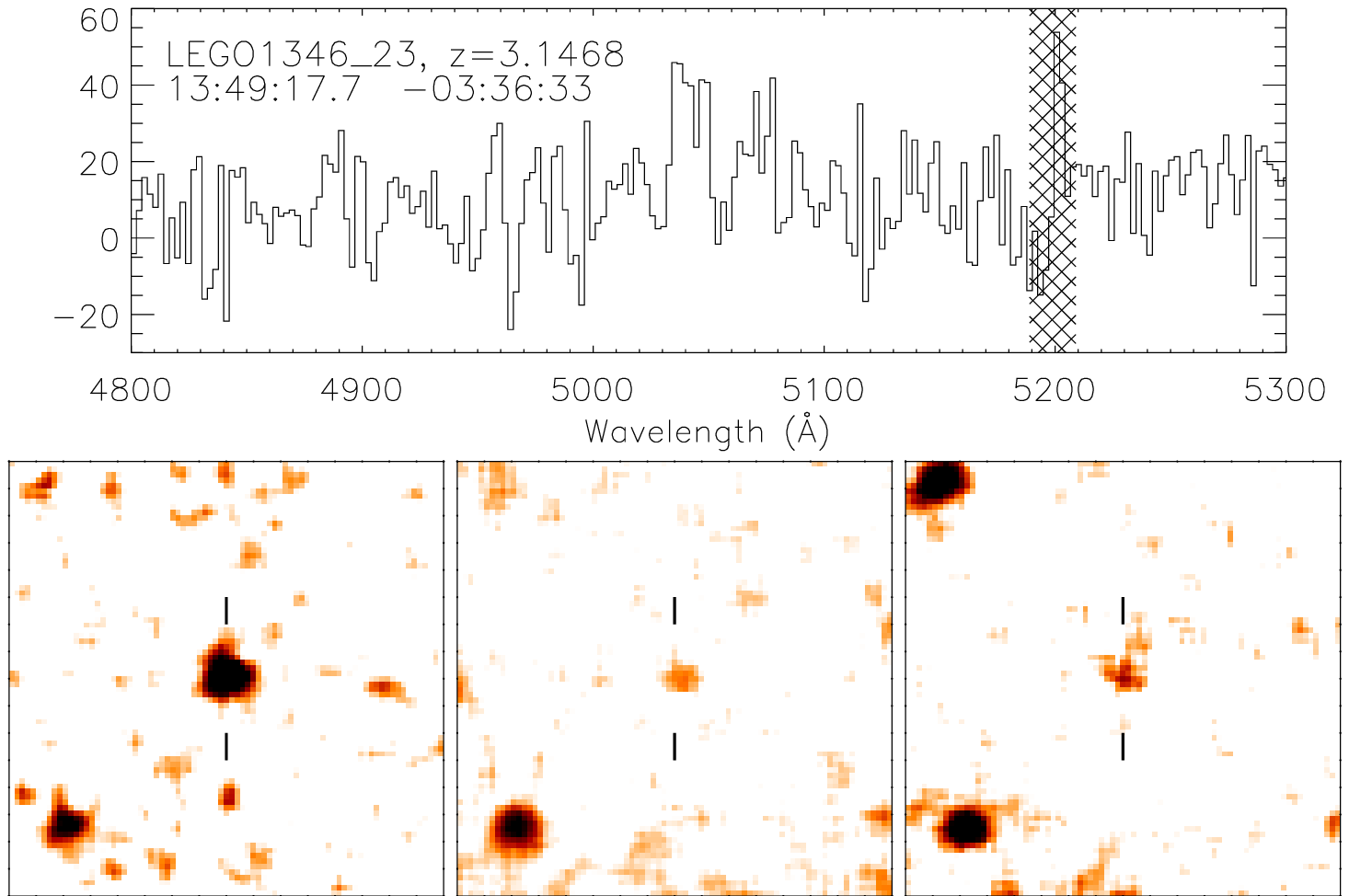, width=5.5cm}
\epsfig{file=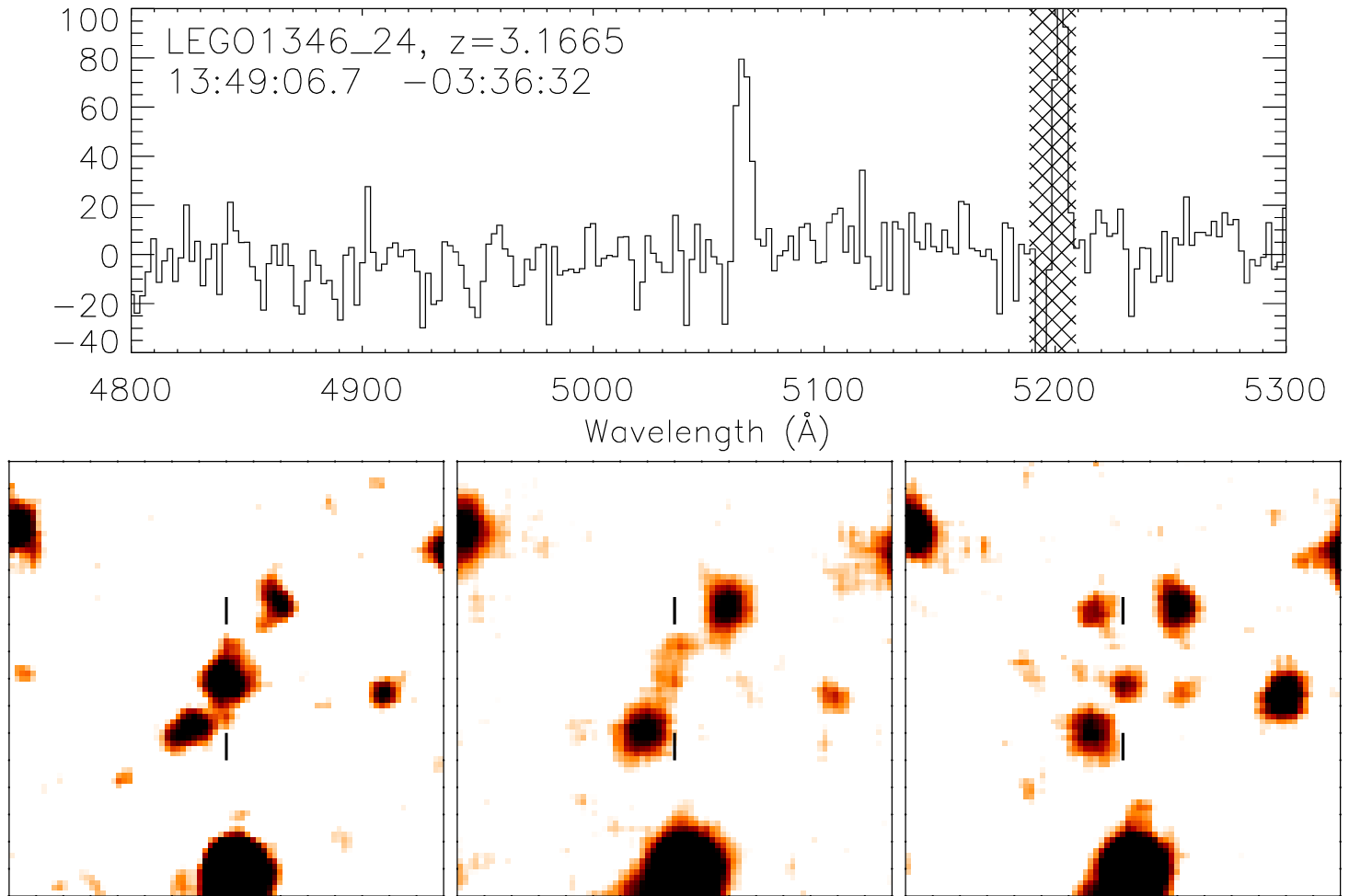, width=5.5cm}
\epsfig{file=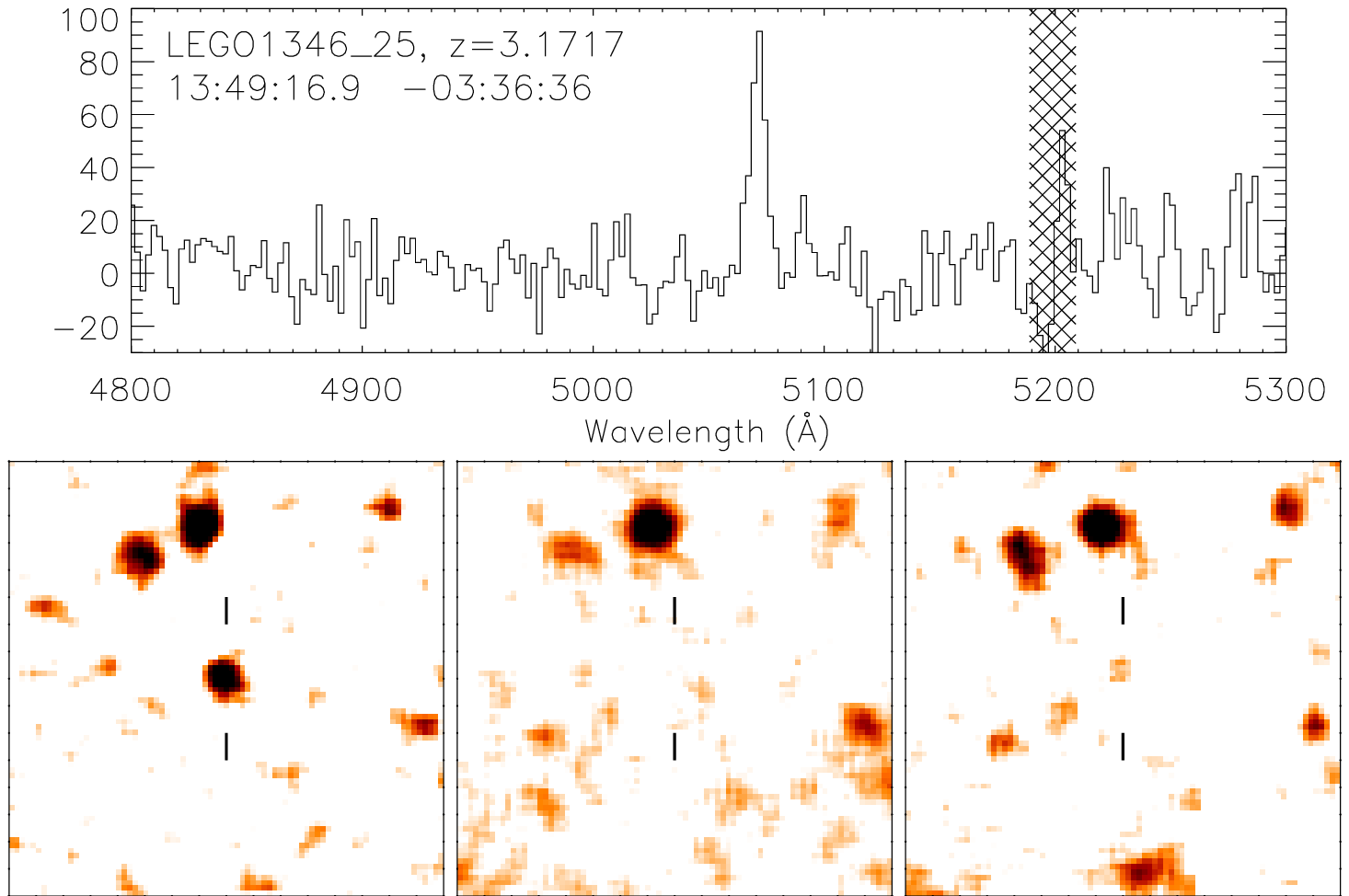, width=5.5cm}
\vskip 0.2cm
\caption{
16$\times$16 arcsec$^2$ images and 1-D spectra of 18 confirmed LEGOs in the
field of BRI\,1346$-$0322. For each candidate, we show images from the
narrow-, B- and R-band filters (from left to right). The units on the
ordinate of the spectra are counts in 1800 sec (as in Fynbo et al. 2001). The
region around the sky line at 5199 \AA\ is covered due to
large sky-subtraction residuals. The name, redshift and coordinates (Epoch
2000) are provided for each object.
}
\label{candfigs1346}
\end{center}
\end{figure*}

We first analysed the slitlets containing the spectra of the LEGO candidates.
The combined spectra are displayed in Figs.~\ref{candfigs1346}
and \ref{candfigs2138}. Out of the 27 candidates in the BRI\,1346$-$0322
field, we confirm 20 as being emission-line objects. We consider a candidate 
confirmed if there is an emission line detected with at least 3$\sigma$
significance at the correct position in the slitlet within the wavelength 
range corresponding to the filter transmission. Five candidates were
not observed and for the remaining two an emission line is not detected. Two
of the confirmed emission-line sources are foreground galaxies with
the [\ion{O}{ii}] line located in the narrow-band filter. The overall
efficiency for detection and confirmation of LEGOs is therefore 
\#(confirmed LEGOS)/\#(observed LEGOs) = 18/22 = 82\% so
far. For the Q\,2138$-$4427 field, the overall efficiency is smaller.
Three candidates were not observed and 9 observed candidates were
not confirmed. For the remaining 25, we confirm the presence of an
emission line. Two of the confirmed emission-line sources are
foreground objects. One is an [\ion{O}{ii}] emitter and the other is
a z=2.03664i4 AGN with \ion{C}{iv} located in the narrow-band filter. Hence,
the fraction of confirmed LEGOs is 23/34=68\%.

\begin{figure*}
\begin{center}
\epsfig{file=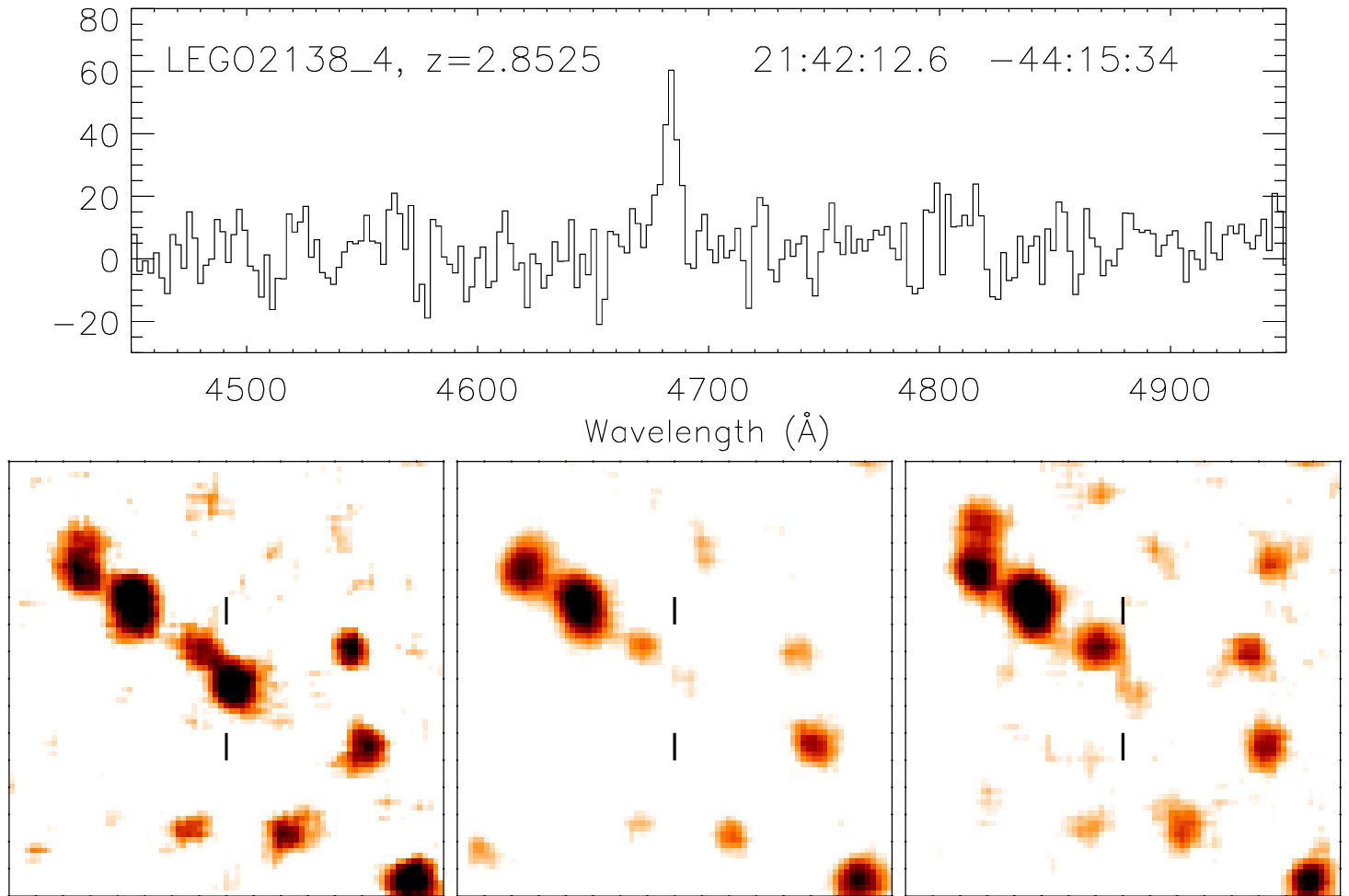, width=5.5cm}
\epsfig{file=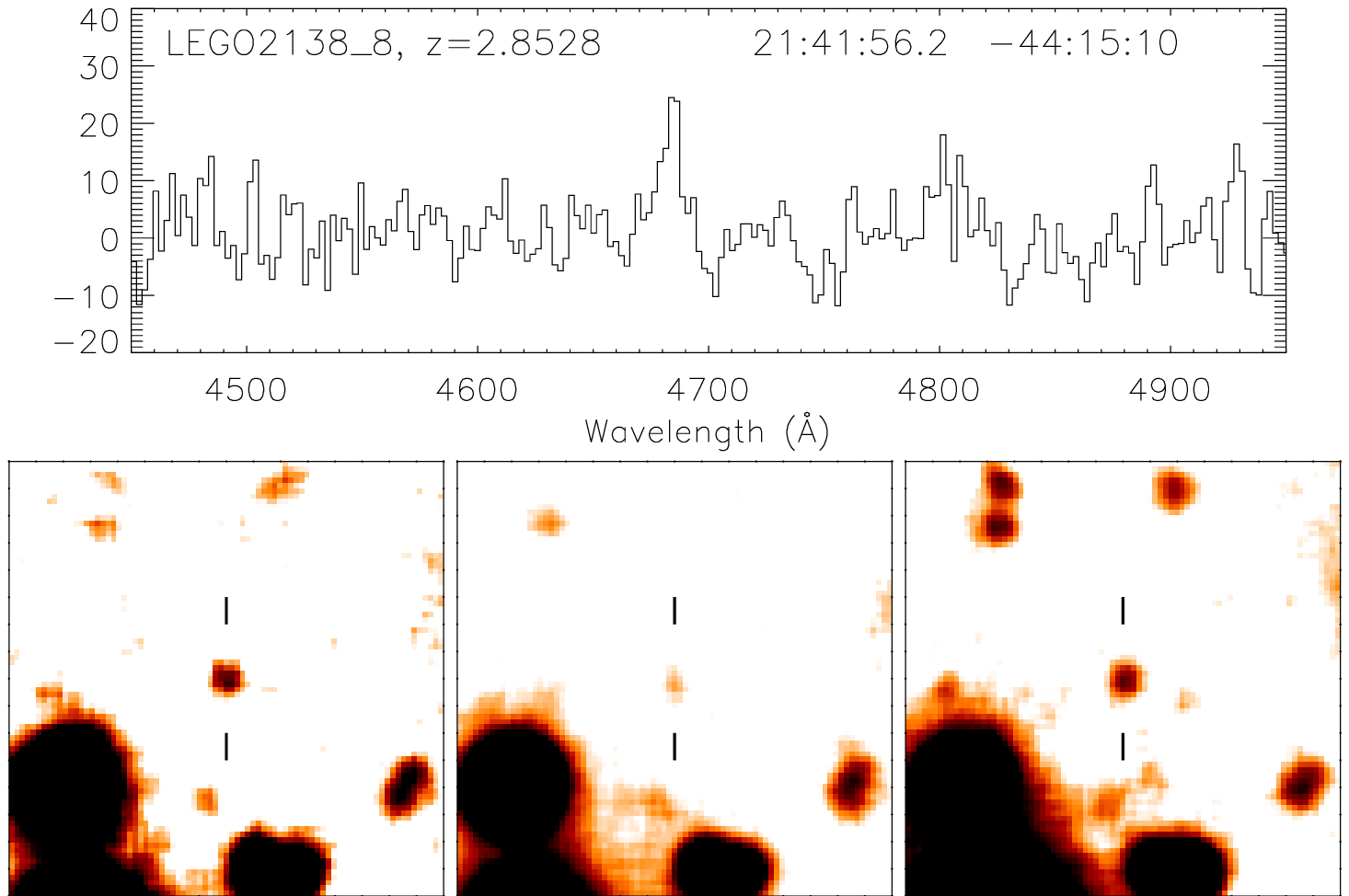, width=5.5cm}
\epsfig{file=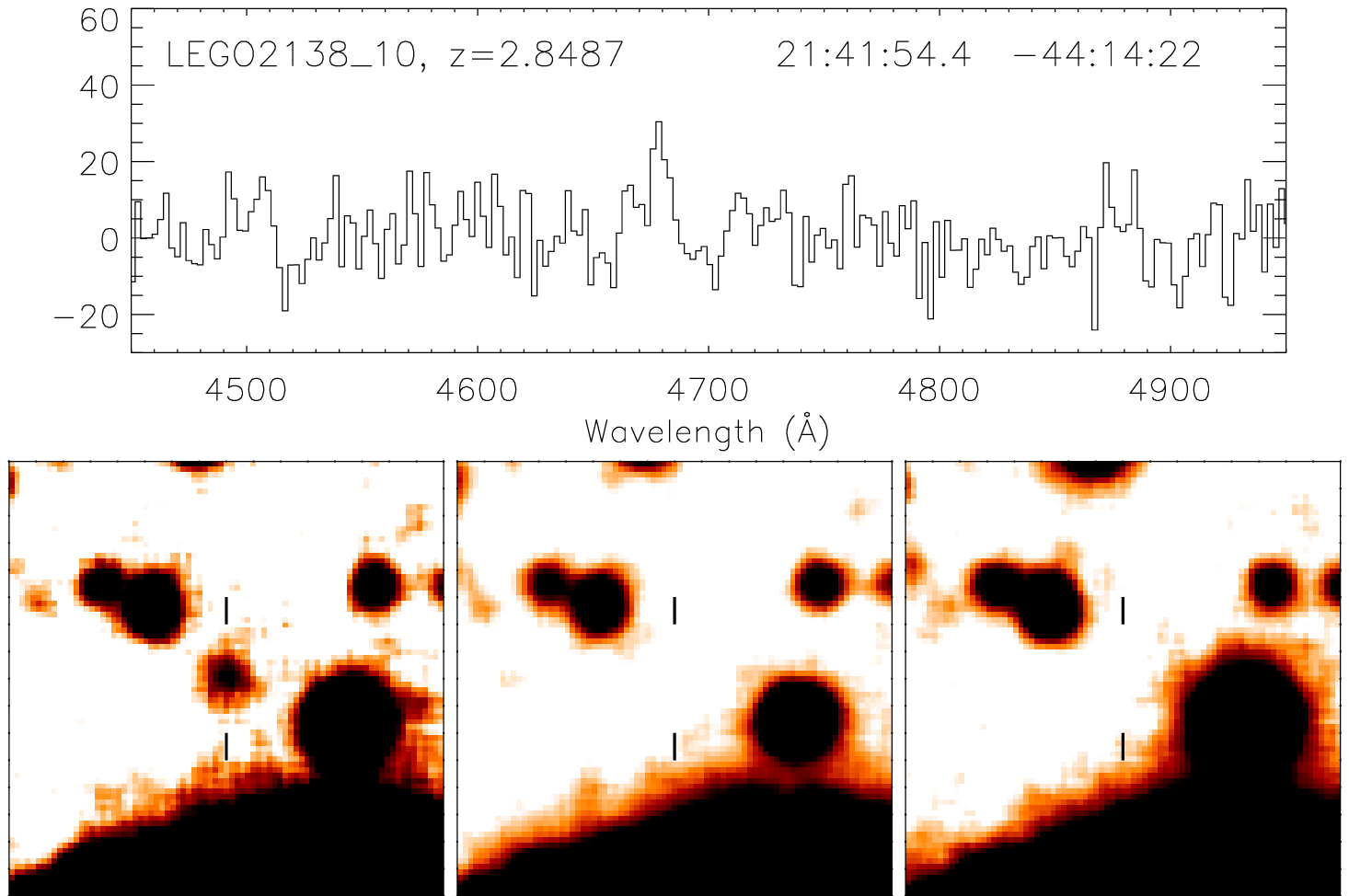, width=5.5cm}\\
\vskip 0.2cm
\epsfig{file=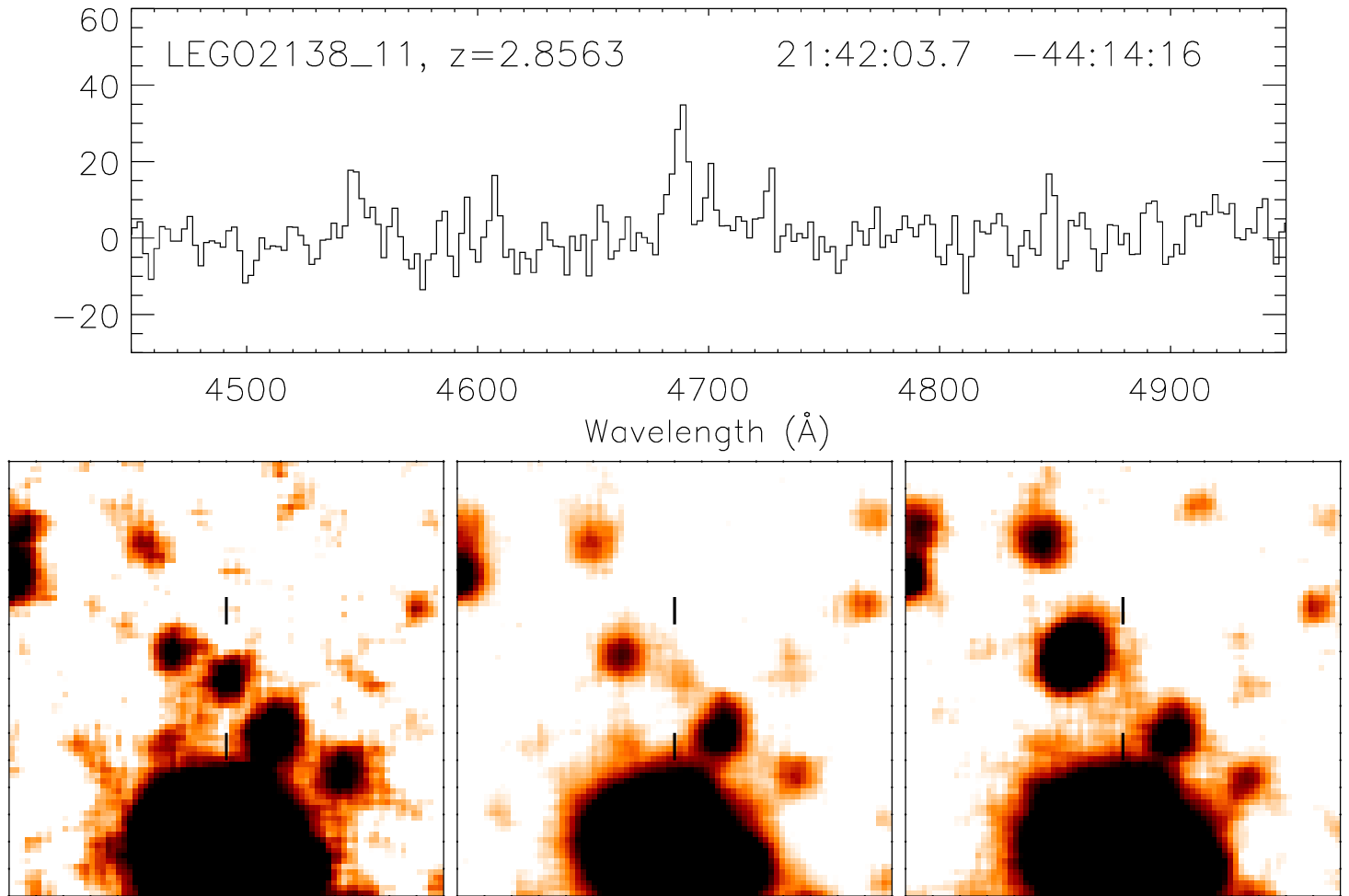, width=5.5cm}
\epsfig{file=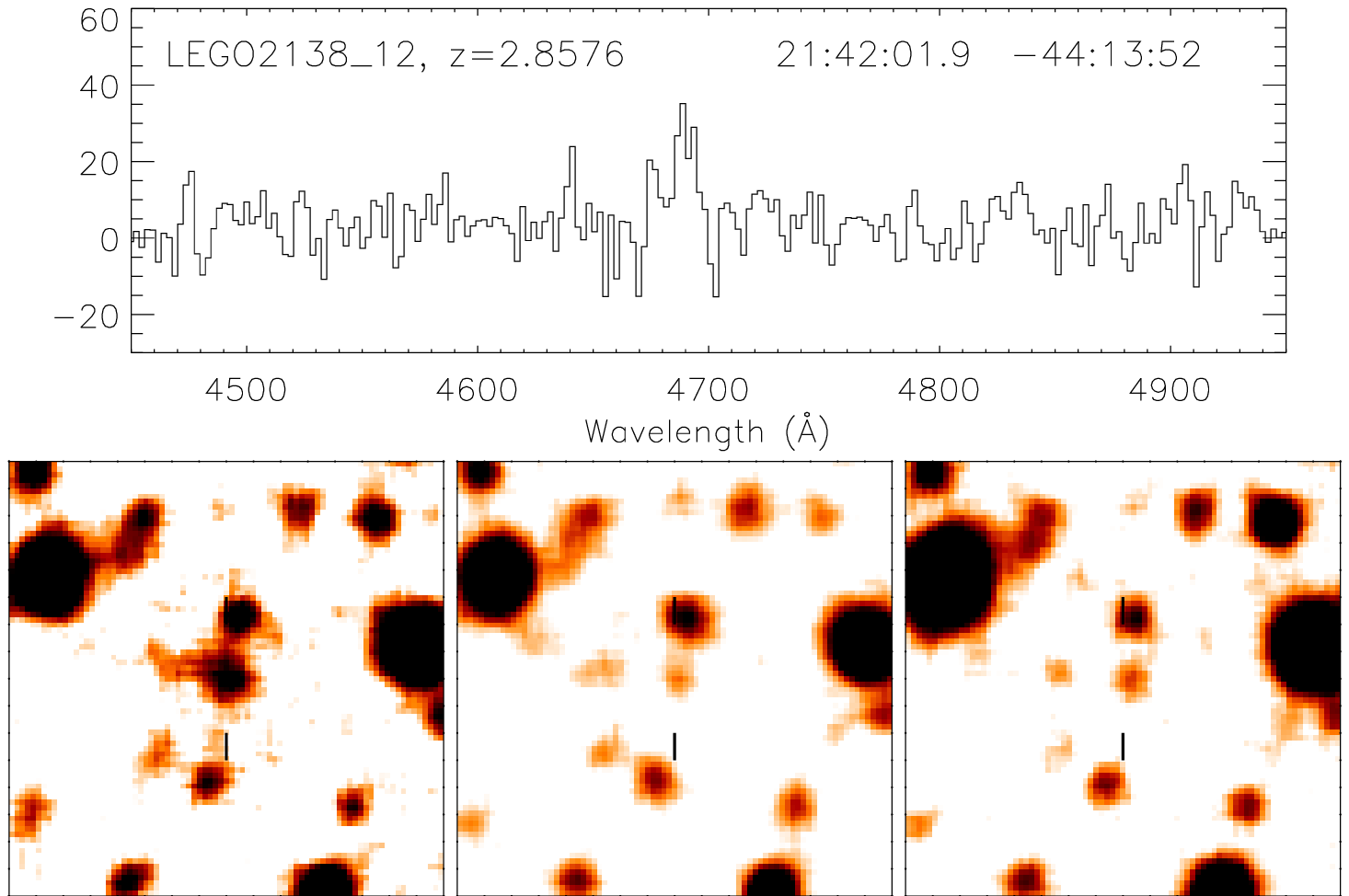, width=5.5cm}
\epsfig{file=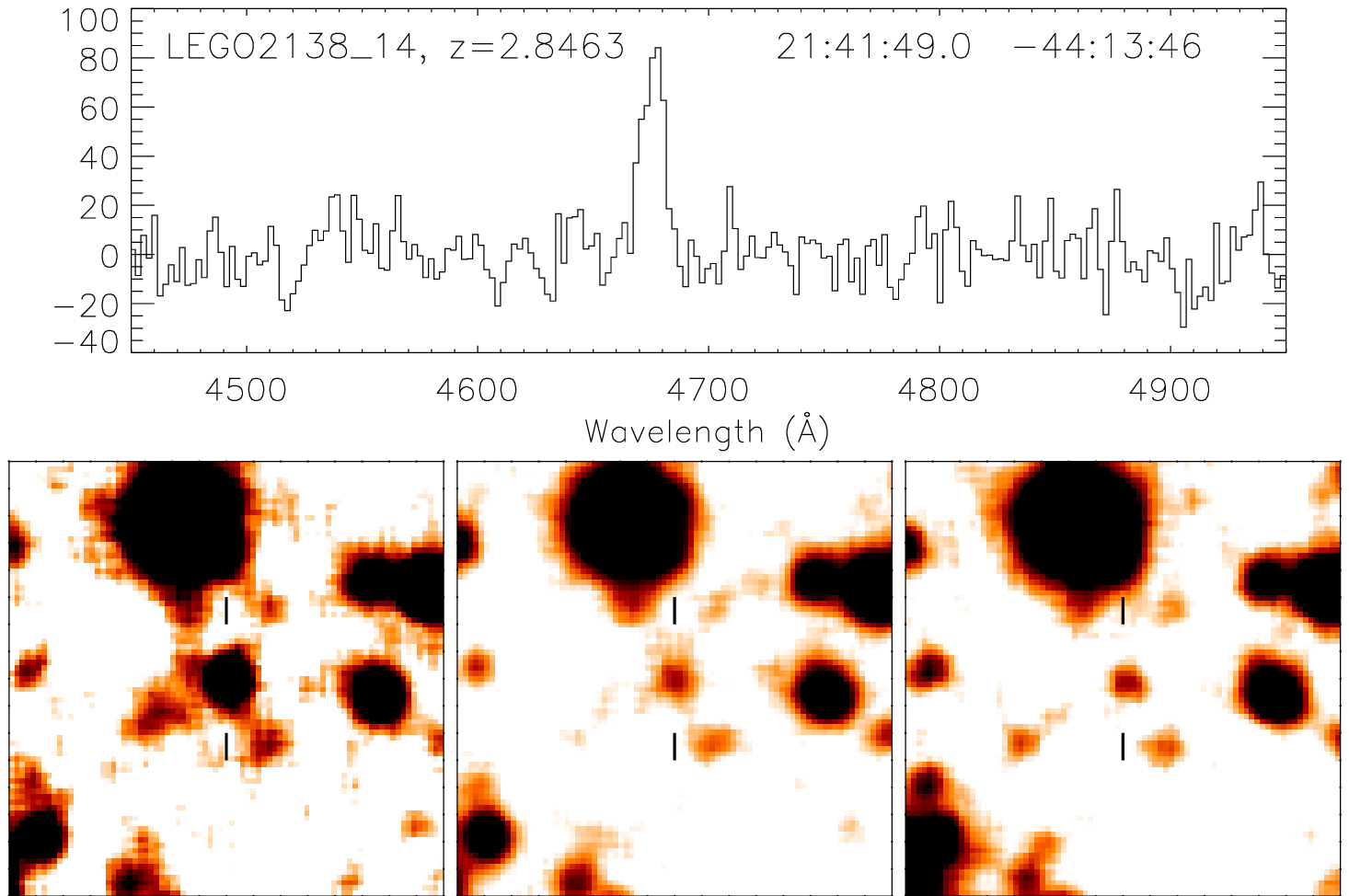, width=5.5cm}\\
\vskip 0.2cm
\epsfig{file=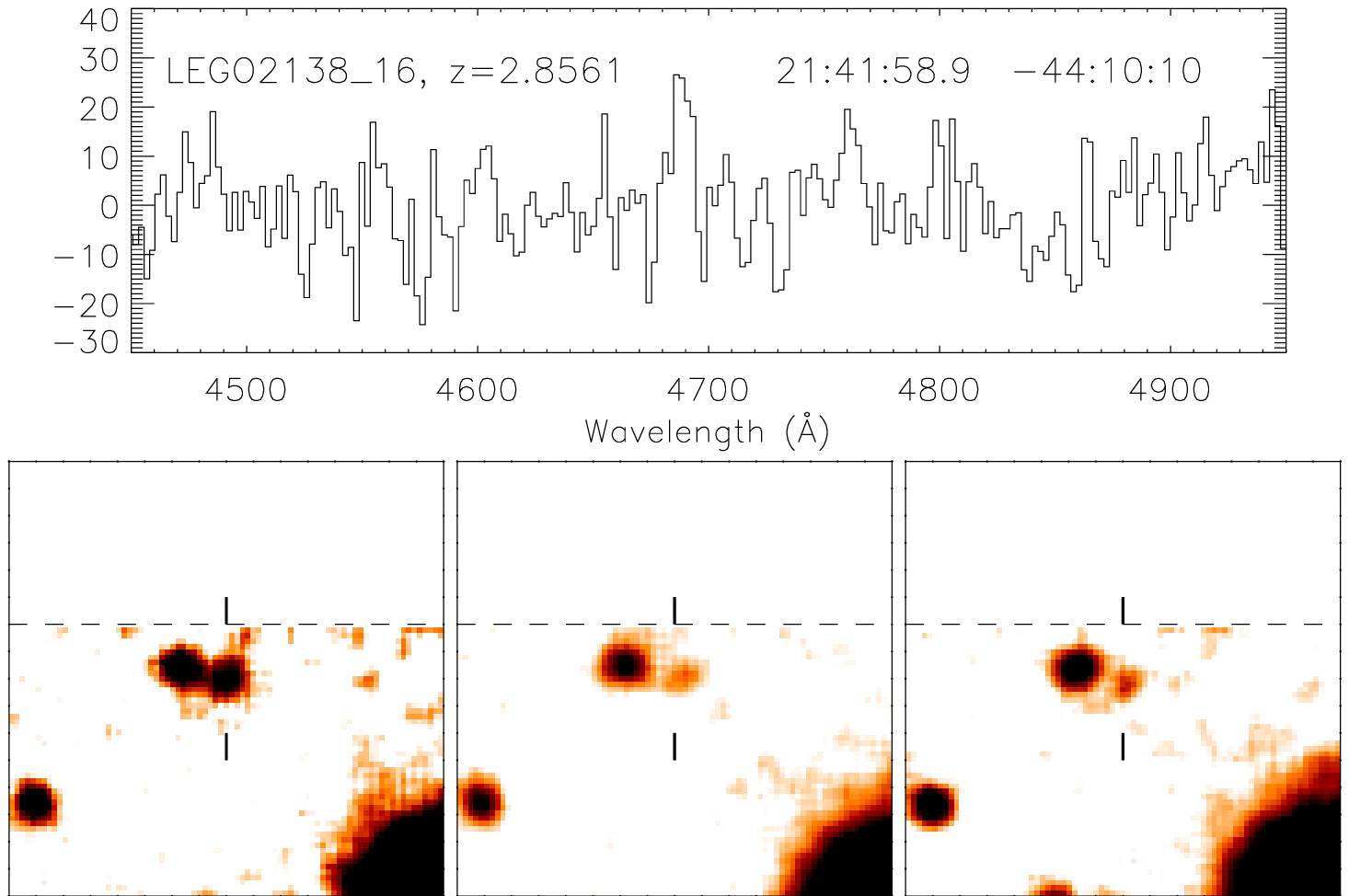, width=5.5cm}
\epsfig{file=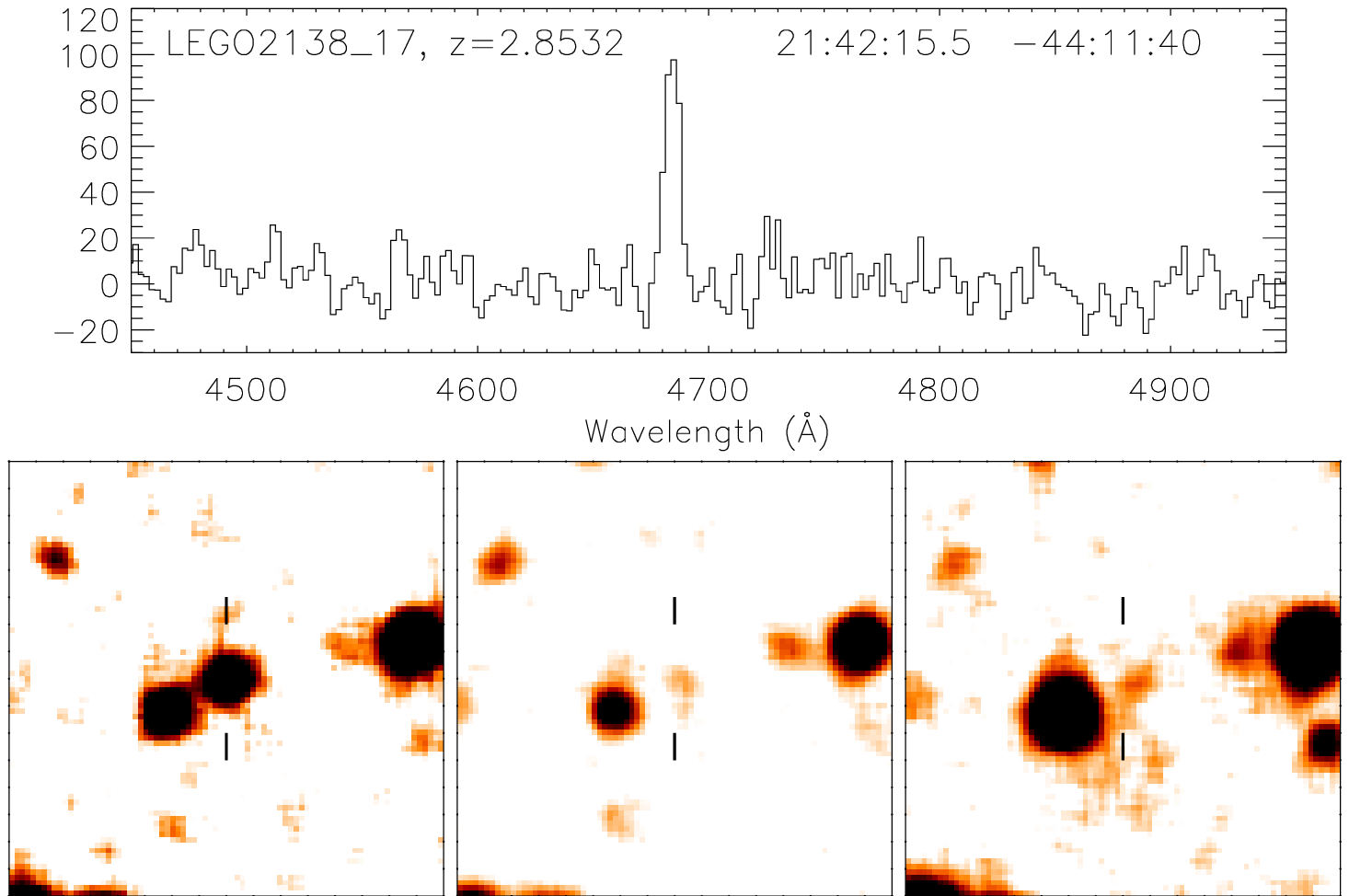, width=5.5cm}
\epsfig{file=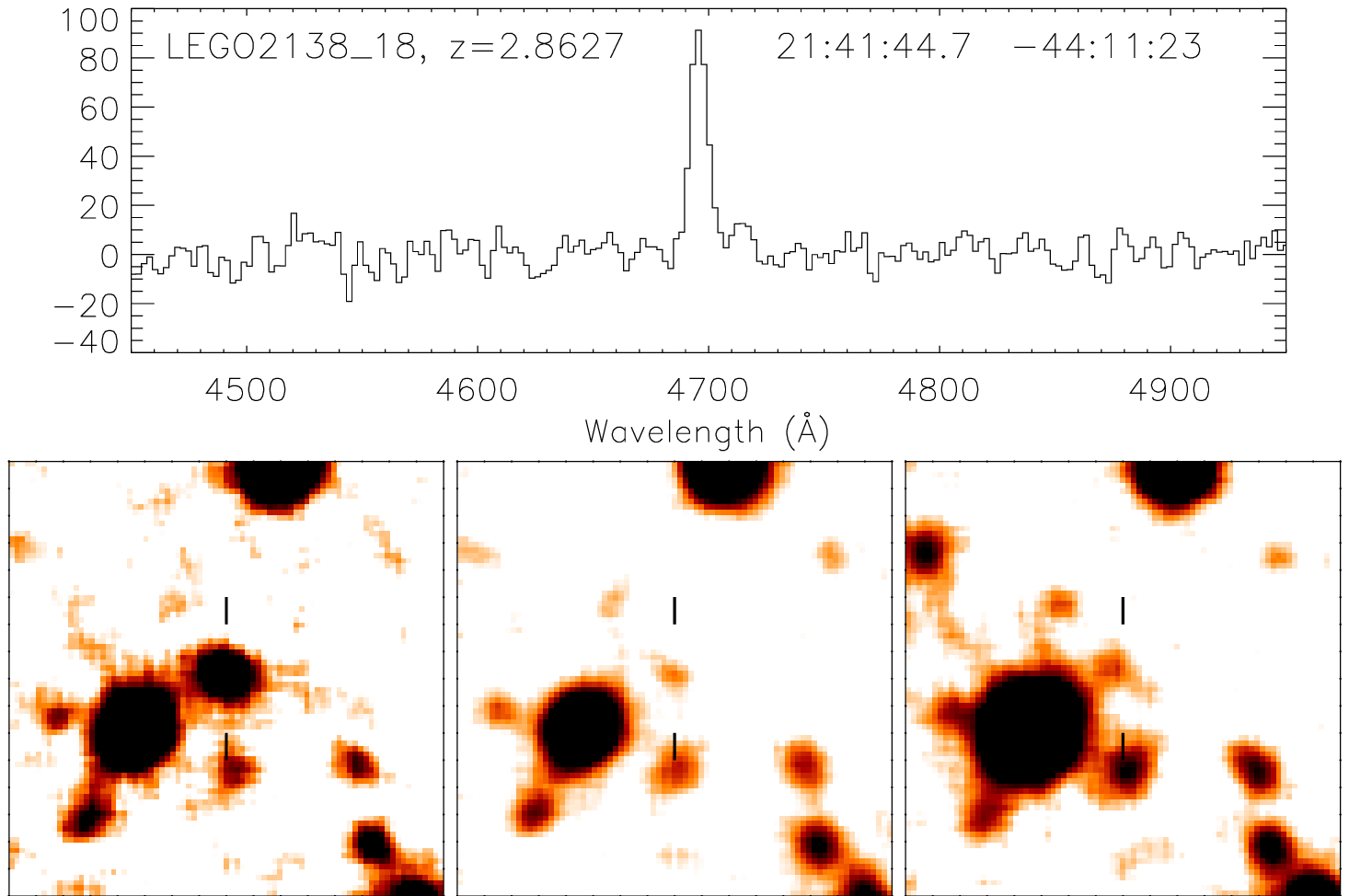, width=5.5cm}\\
\vskip 0.2cm
\epsfig{file=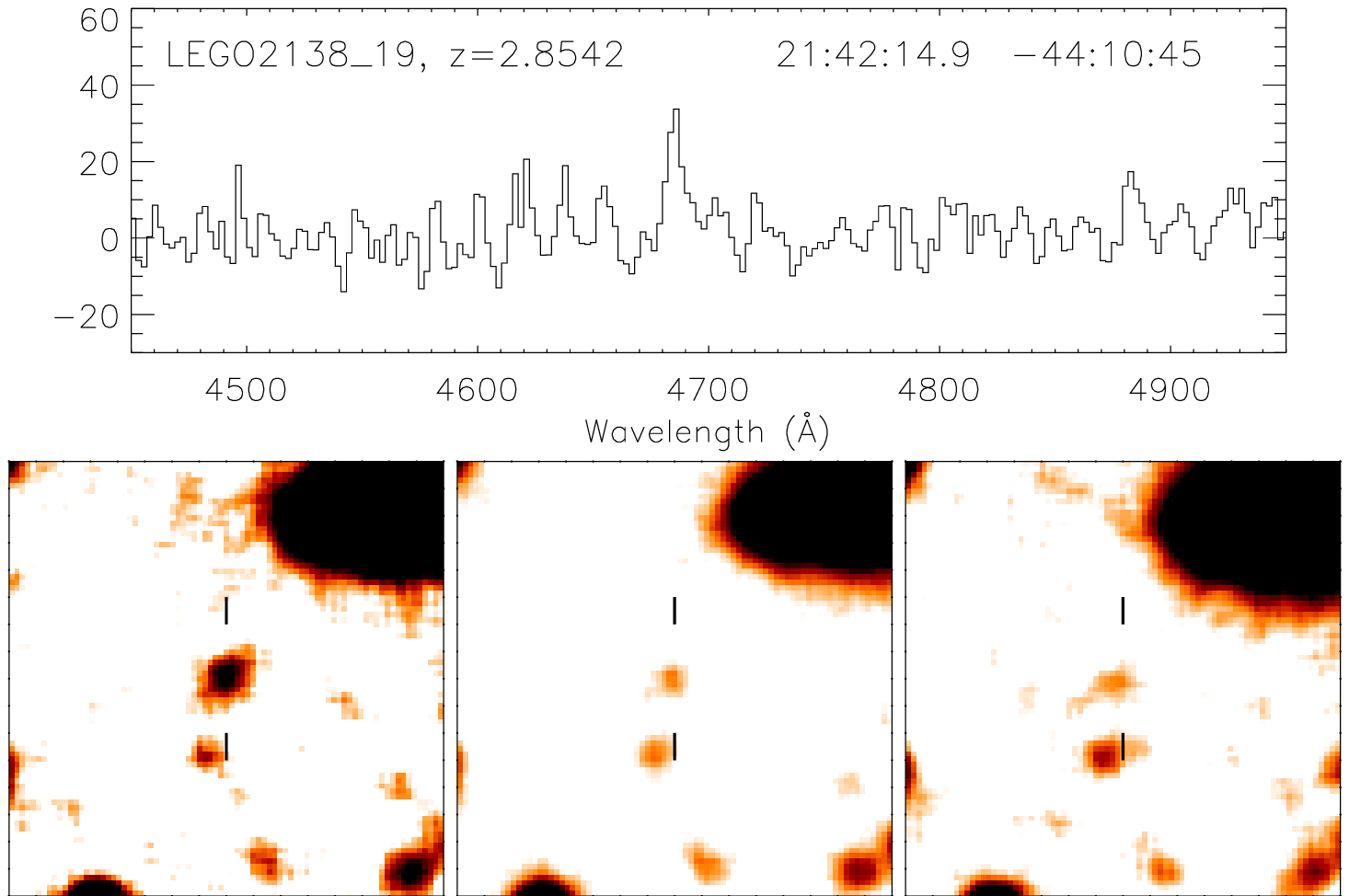, width=5.5cm}
\epsfig{file=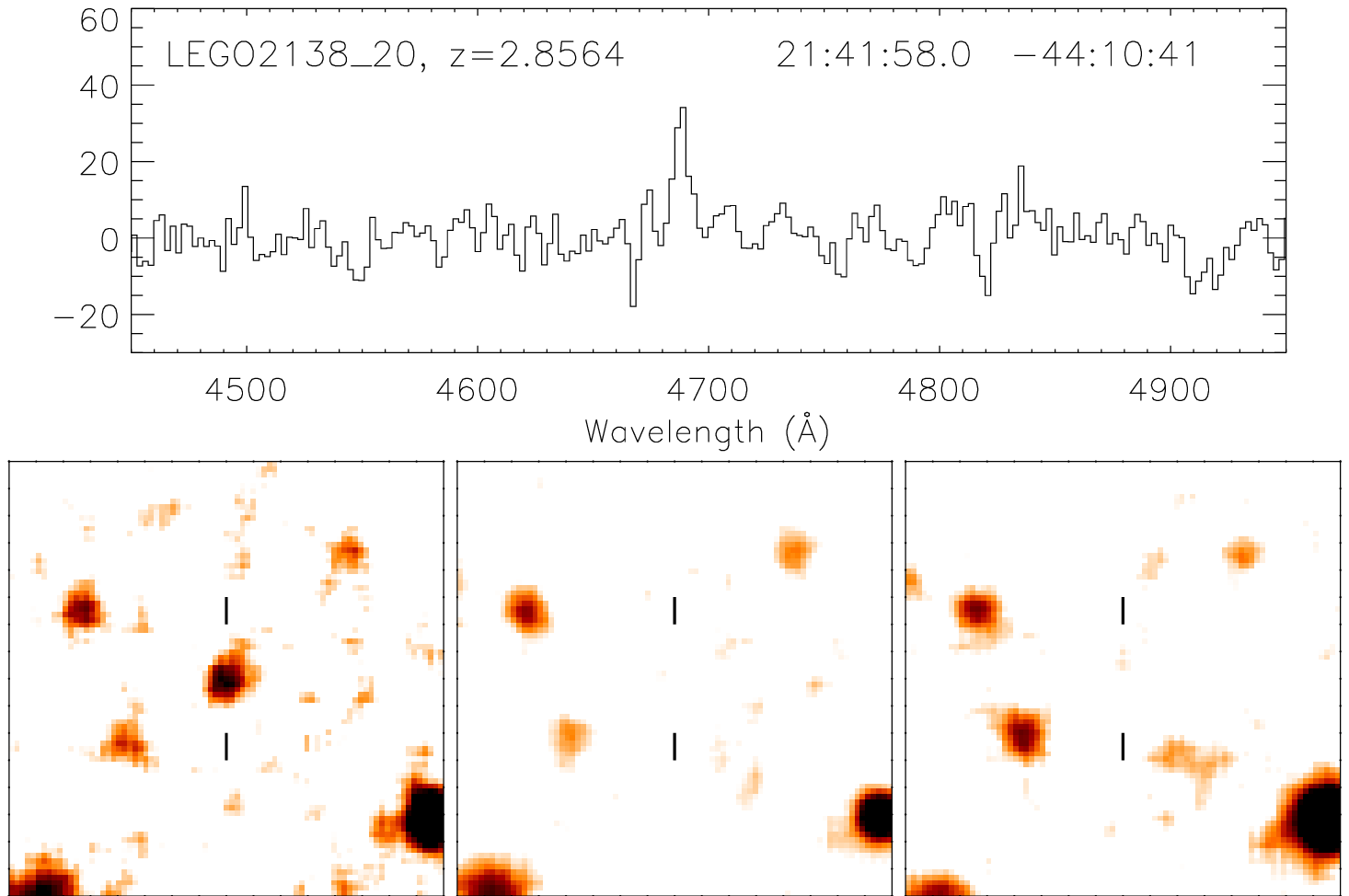, width=5.5cm}
\epsfig{file=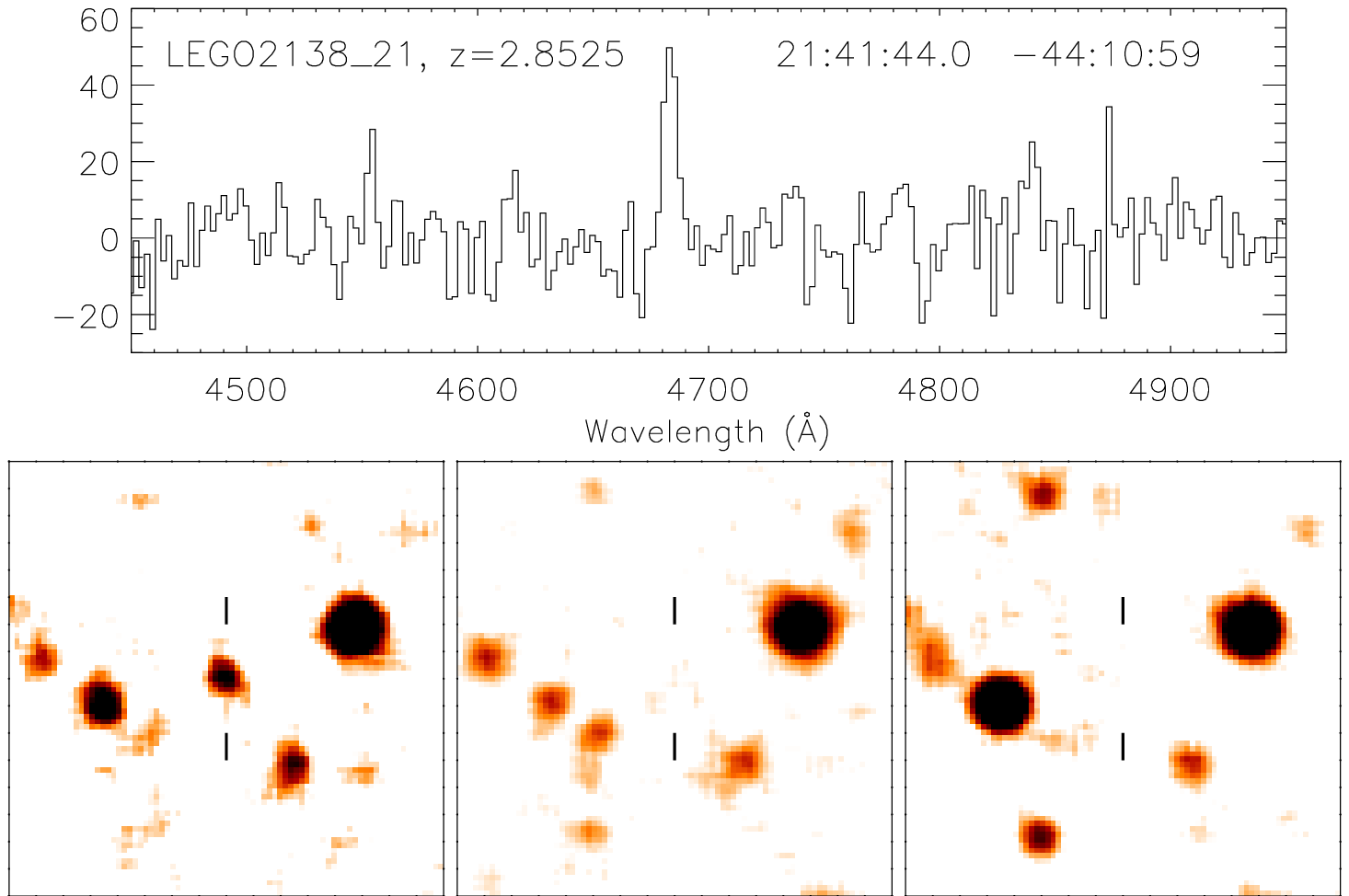, width=5.5cm}\\
\vskip 0.2cm
\epsfig{file=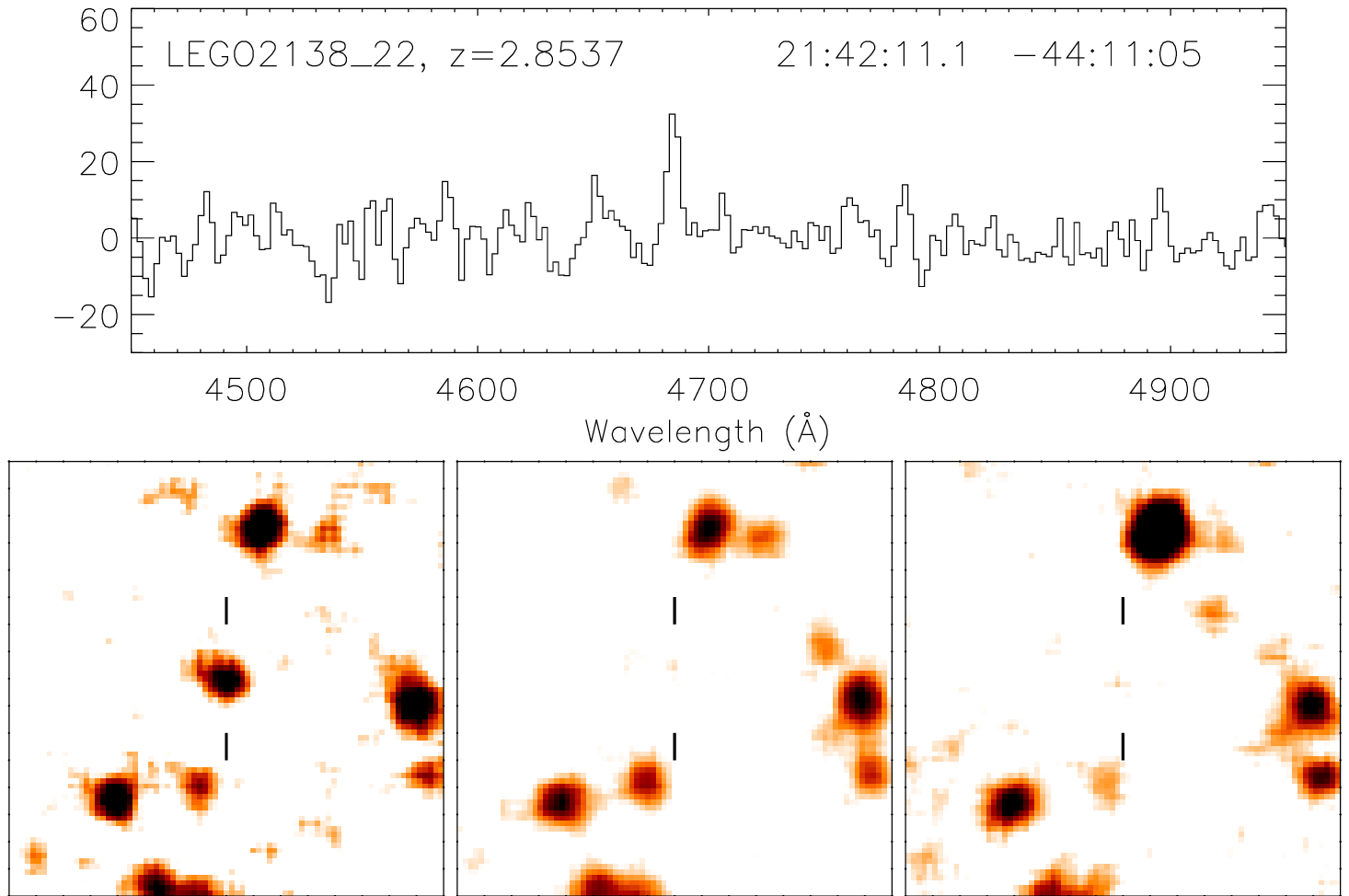, width=5.5cm}
\epsfig{file=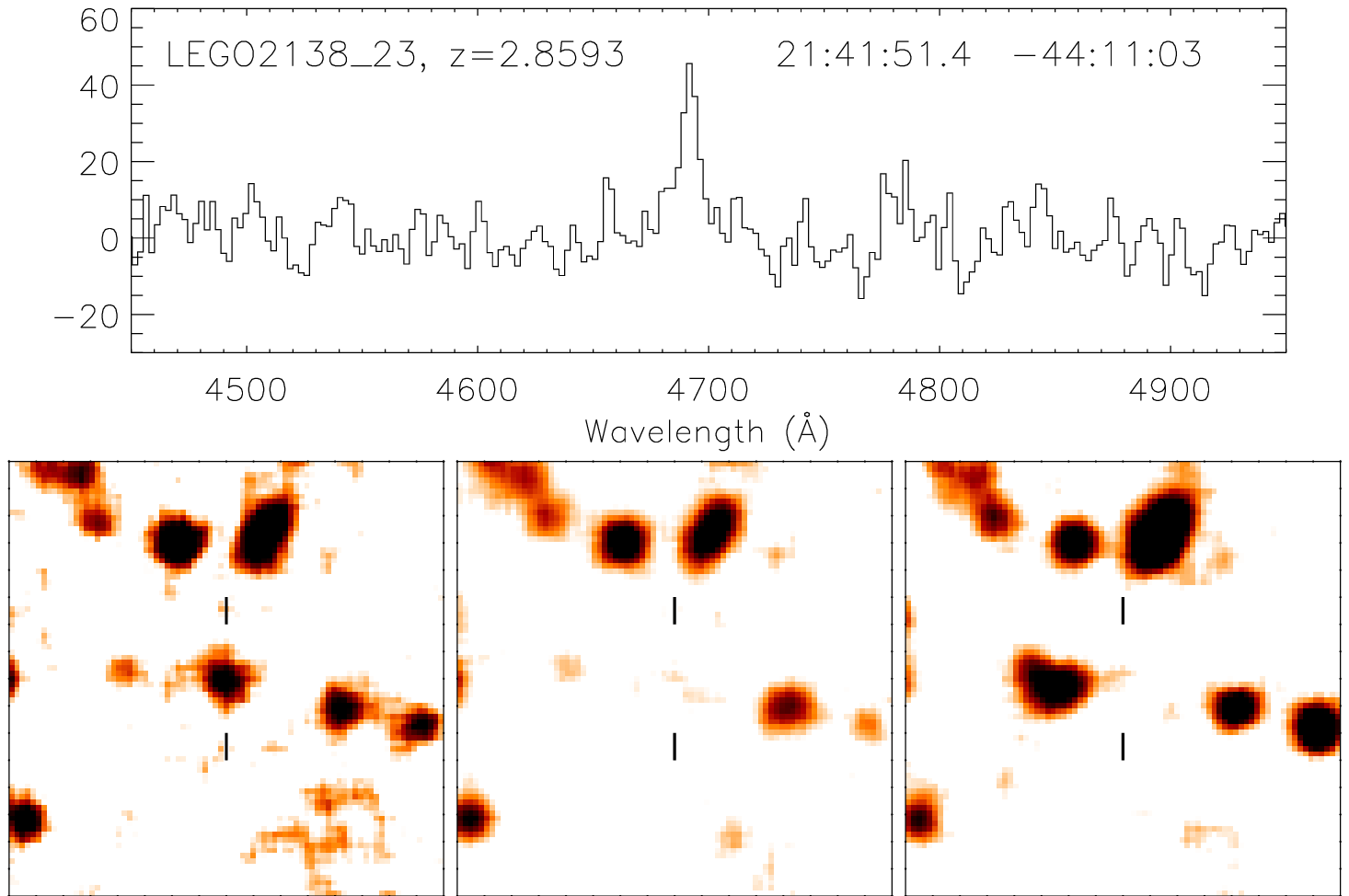, width=5.5cm}
\epsfig{file=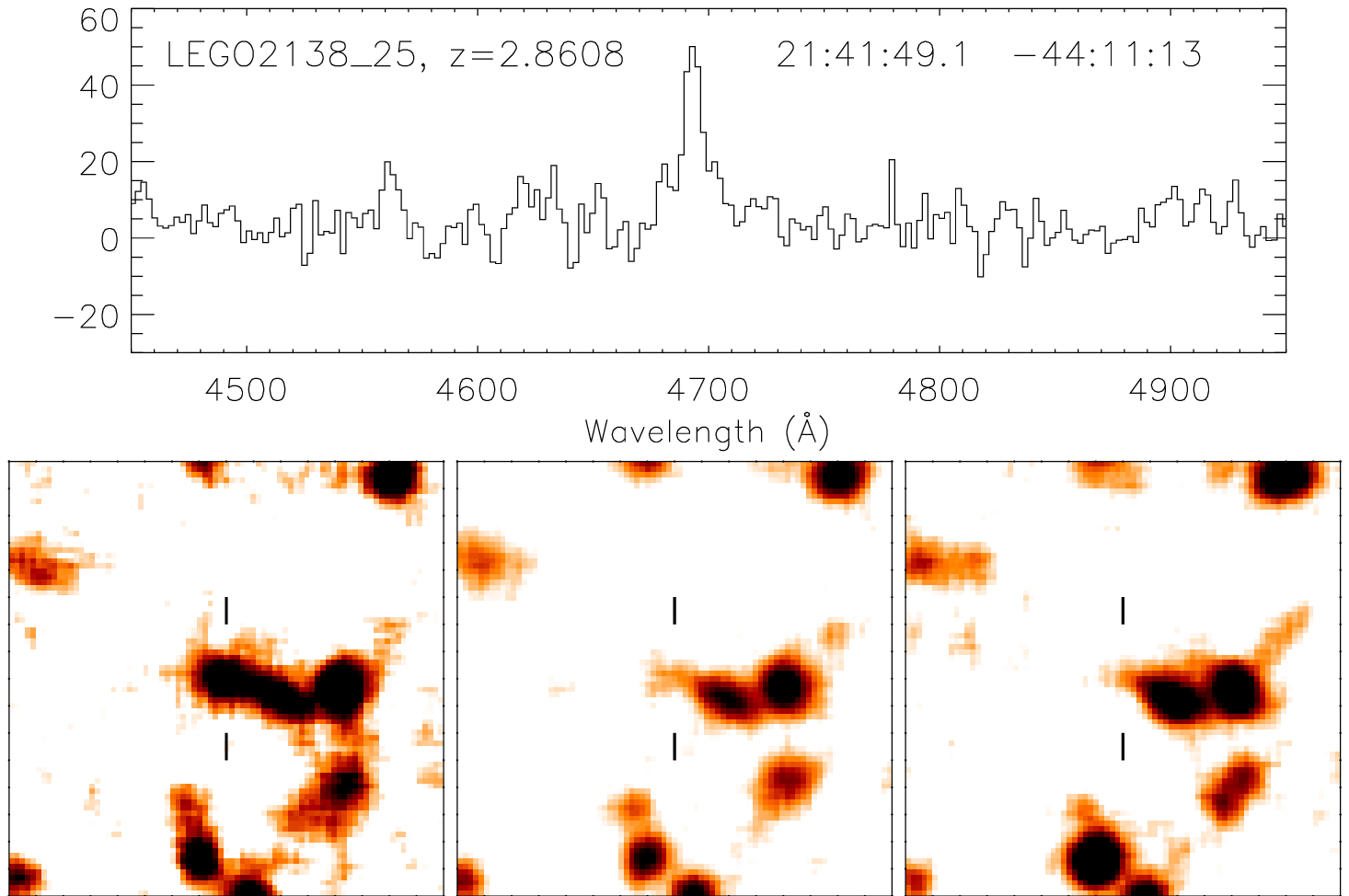, width=5.5cm}\\
\vskip 0.2cm
\epsfig{file=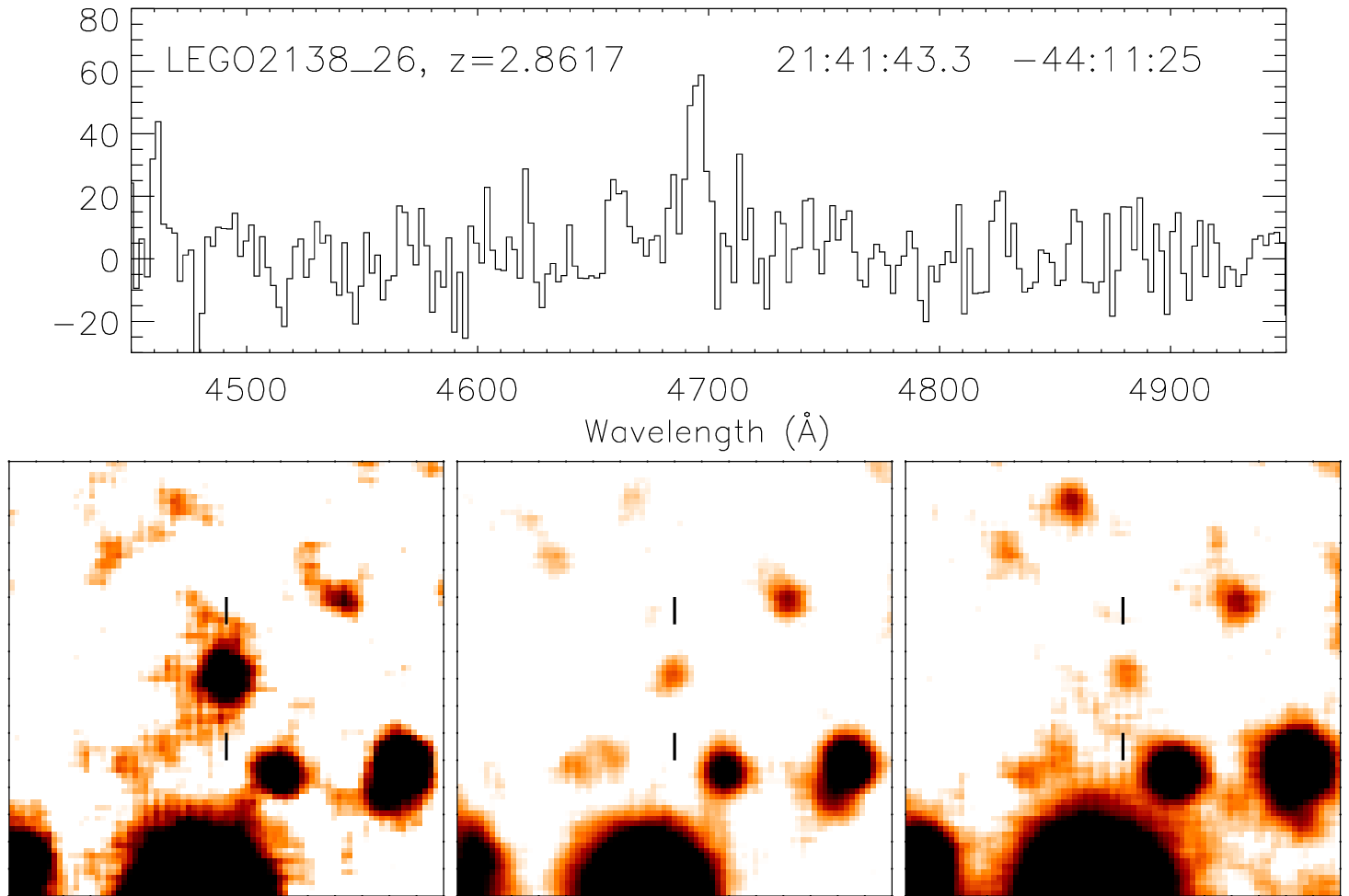, width=5.5cm}
\epsfig{file=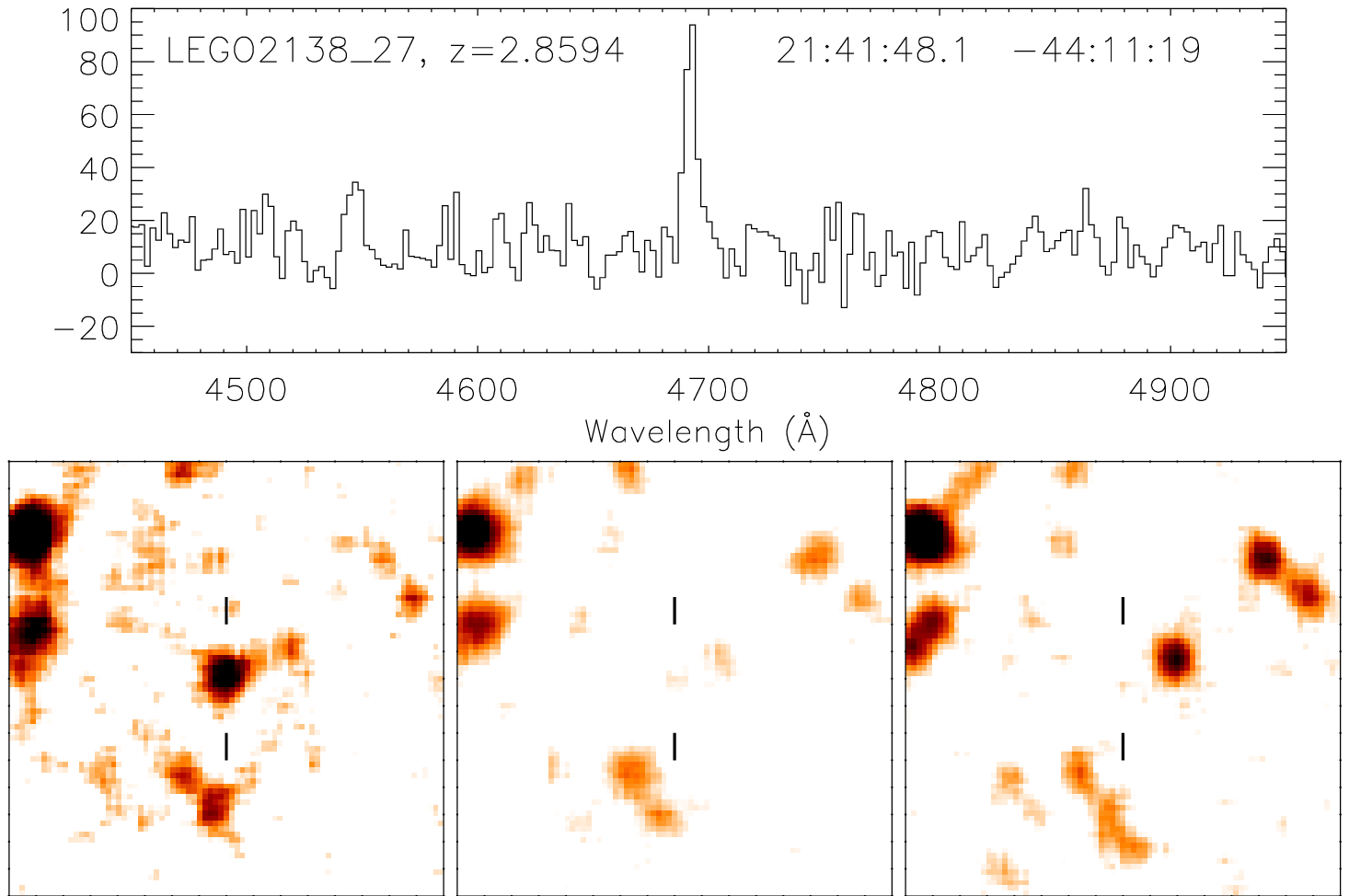, width=5.5cm}
\epsfig{file=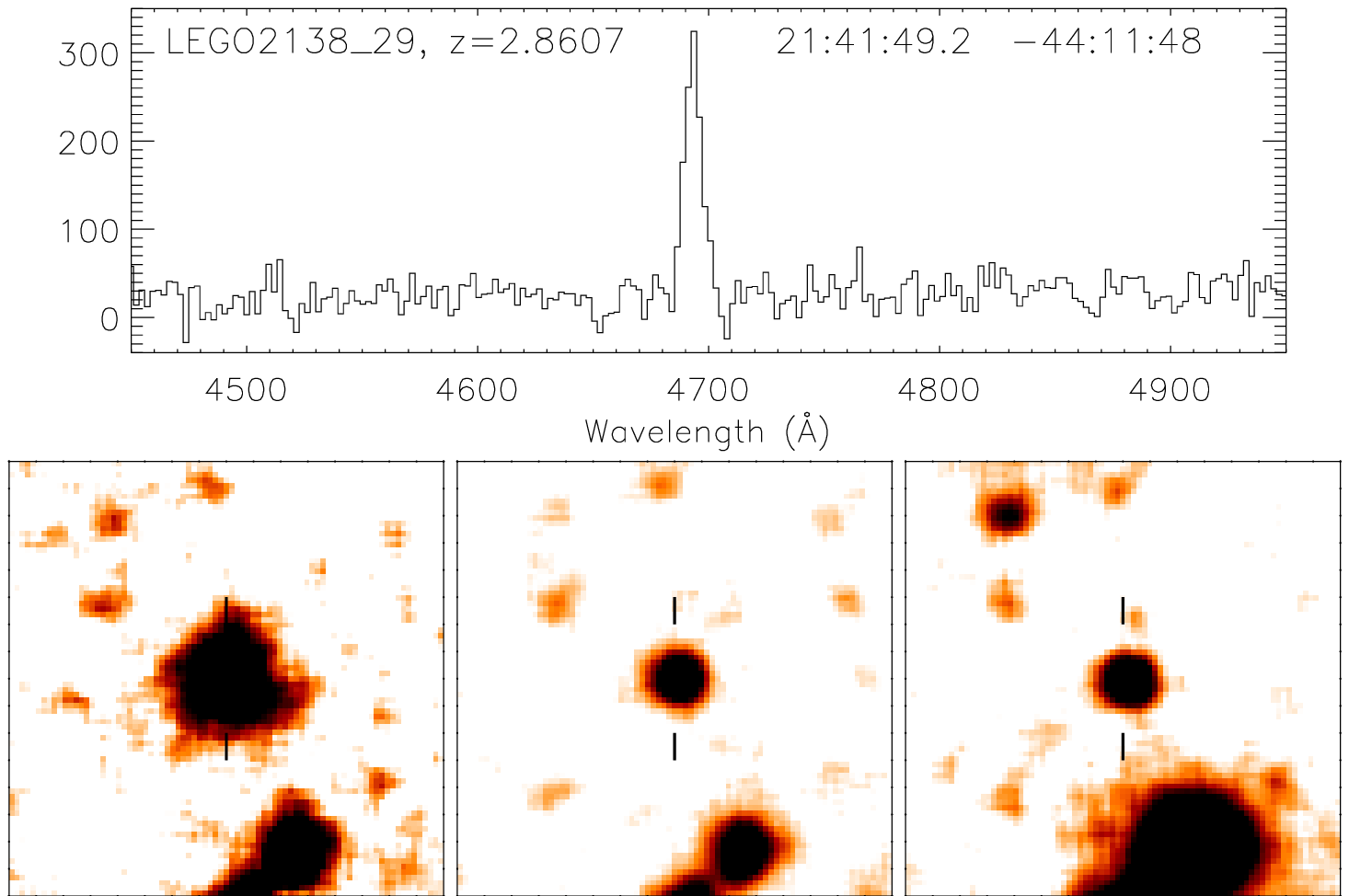, width=5.5cm}\\
\caption{
This figure is continued on the next page.
}
\end{center}
\end{figure*}

\addtocounter{figure}{-1}

\begin{figure*}
\begin{center}
\epsfig{file=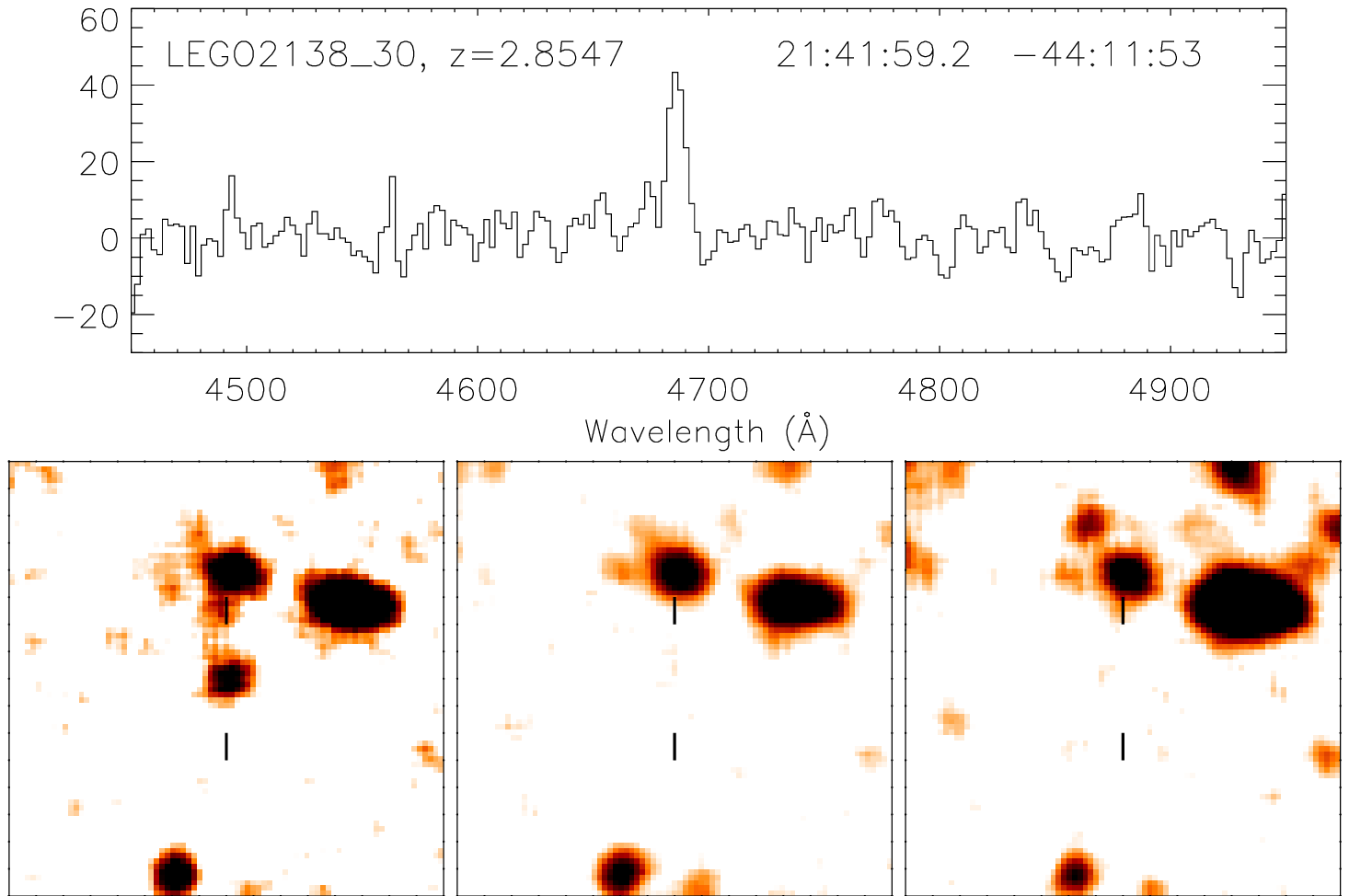, width=5.5cm}
\epsfig{file=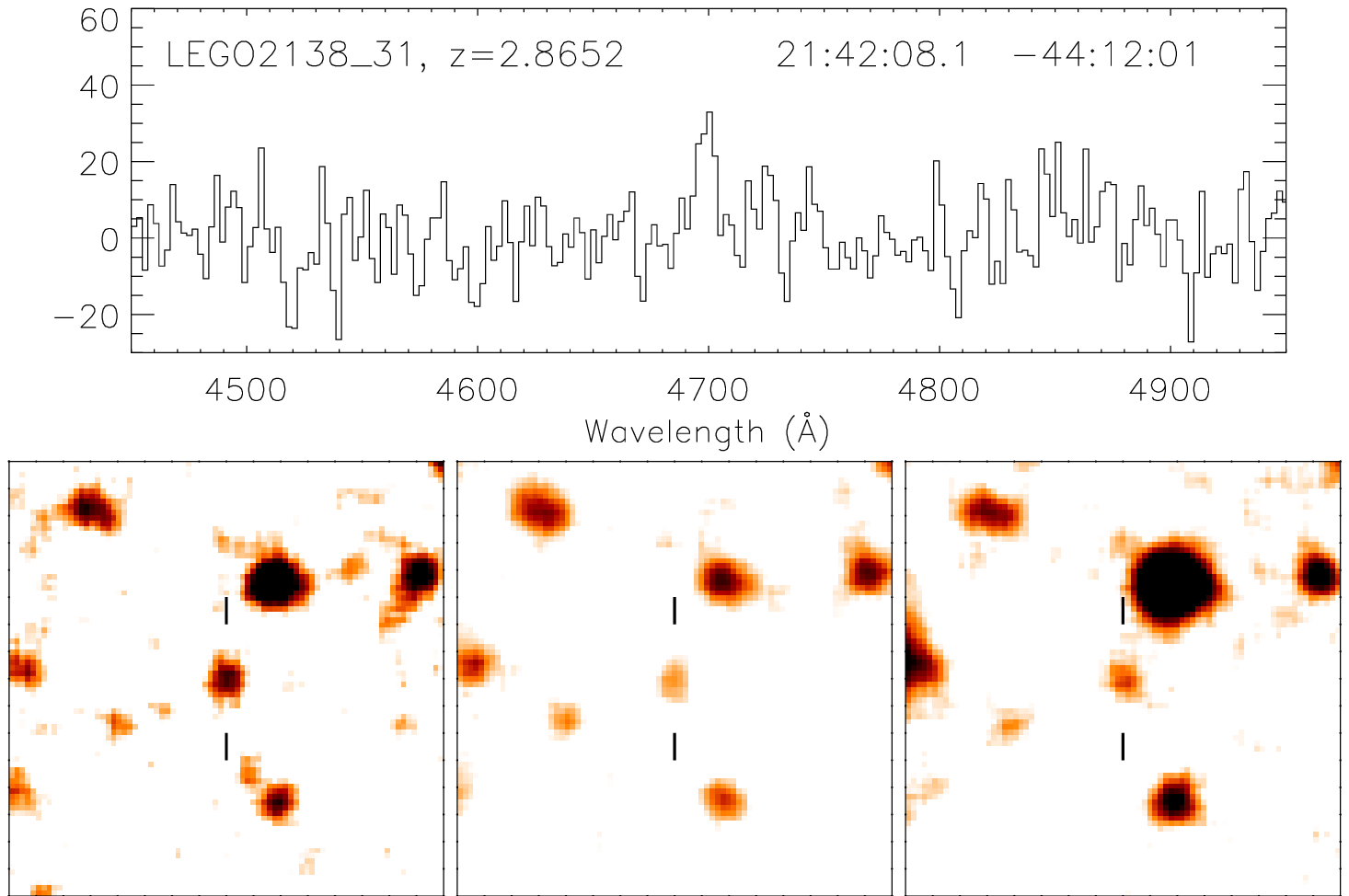, width=5.5cm}
\epsfig{file=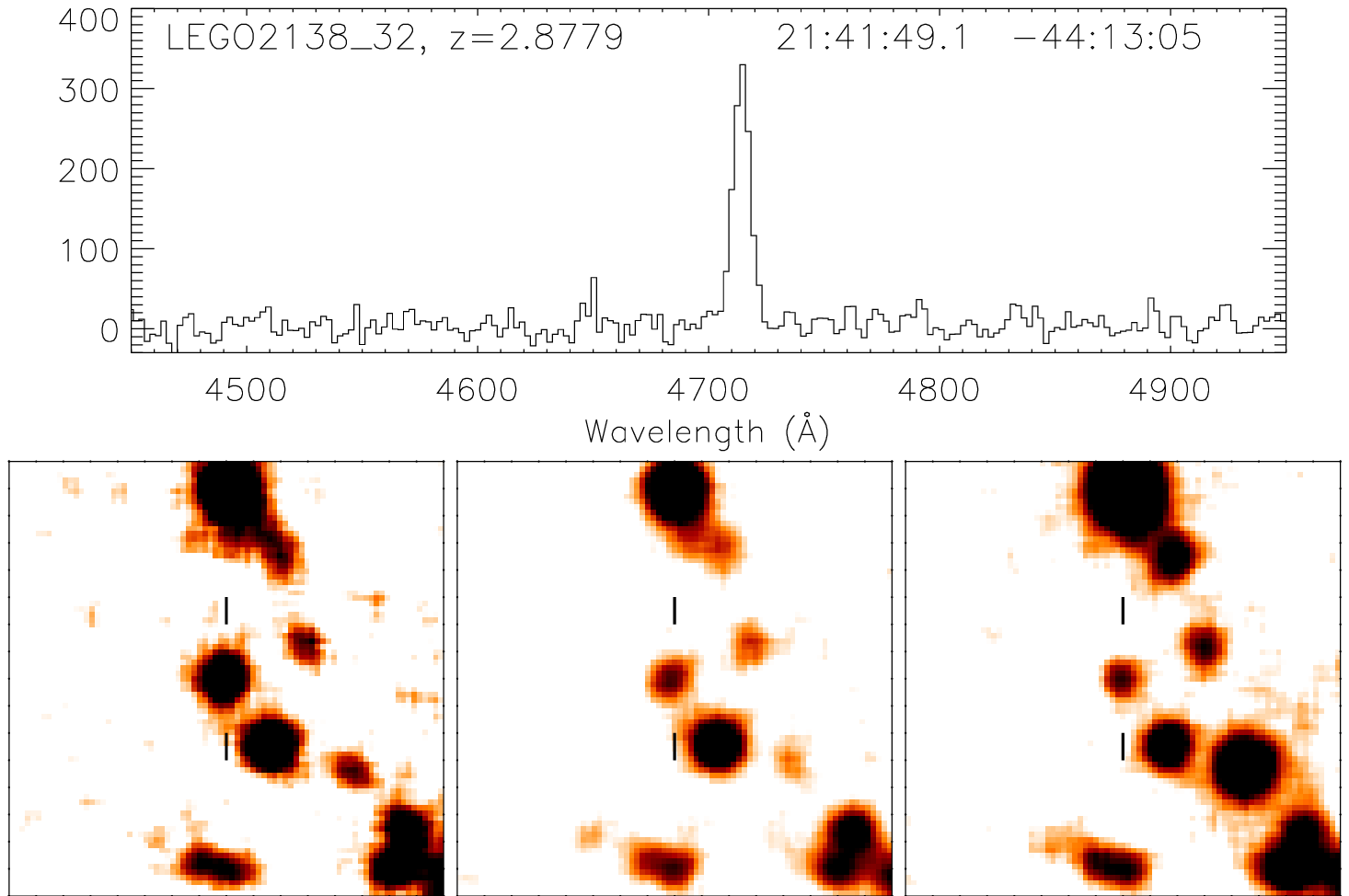, width=5.5cm}\\
\epsfig{file=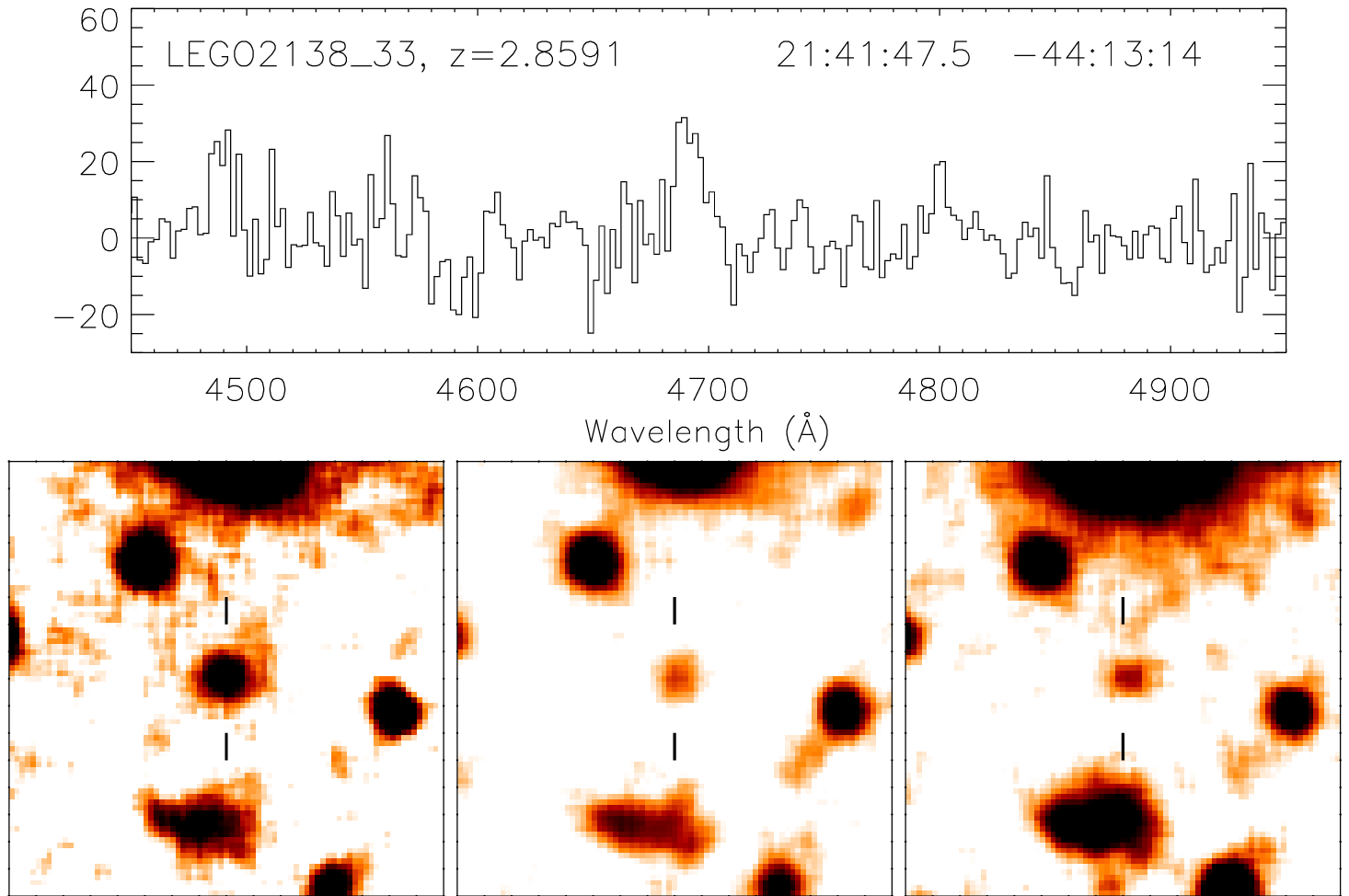, width=5.5cm}
\epsfig{file=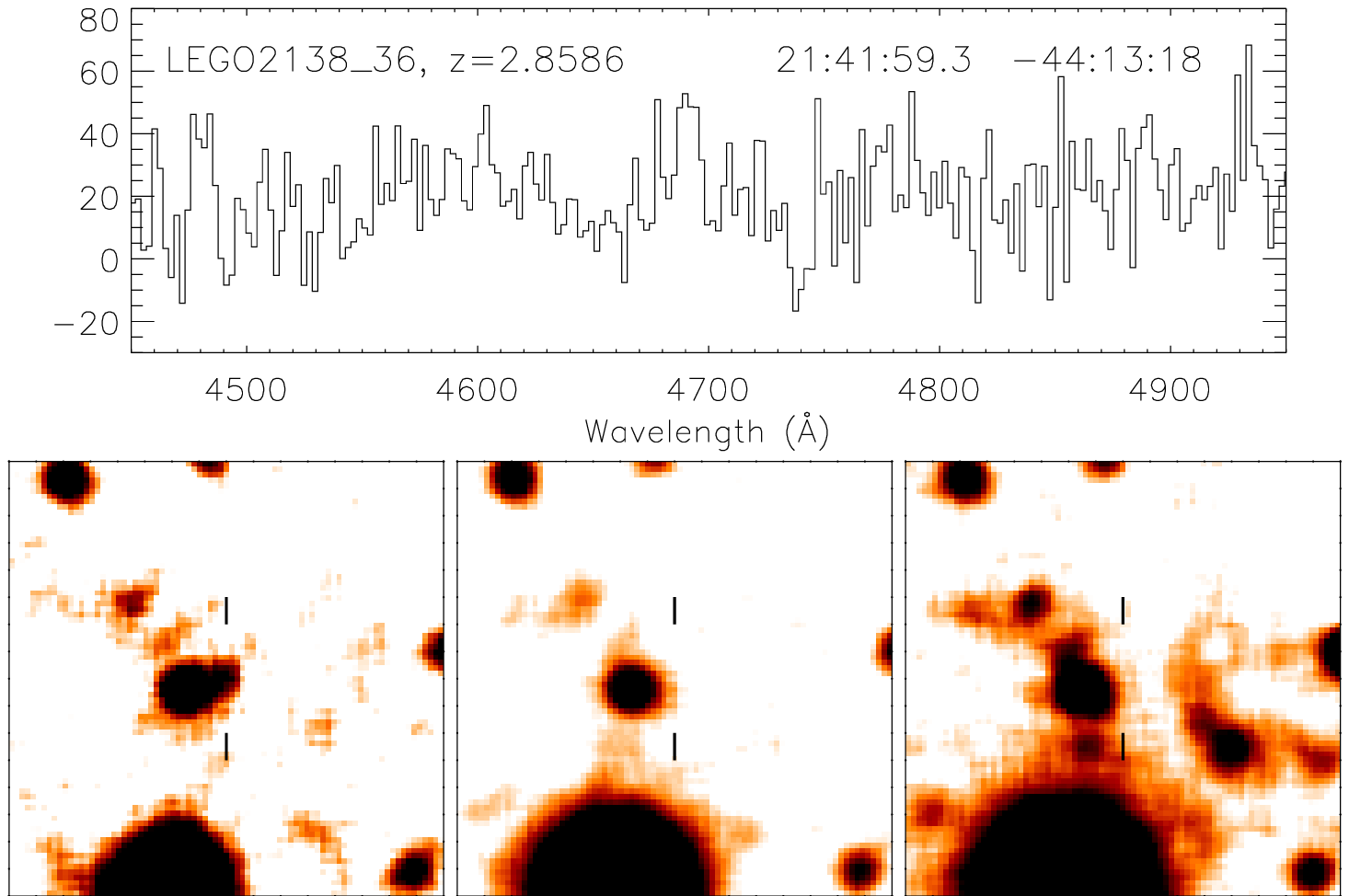, width=5.5cm}
\caption{
16$\times$16 arcsec$^2$ images and 1-D spectra of 23 confirmed LEGOs in the
field of Q\,2138$-$4427. For each candidate, we show images from the
narrow-, B- and R-band filters (from left to right). The units on the
ordinate of the spectra are counts in 1800 sec (as in Fynbo et al. 2001). The
name, redshift and coordinates (Epoch 2000) are provided for each object.
}
\label{candfigs2138}
\end{center}
\end{figure*}

For the confirmed LEGOs, we derived Ly$\alpha$ fluxes, equivalent
widths (EWs) and star-formation rates (SFRs) as described in detail
in Fynbo et al. (2002).

\subsection{Basic properties of LEGOs}
\begin{figure}
\begin{center}
\epsfig{file=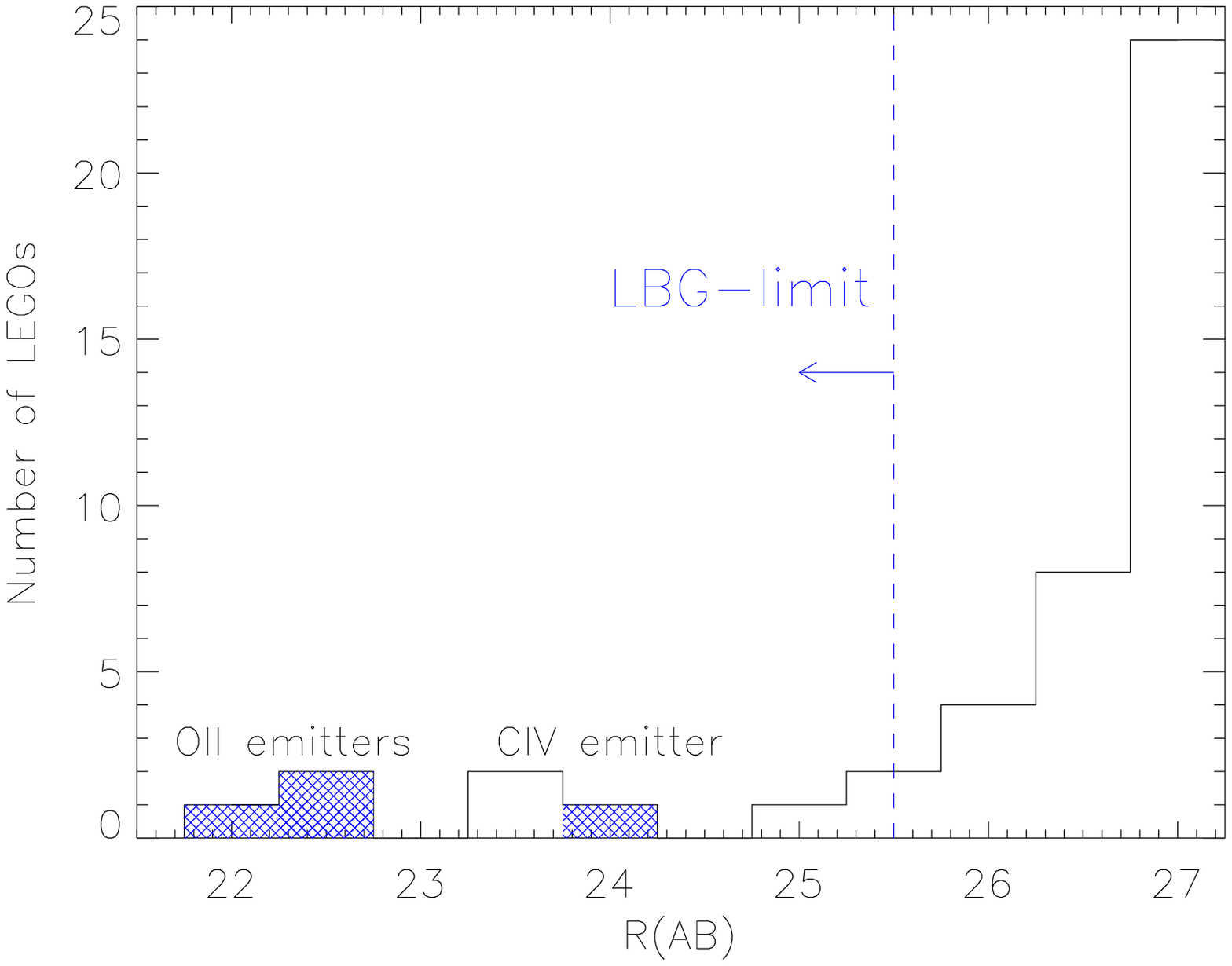, width=8.5cm}
\caption{Histogram of R(AB) magnitudes of 45 confirmed emission-line sources.
Objects detected at less than 5$\sigma$ in the R band are located in
the last bin, corresponding to R=27, and the 4 foreground emission-line sources
(three [\ion{O}{ii}] emitters and one \ion{C}{iv} emitter) are indicated with
the double-hatched pattern at the bright end. {\it 85\% of the LEGOs are
fainter than the spectroscopic limit of R(AB)=25.5 for LBGs}.
}
\label{lumfunc}
\end{center}
\end{figure}

The confirmed LEGOs are in general very faint. In Fig.~\ref{lumfunc}, we
show an histogram with the R(AB) magnitudes measured for confirmed
emission-line sources. At the bright end (R$<$24), we find that 4 out of 6
sources are foreground emission-line galaxies (three [\ion{O}{ii}] emitters
and a z=2.0364 AGN with \ion{C}{iv} located in the narrow-band filter). 85\%
of the confirmed LEGOs are fainter than the R(AB)=25.5 spectroscopic limit
for LBGs in current ground-based surveys (see e.g. Fig.~6 of
Shapley et al. 2003). The few LEGOs in our sample that are so bright that
they would have made it into the LBG samples are quite remarkable objects.
LEGO2138\_29 (see Fig.~\ref{candfigs2138}) has Ly$\alpha$ emission that is
much more extended than its continuum emission. This has already been seen
for other LEGOs (e.g. M\o ller \& Warren 1998; Fynbo et al. 2001) but not as
clearly as in this case. LEGO1346\_17 (see Fig.~\ref{candfigs1346}) seems
to be part of a complex system with several components, although we
cannot exclude chance alignment.

\begin{figure}
\begin{center}
\epsfig{file=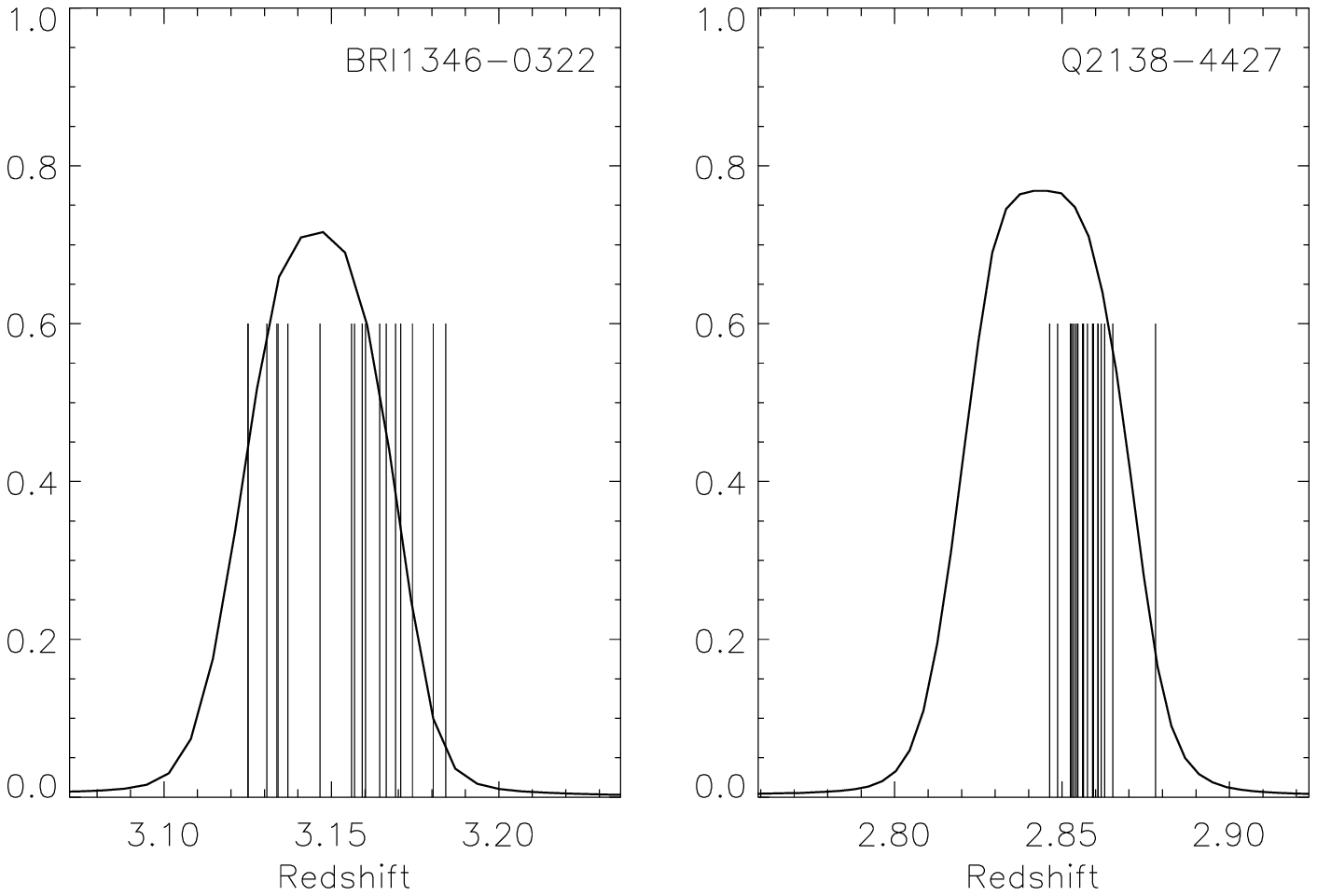, width=8.5cm}
\caption{Redshift distributions of LEGOs in the two fields relative to
the filter response curves. The redshifts of LEGOs in the field of
BRI\,1346$-$0322 fill out the volume probed by the filter, whereas LEGOs in
the field of Q\,2138$-$4427 have a very narrow redshift distribution.
}
\label{zdist}
\end{center}
\end{figure}

Based on observed Ly$\alpha$ fluxes, the range of SFRs for the confirmed LEGOs
is 0.20--15 M$\odot$ yr$^{-1}$ if the extinction is negligible. The observed
EWs range from less than 100 \AA\ to more than 1000 \AA . This corresponds to
about 20 \AA\ to 250 \AA\ in the rest-frame and is consistent with
the theoretically expected Ly$\alpha$ EWs for metal-poor starburst galaxies
(Charlot \& Fall 1993; Valls-Gabaud 1993; Schaerer 2003).

The surface density of confirmed sources is of the order of 10 arcmin$^{-2}$
per unit redshift down to a Ly$\alpha$ flux limit of 
$\sim$7$\times$10$^{-18}$ erg s$^{-1}$ cm$^{-1}$ and the EW limit shown in
Fig.~\ref{select}. This is about a factor of five higher than the
surface density of LBGs down to R=25.5, $\sim$2$\times$1.2 (Steidel et 
al. 1999) even if most LBGs are not themselves Ly$\alpha$ emitters. In other 
words, the LBGs are the tip of an iceberg consistent with the conclusion of 
Fynbo et al. (1999). This reflects the steepness of the luminosity 
function for z=3 galaxies.

In future papers, we will address the morphology, clustering properties
and luminosity function of the LEGOs from this and complementary VLT surveys.

\subsection{Redshift distributions}
In Fig.~\ref{zdist}, we show the redshift distributions of LEGOs in the
two fields relative to the filter response curves. The redshifts of LEGOs
observed in the field of BRI\,1346$-$0322 fill out the sampled volume, with
a mean redshift of 3.155 and a standard deviation of 0.019. On the contrary,
the redshifts of LEGOs in the field of Q\,2138$-$4427 have a mean of 2.858
and a standard deviation of only 0.006, corresponding to a velocity dispersion
of 470 km s$^{-1}$. A significant part of this velocity spread must be
caused by peculiar velocities or offsets between the Ly$\alpha$ and systemic
redshifts, and therefore, the Hubble flow depth of the structure should be
even less. The mean observed redshift is close to the redshift of the DLA
absorber toward Q\,2138$-$4427 (z=2.851). This indicates the presence of
a large-scale structure of galaxies, e.g. a pancake-like structure at the
redshift of the DLA absorber, surrounded by voids. Independent evidence
for this comes from the observation of strong metal absorption lines at
the same redshift in two nearby QSOs (Francis \& Hewitt 1993; D'Odorico et 
al. 2002). The redshift distribution of LEGOs in the field of Q\,2138$-$4427 
is similar to that of LEGOs in the fields of radio galaxies (Pentericci et 
al. 2000; Venemans et al. 2002). In a subsequent paper (Ledoux et al., in 
prep.), we will address the properties of the DLA absorber.

\subsection{Unconfirmed candidates}
Most of the candidates that we did not confirm are faint in
the narrow-band images and/or have low EWs (see Fig.~\ref{select}). 
These candidates could either have been missed by the slitlets, be
too faint for the follow-up spectroscopy (due to the bad seeing we did not
reach the planned detection limit), or simply not be LEGOs.  As
seen in Fig.~\ref{colcol}, the unconfirmed candidates tend to have
redder colours (B(AB)$-$R(AB) = 0.9--1.8) than the confirmed 
candidates (see Fig.~\ref{colcol}). The confirmed emission-line sources 
detected in the broad bands indeed have blue colours (typically 
B(AB)$-$R(AB)$<$0.8). However, the fact that one of the confirmed 
candidates, LEGO2138\_31, has colours and flux very similar to the 
unconfirmed candidates makes it plausible that at least some of these 
will be confirmed with deeper spectroscopy under better seeing 
conditions. This we plan to check with future observations.

\subsection{Interloopers}
\begin{figure*}
\begin{center}
\epsfig{file=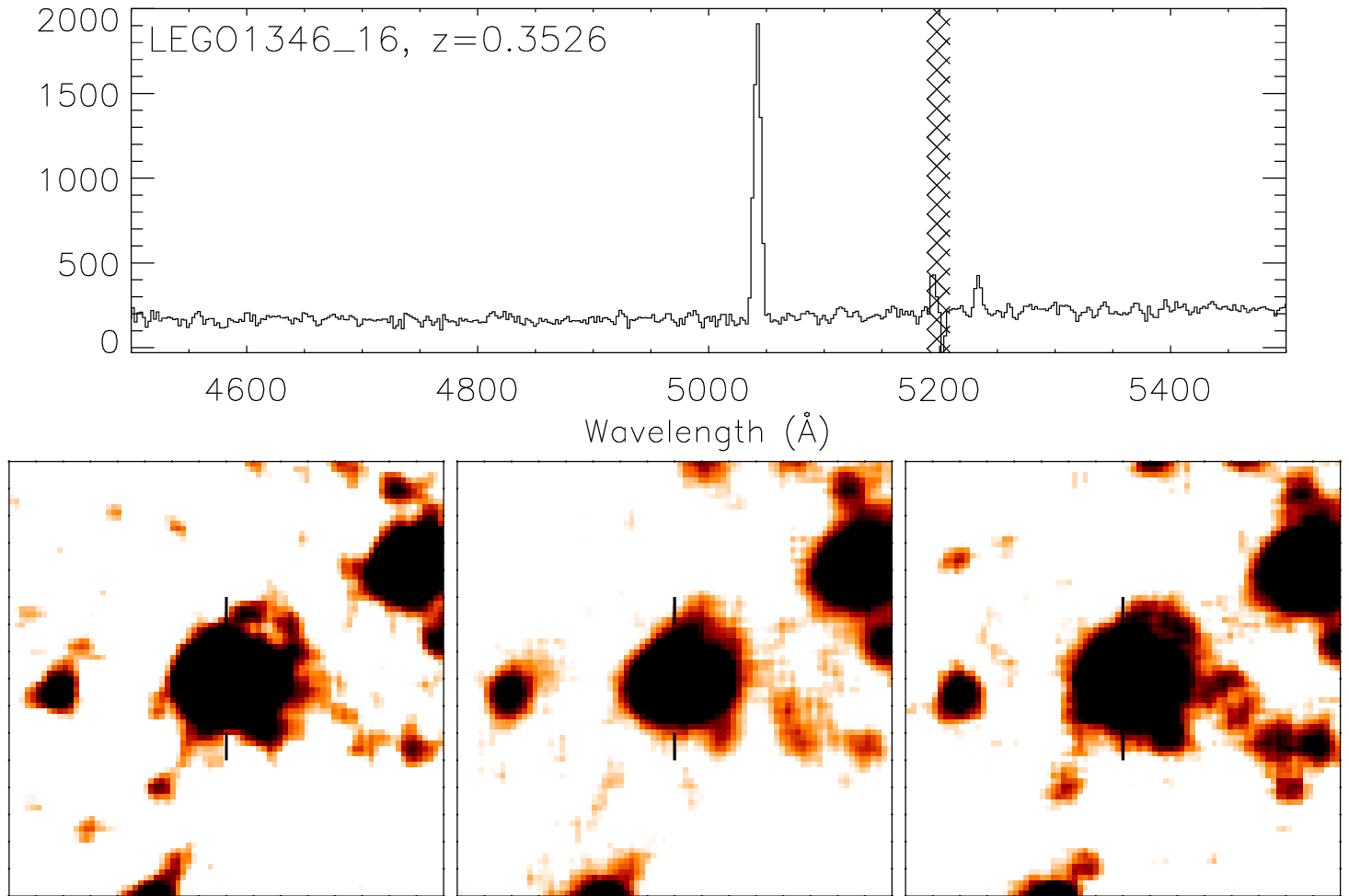, width=5.5cm}
\epsfig{file=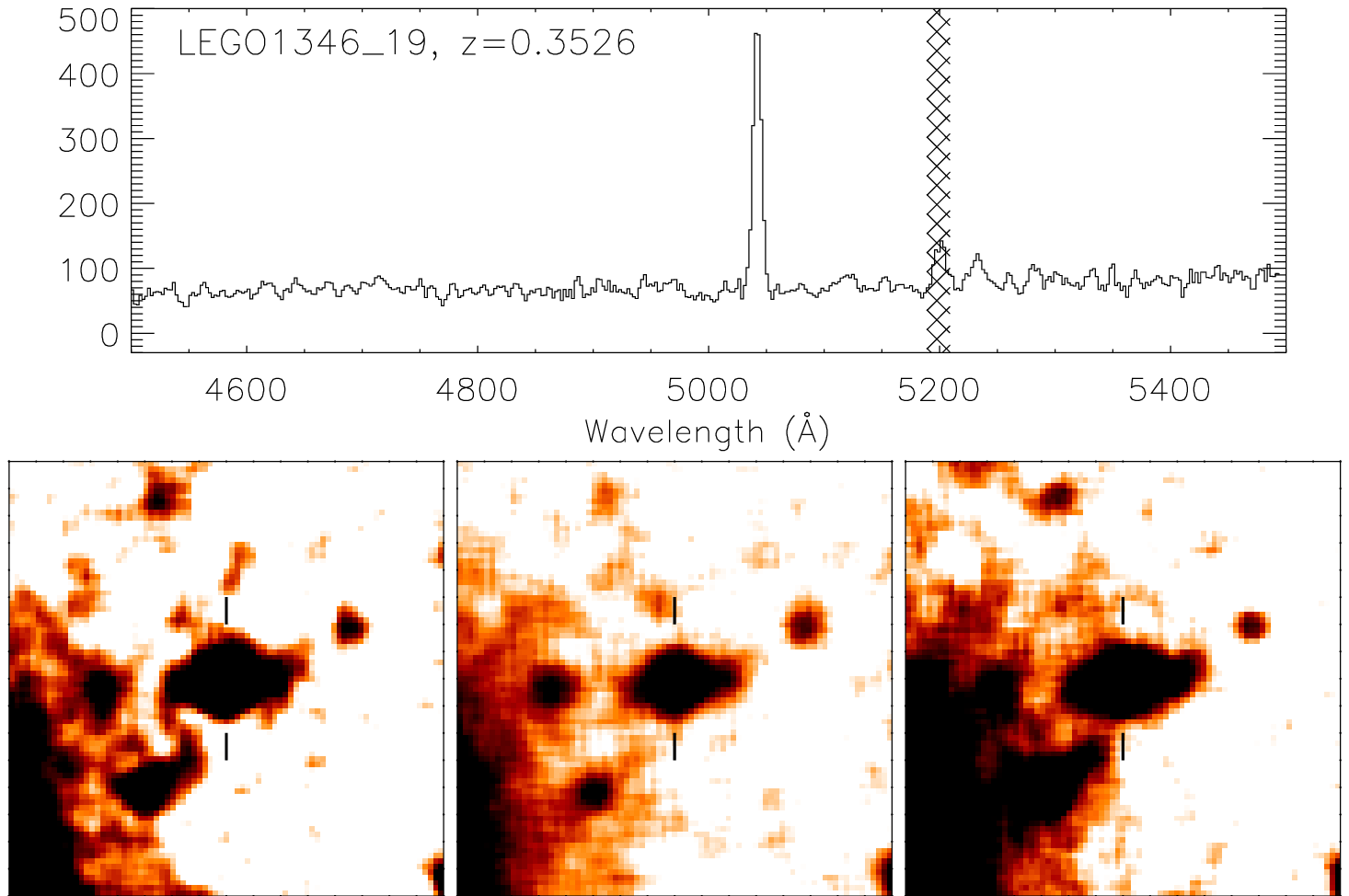, width=5.5cm}
\epsfig{file=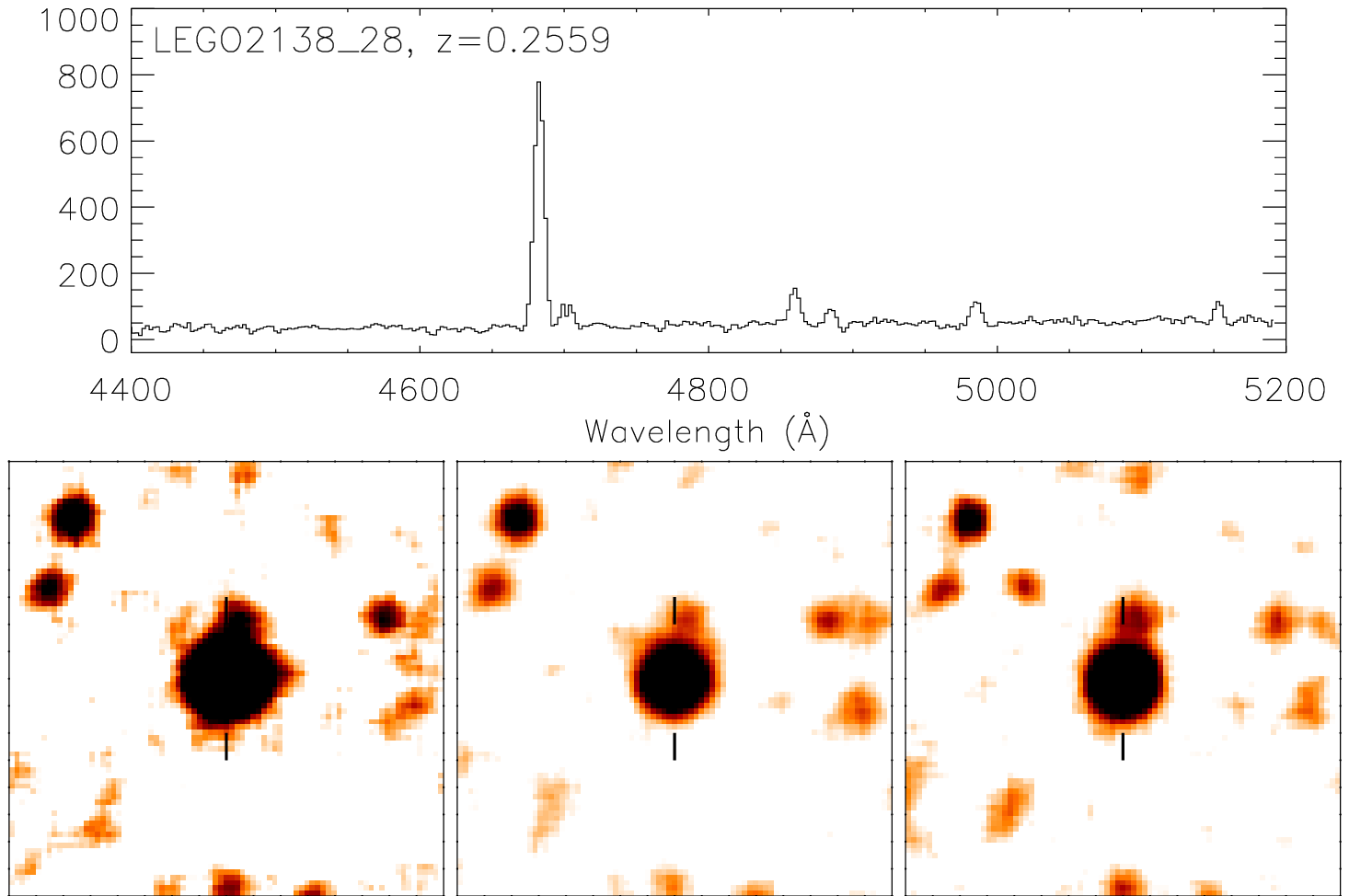, width=5.5cm}\\
\vskip 0.2cm
\epsfig{file=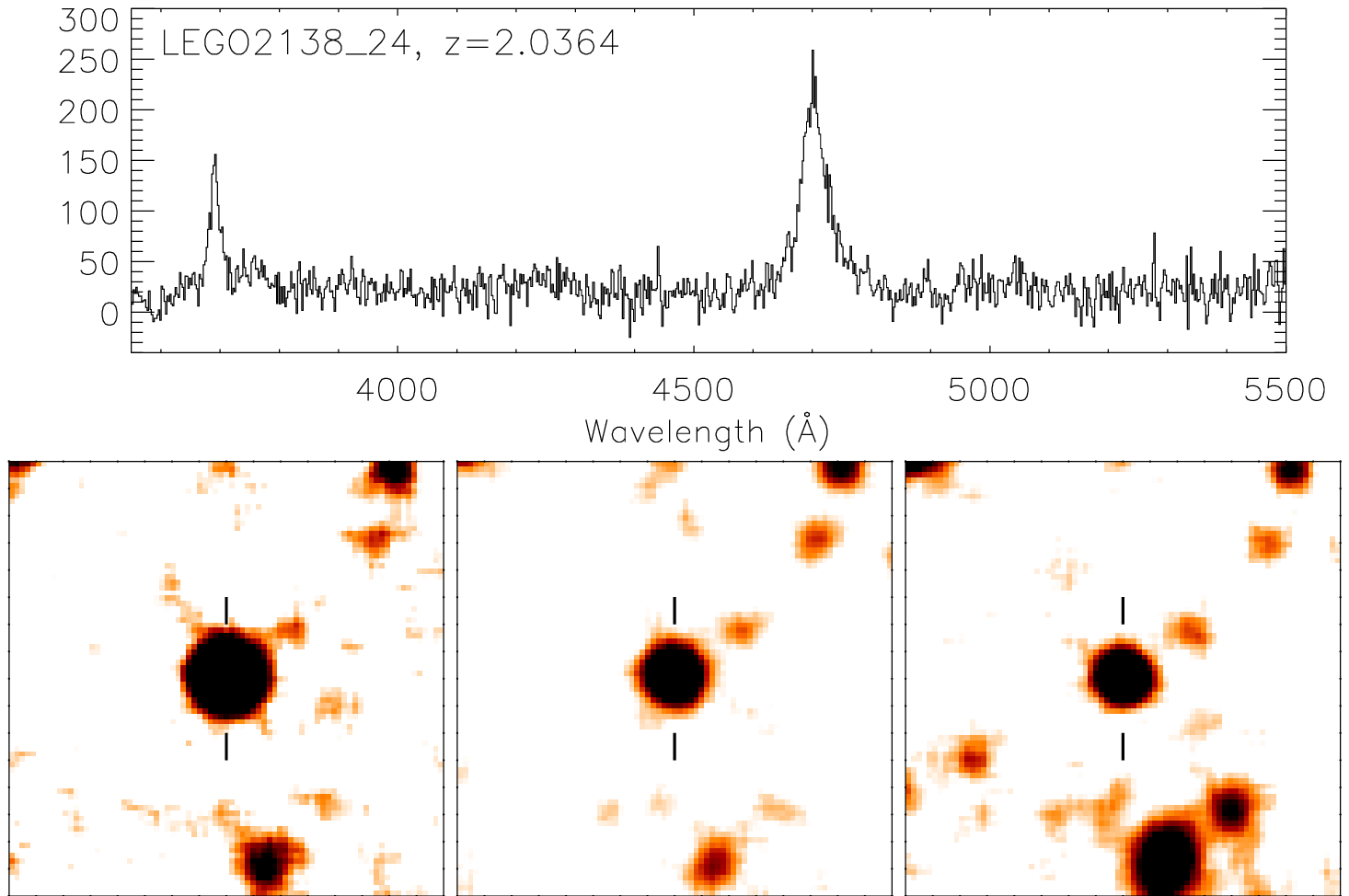, width=5.5cm}
\caption{
Shown are 16$\times$16 arcsec$^2$ images and 1-D spectra of the
four candidates that turned out to be foreground objects with other lines
than Ly$\alpha$ observed in the narrow-band filters (three [\ion{O}{ii}]
emitters and an AGN with its \ion{C}{iv} emission line). The spectra are not
flux calibrated.}
\label{interlopers}
\end{center}
\end{figure*}

Four of our colour selected candidates are objects at lower redshifts
with other emission lines in the narrow filters (three [\ion{O}{ii}]
emitter and a faint (R(AB)=23.9) AGN with \ion{C}{iv} in the narrow filter. 
Similar AGNs are common as found e.g. in the Chandra Deep Field South
(Mainieri, private communication). Subtraction of the Point-Spread-Function
of the AGN reveals no obvious host galaxy and we conclude that the 
host galaxy must be either very faint or very compact. The 
Interloopers are all characterized by small observed EWs ($<100$ \AA ),
bright continuum magnitudes, and the presence of other lines in the spectra. 
Their spectra are shown in Fig.~\ref{interlopers}.

\subsection{Serendipitously detected LEGOs at other redshifts}

\begin{figure*}
\begin{center}
\epsfig{file=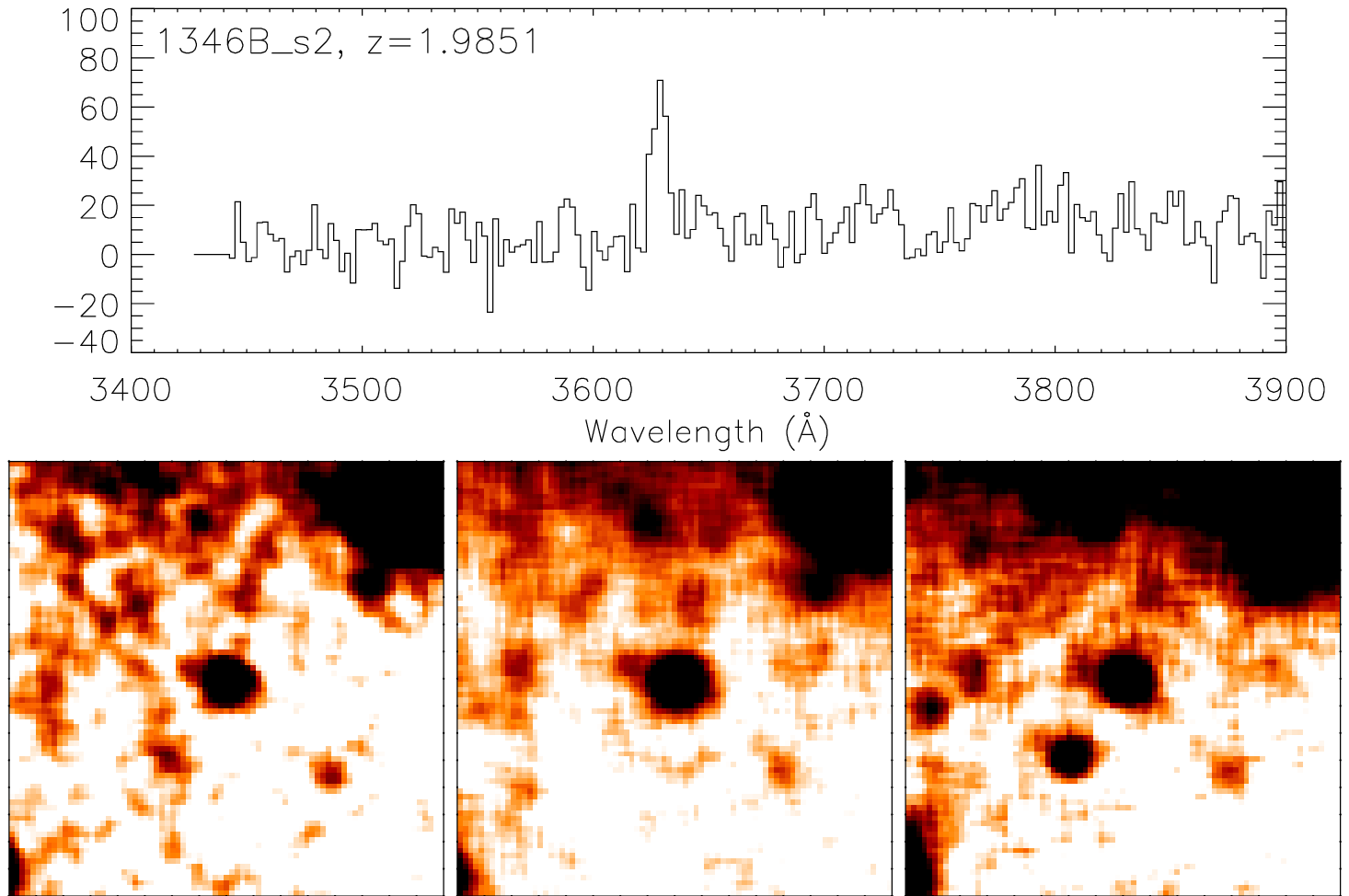, width=5.5cm}
\epsfig{file=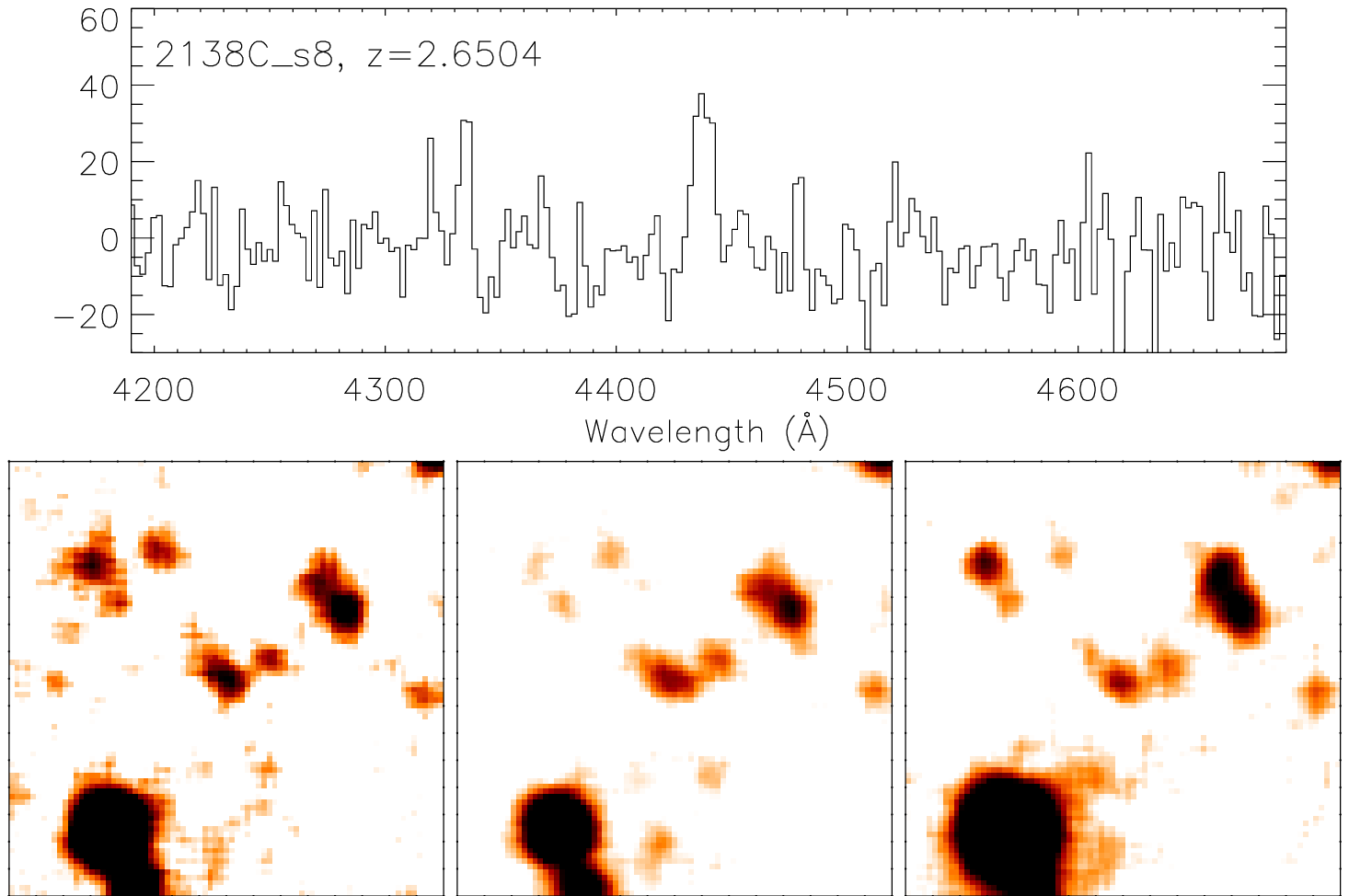, width=5.5cm}
\epsfig{file=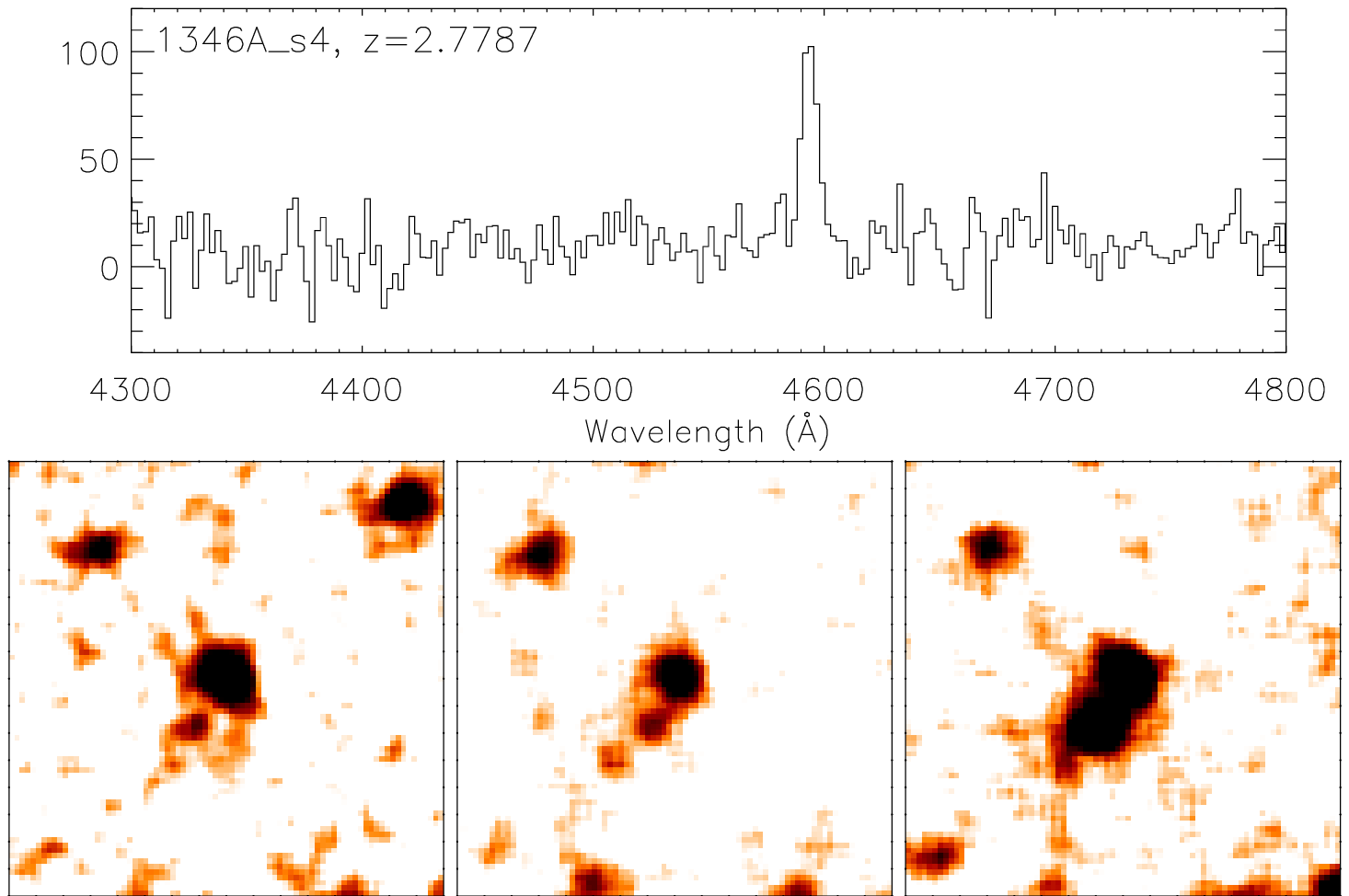, width=5.5cm}\\
\vskip 0.2cm
\epsfig{file=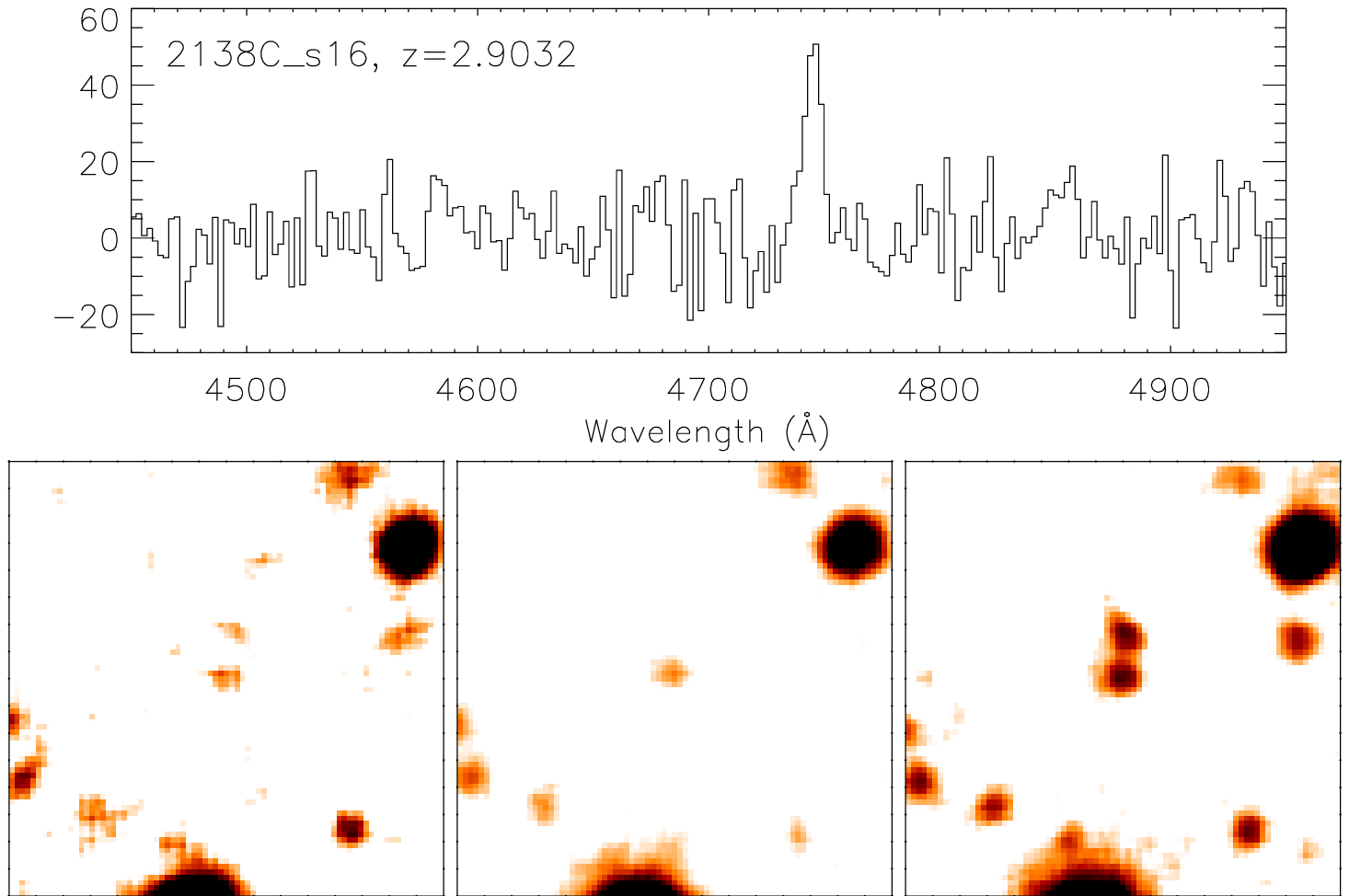, width=5.5cm}
\epsfig{file=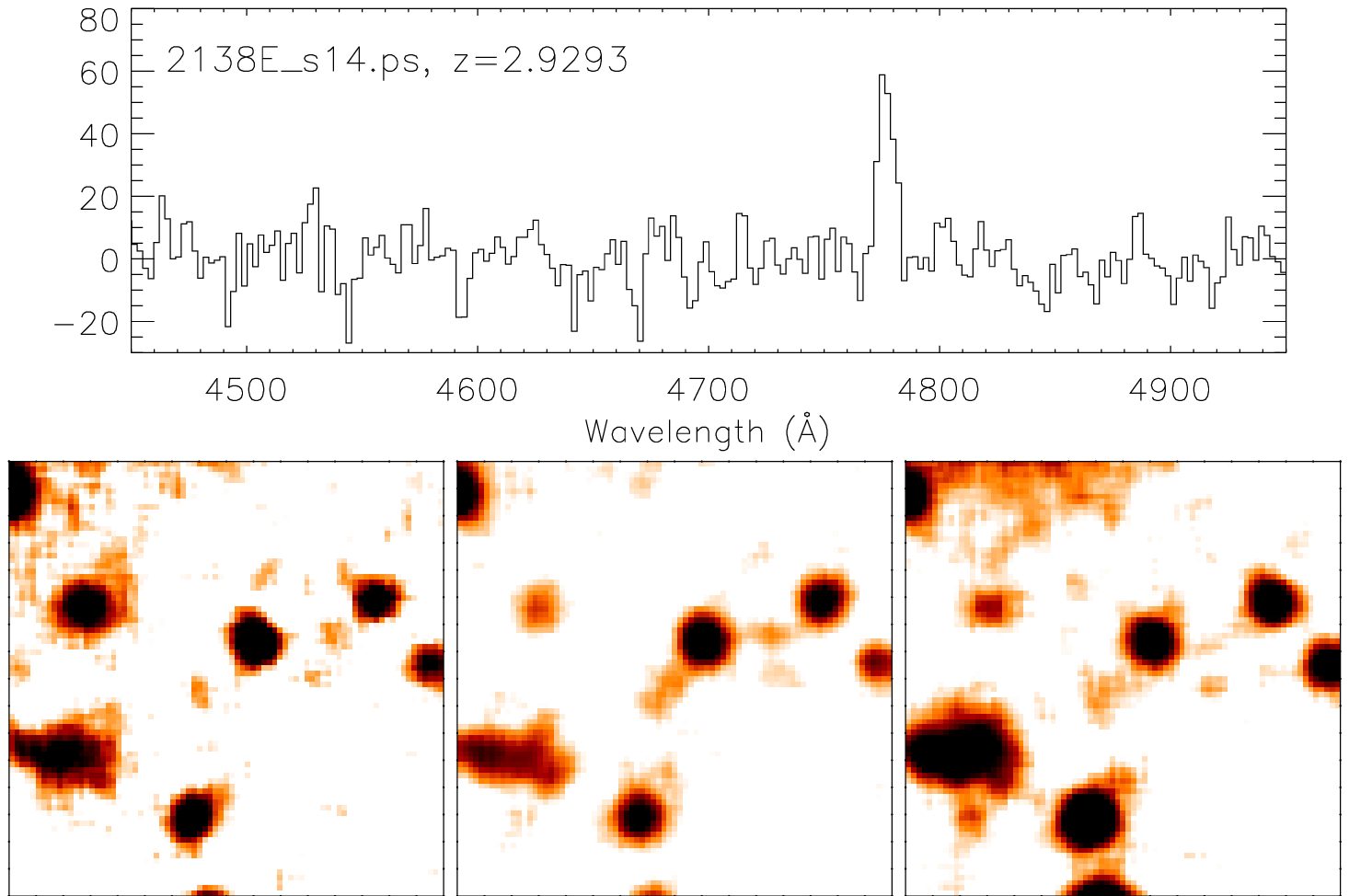, width=5.5cm}
\epsfig{file=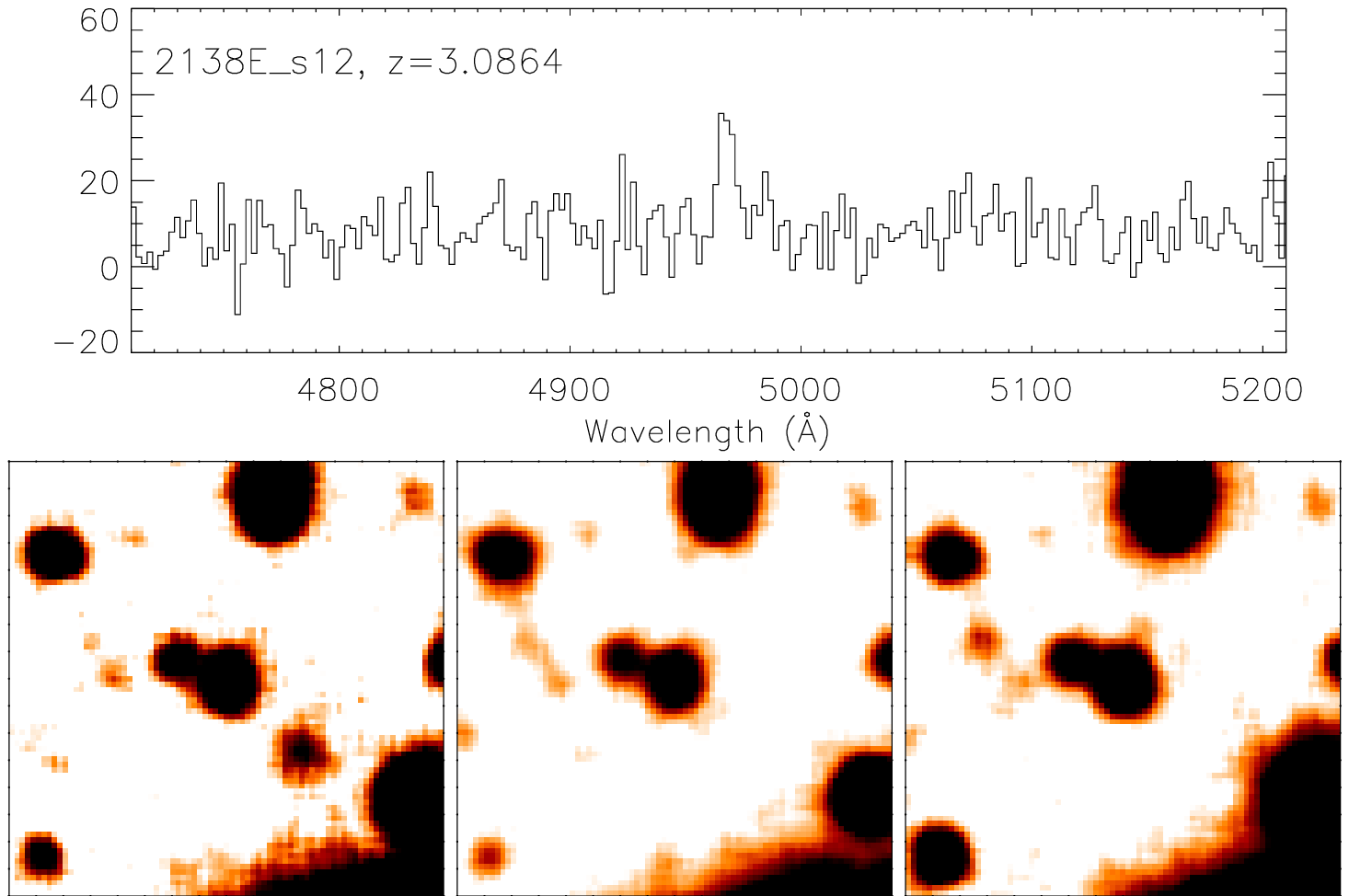, width=5.5cm}\\
\vskip 0.2cm
\epsfig{file=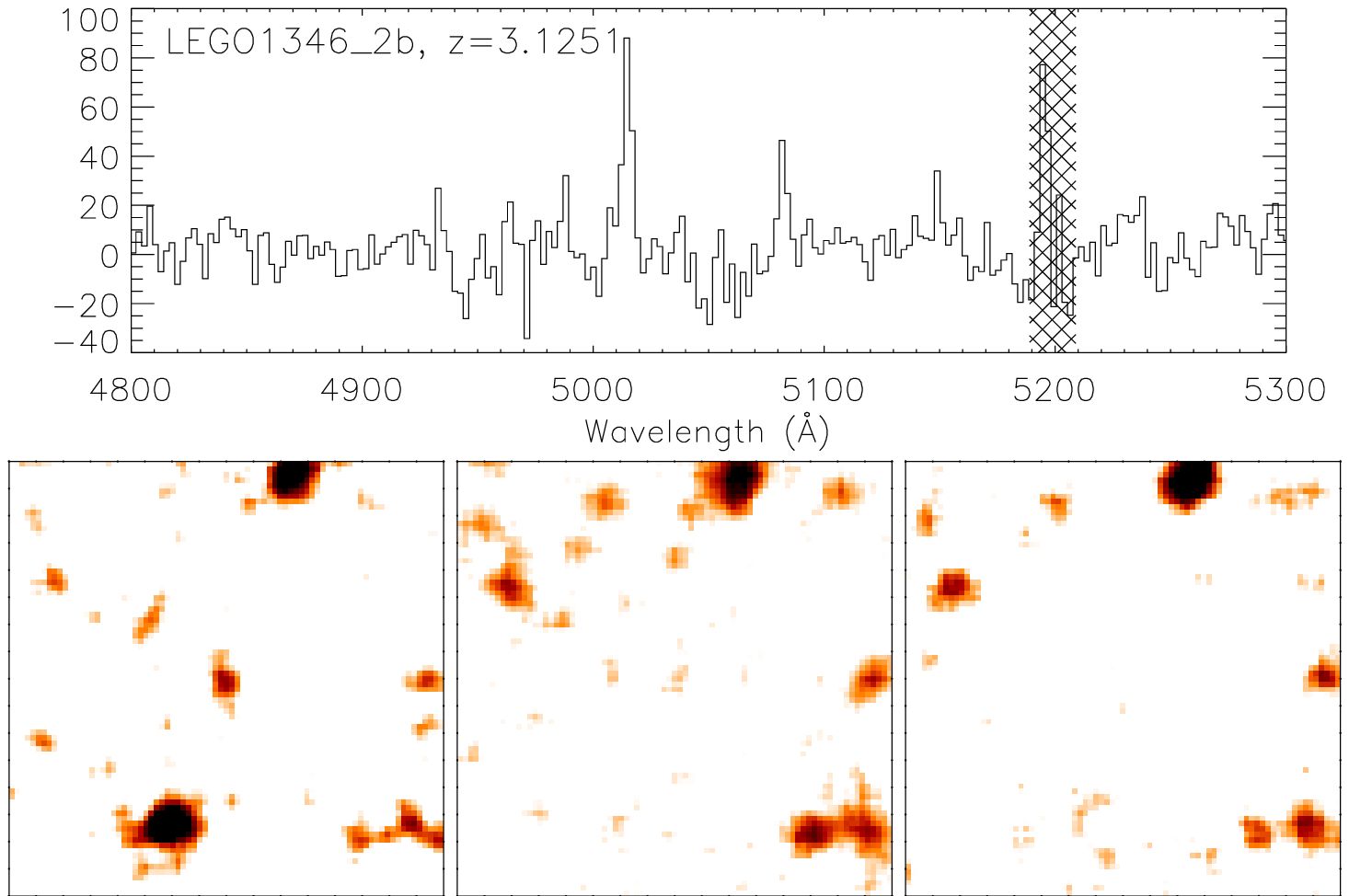, width=5.5cm}
\epsfig{file=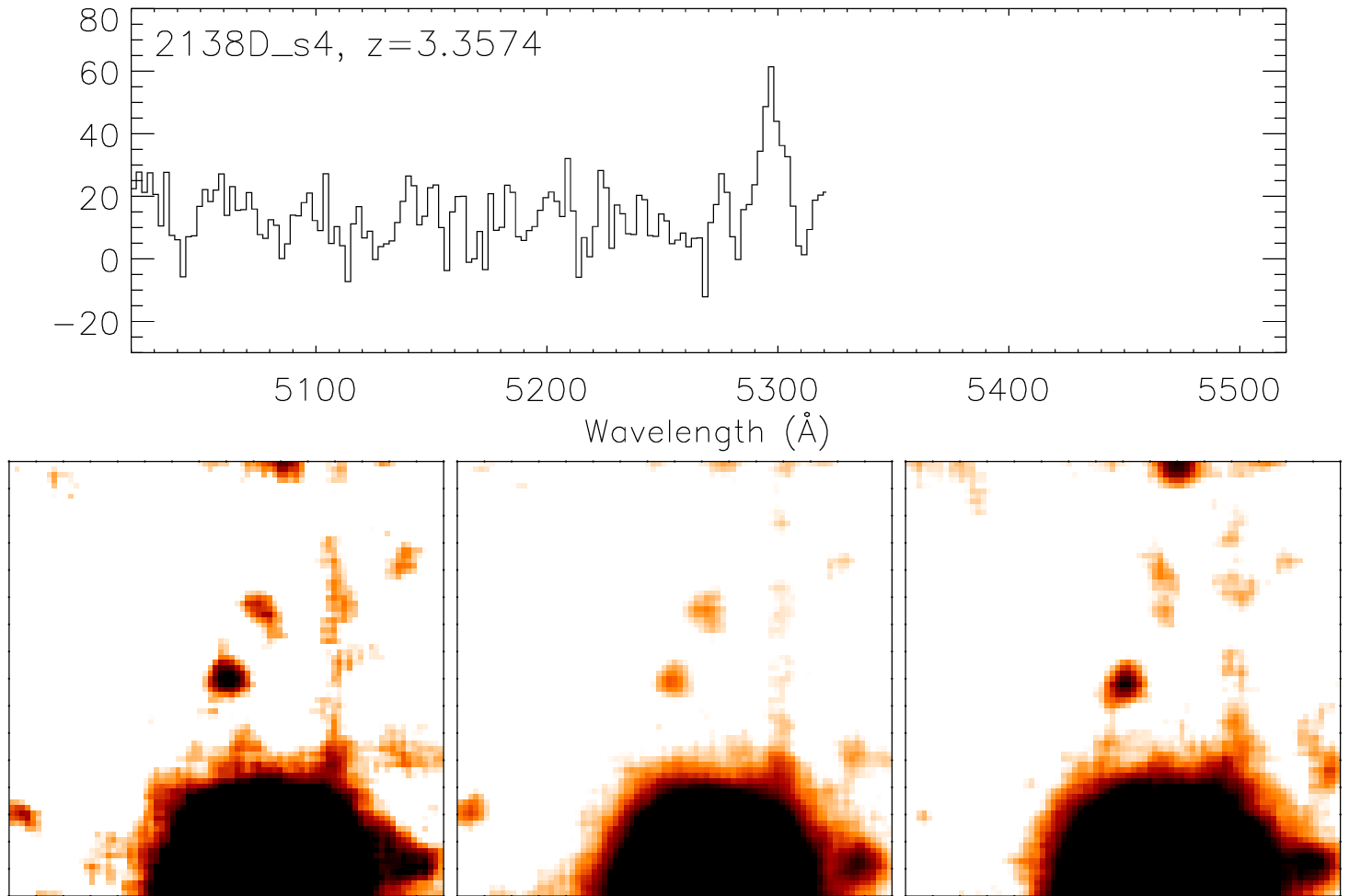, width=5.5cm}
\epsfig{file=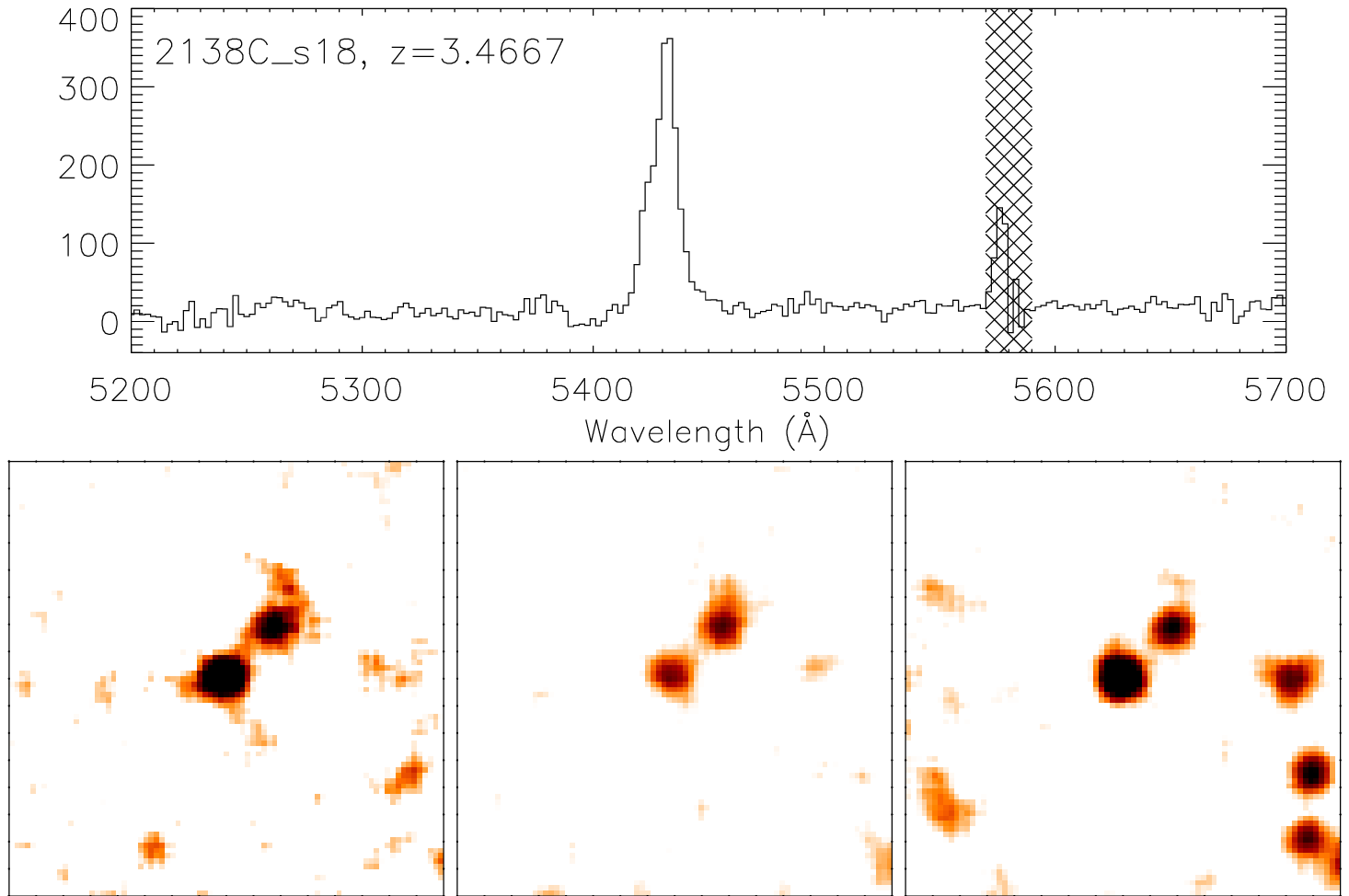, width=5.5cm}
\caption{
Shown are 16$\times$16 arcsec$^2$ images and 1-D spectra of serendipitously
detected, probable LEGOs. The spectra are ordered by redshifts that range from
z=1.98 to z=3.47. Note that the object LEGO1346\_2b located at z=3.1251 is a
neighbour to the z=3.1301 LEGO1346\_2 seen in the lower left corner of the
same image.
}
\label{serendipitous}
\end{center}
\end{figure*}

In each mask about half of the available 19 slitlets could be used
on candidates. The remaining slitlets were placed either on stars to check 
the slit alignment, on faint galaxies, or on regions of blank sky.
We have carefully analyzed all the 2D-spectra for other emission line
sources that could be serendipitously detected LEGOs at
other redshifts (see e.g. Manning et al. 2000 for other such
cases). In addition to a number of [\ion{O}{ii}] emitters we detect
a significant number of emission line sources that are most likely
Ly$\alpha$ emitters based on their extremely faint continua and
the lack of other lines in the spectra. The spectra and images of these
are shown in Fig.~\ref{serendipitous}. One of the sources, called
LEGO1346\_2b, a is a neighbour to the z=3.1301 LEGO1346\_2 and has a 
redshift within the narrow filter used for the BRI\,1346$-$0322 field, 
z=3.1251. 
However, in the narrow-band image it is fainter than our 5$\sigma$ cut. 
The serendipitously detected LEGOs have redshifts ranging from 1.98 to 
3.47.

\section{Discussion}
\label{discuss}
The main result of this paper is that the selection method we have 
applied has proven to work. Clearly, deep narrow-band imaging combined 
with follow-up MOS spectroscopy is an efficient method to build up a 
significant sample of spectroscopicallly confirmed high-redshift galaxies 
below the flux-limit in ground based surveys for LBGs. 
Compared to selecting U-dropout galaxies from 
deep HST images such as the Hubble Deep Fields or the Chandra
Deep Fields this technique is much cheaper and can be applied 
over much larger fields securing that large samples can be built
with fairly modest amounts of observing time. Obviously, we only 
select objects with strong emission lines and samples selected
based on the Ly$\alpha$ will therefore not be complete to the 
continuum limit of the survey. For the bright LBGs in ground 
based surveys only 20--30\% are Ly$\alpha$ emitters with a 
rest-frame EW above 20 \AA . The fraction of strong 
Ly$\alpha$ emitters could be larger at the faint end of the 
luminosity function. However, Steidel et al. (2000) found that the fraction 
of Ly$\alpha$ emitters was not significantly higher at the faint
end of a sample of LBGs and narrow-band selected galaxies at z=3.09.
Shapley et al. (2003) also argue that a constant fraction of 
Ly$\alpha$ emitters down to $\textrm{R}=25.5$ is consistent with
the data when selection effects are taken into account.

It is interesting to ask how the LEGOs compare with the LBGs at the 
bright end of the luminosity function. Given the importance of dust for the
escape of Ly$\alpha$ emission from galaxies the naive expectation 
would be that Ly$\alpha$ emitters are drawn from the younger, less 
chemically evolved and in general faint end of the population of 
high-redshift galaxies (Fynbo et al. 2001; Malhotra \& Rhoads 2002).
Contrary to this expectation, Shapley et al. (2001) suggest, based on 
rest-frame optical properties of LBGs, that the Ly$\alpha$ emitting 
sub-sample of the LBGs are the oldest objects among the Lyman-break 
selected galaxies. In this scenario LBGs start out in a dusty burst 
phase for 50--100 Myr followed by a more quiescent phase with less 
extinction for several hundred Myr up to a Gyr. However,
most of the Ly$\alpha$ emitters in our samples are too faint to be included
in the LBG samples and it is therefore likely that the properties of the
Ly$\alpha$ emitting LBGs are different from LEGOs in general. 

During most of the previous decade Ly$\alpha$ emission was considered
an inefficient survey method for high-redshift galaxies due to a number
of unsuccessful surveys (e.g. Prichet 1994 and references therein). It is
now clear that the first surveys for Ly$\alpha$ emitters were unsuccessful 
mainly because they reached significantly too shallow detection limits. 
The theoretical expectation was that todays large ellipticals formed 
in a fast, monolithic collapse (e.g. Patridge \& Peebles 1967). In the
hierarchical picture of galaxy formation the high-redshift galaxies are
smaller and hence fainter than expected when the first surveys were 
planned.

The main advantage of LBG surveys is that 
they probe a very large volume and hence provide a large number of galaxies 
per field.  However, there is a number of studies that are most efficiently
done with LEGOs as probes: {\it i)} LEGOs can be used to probe the
faint end of the luminosity function (Fynbo et al. 2001); {\it ii)} 
LEGOs can be detected and spectroscopically confirmed at both lower 
(Fynbo et al. 1999, 2002) and higher redshifts (Dey et al. 1998; Ellis 
et al. 2001; Venemans et al. 2002; Hu et al. 2002; 
Taniguchi et al. 2003) than is currently possible with techniques
based on the continuum; {\it iii)} the large space density reachable with 
surveys for LEGOs allows a detailed study of the underlying large scale 
structure and to probe the environments of other high-redshift objects such 
as radio galaxies 
(Kurk et al. 2000; Venemans et al. 2002), Gamma Ray Burst host galaxies 
(Fynbo et al. 2002) or QSO absorbers (e.g. M\o ller \& Warren 1993, 
Francis et al. 1995; and this paper).

The optimal way to proceed with Ly$\alpha$ surveys seems to be the use 
of large area cameras on
8-m class telescopes. First results regarding the luminosity function and 
clustering properties of LEGOs using the Suprime Camera (Miyazaki et al. 
2002) on SUBARU has been reported by
(Ouchi et al. 2003), however that survey is not as deep as the survey 
presented here and it targets a higher redshift (z=4.86). A narrow-band image 
targeting z=3 LEGOs to the same depth as in the present survey, but obtained 
with the Suprime Camera will provide of order 500 candidates per field. 
Furthermore, it is mandatory that candidates based on narrow-band imaging are 
subsequently confirmed (or rejected) based on spectroscopy to make sure that 
conclusions based on surveys of LEGOs can be trusted. A major reason for the 
success of the Lyman-Break surveys is the spectroscopic confirmation of most 
of their candidates. Studies of LEGOs should follow this good example.
Furthermore, the measurement of several hundred redshifts in one field can be 
used to map out the underlying filamentary structure  
(M\o ller \& Fynbo 2001) and even provide an independent measurement of the 
cosmological constant (Weidinger et al. 2002).

\section*{Acknowledgments}
We thank our referee C. Steidel for comments that help us improve the 
discussion and the Paranal staff for excellent support during
the visitor run in July 2002. JPUF and CL gratefully acknowledge the 
receipt of an ESO research fellowship. We acknowledge helpful discussions
with Bruno Leibundgut, Nicolas Bouche, Masami Ouchi and Vincenzo Mainieri. 
JPUF gratefully acknowledge support from the Carlsberg Foundation.

\end{document}